\documentclass[a4paper,fleqn,usenatbib,useAMS]{mnras}

\usepackage{graphicx}	% Including figure files
\usepackage{amsmath}	% Advanced maths commands
\usepackage{amssymb}	% Extra maths symbols
\usepackage{multicol}  
\usepackage{booktabs}
\usepackage{bm}		% Bold maths symbols, including upright Greek
\usepackage{pdflscape}	% Landscape pages

\newcommand{\ra}[1]{\renewcommand{\arraystretch}{#1}}

\usepackage[T1]{fontenc}
\usepackage{ae,aecompl}

\usepackage{newtxtext,newtxmath}
\title[Radial Tidal Alignment]{A Radial Measurement of the Galaxy Tidal Alignment Magnitude with BOSS Data}

\author[Martens et al.]{Daniel Martens$^{1}$, Christopher M. Hirata$^{1}$, Ashley J. Ross$^{1}$, Xiao Fang$^{1}$
\\
$^{1}$Center for Cosmology and AstroParticle Physics, Department of Physics, The Ohio State University, 191 W Woodruff Ave, \\  Columbus OH 43210, U.S.A.
}

\date{21 February 2018} % 

\pubyear{2018}

\begin{document}
\label{firstpage}
\pagerange{\pageref{firstpage}--\pageref{lastpage}}
\maketitle

\begin{abstract}
The anisotropy of galaxy clustering in redshift space has long been used to probe the rate of growth of cosmological perturbations. However, if galaxies are aligned by large-scale tidal fields, then a sample with an orientation-dependent selection effect has an additional anisotropy imprinted onto its correlation function.
%Cosmological parameters can be measured through the observation of redshift space distortion (RSD) effects using galaxy clustering analyses. However, gravitational tidal fields can influence galaxy intrinsic alignments in ways that can mimic redshift space distortions. Specifically, observational selection effects acting on tidally aligned galaxies can cause a difference in Fourier modes measured parallel and perpendicular to the line of sight, through the projection of galaxy orientation.
We use the LOWZ and CMASS catalogs of SDSS-III BOSS Data Release 12 to divide galaxies into two sub-samples based on their offset from the Fundamental Plane, which should be correlated with orientation. These sub-samples must trace the same underlying cosmology, but have opposite orientation-dependent selection effects. We measure the clustering parameters of each sub-sample and compare them in order to calculate the dimensionless parameter $B$, a measure of how strongly galaxies are aligned by gravitational tidal fields. We found that for CMASS (LOWZ), the measured $B$ was $-0.024 \pm 0.015$ ($-0.030 \pm 0.016$). This result can be compared to the theoretical predictions of \cite{hirata2009}, who argued that since galaxy formation physics does not depend on the direction of the ``observer,'' the same intrinsic alignment parameters that describe galaxy-ellipticity correlations should also describe intrinsic alignments in the radial direction. We find that the ratio of observed to theoretical values is $0.51\pm 0.32$ ($0.77\pm0.41$) for CMASS (LOWZ). We combine the results to obtain a total ${\rm {Obs}/{Theory}} = 0.61\pm 0.26$. This measurement constitutes evidence (between 2 and 3$\sigma$) for radial intrinsic alignments, and is consistent with theoretical expectations ($<2\sigma$ difference).
%This can be compared to a measurement of $-0.0348 \pm 0.0038$ from \cite{singh2015}, who use galaxy-galaxy lensing to obtain the alignment magnitude.
\end{abstract}


%%%%%%%%%%%%%%%%%%%%%%%%%%%%%%%%%%%%%%%%%%%%%%%%%%
%%%%%%%%%%%%%%%%% BODY OF PAPER %%%%%%%%%%%%%%%%%%
\section{Introduction}
\label{sec:introduction}

%\flagthis{Introductory paragraphs much improved; there are still a couple minor things to discuss.}
% \flagthis{Re-phrase introductory paragraphs, clearly distinguishing peculiar velocities (an effect on specific galaxies) from RSD (an effect on maps of large scale structure). You have part of the history here, e.g.\ the Kaiser paper showing how to use RSDs as a cosmological probe, but in the modern era RSDs are no longer a competitive way to measure $\Omega_m$ (and may never be); starting in the late 2000s the main emphasis was on testing the standard model of growth of structure vs.\ e.g.\ modified gravity.} Fixed - please comment if it needs more work! -Daniel

% \xiao{This is a general comment. Throughout the paper there are terms ``medians'' and ``means'' being used, but do you only mean ``means''? I know they are the same for symmetric distributions, just want to make sure..}
%Yes xiao, they are being used intentionally. we subtract the median from the fits before setting up our groups. 

Galaxy peculiar velocities have long been known as a source of noise in using the redshift of a galaxy to infer its distance from the observer. When redshift surveys are used to map the large-scale structure of the Universe, these peculiar velocities leave artifacts known as redshift space distortions (RSDs). On small scales, RSDs lead to ``fingers of God'' -- structures that are smeared in redshift space by their internal velocity dispersion rather than by their physical size, and hence appear to be pointed at the observer \citep{jackson1972}. This smearing increases the difficulty of making precise measurements in redshift space. However, on large scales, RSDs are also on a short list of invaluable cosmological probes. Matter overdensities cause infall of galaxies on these scales, which results in a ``squashing'' effect when viewed in redshift space. Measurements on the magnitude of this large-scale infall can reveal information on the clustering of matter in the Universe, even out to linear scales. 

Early galaxy surveys \citep{kirshner1981,bean1983} lacked the number density of sources to well quantify the effects of RSDs, as they were limited to a few thousand galaxies, although attempts were made to model them \citep{davis1983}. By the 1990's, however, improved spectroscopic techniques led to a higher number density in galaxy surveys, which provided greater empirical evidence for mapping nonlinear effects \citep{kaiser1986B}. The uniformity of the Universe on scales above approximately 100$h^{-1}$ Mpc was seen in tandem with a complicated network of non-linear behaviors on smaller scales. While on small scales, RSDs trace the velocity dispersion of the galaxies, the RSD signal on scales large compared to the Finger of God length does not become negligible; instead it traces linear-regime infall into potential wells. The distortions on these larger scales can measure the matter density of the Universe, $\Omega_{m}$, as described in detail by \cite{kaiser1987}. Following this, \cite{hamilton1992} derived formulas to measure $\Omega_{m}$ given the characteristic anisotropic quadrupole of the correlation function.

These methods have been applied to increasingly larger surveys, primarily producing measurements for the RSD parameters themselves. \cite{hamilton1993} applied these to the Infrared Astronomical Satellite (IRAS) 2 Jy, observing 2,658 galaxies, and finding that $\Omega_{m} = 0.5^{+0.5}_{-0.25}$. \cite{cole1995} measured redshift-space distortions for 1.2-Jy and QDOT surveys, and compared their results to N-body simulations to find points of breakdown with linear theory, further studied by \cite{loveday1996} on the Stromlo-APM redshift survey. The optically selected Durham/UKST Galaxy Redshift Survey pursued similar measurements \citep{ratcliffe1998}. Within recent years, galaxy samples have become larger, and errors on measured parameters have become correspondingly smaller. \cite{peacock2001} used the 2dF Galaxy Redshift Survey to measure redshift-space distortions from 141,000 galaxies, and detected a large-scale quadrupole moment at greater than 5-sigma significance. This result was expanded by \cite{verde2002} through incorporating the galaxy bispectrum to place a measurement on the matter density, finding $\Omega_{m} = 0.27 \pm 0.06$. Currently, RSD clustering measurements serve to test the standard models of the growth of structure. In this endeavor, larger samples of galaxies have been analyzed using higher redshift surveys \citep{ross2007,guzzo2008}, 6dFGS \citep{beutler2012}, WiggleZ \citep{blake2012}, and SDSS and BOSS \citep{tegmark2004,okumura2009,dawson2013,chuang2016,satpathy2016,ross2016,marin2016,beutler2017}. RSD measurements have also been conducted for high-redshift Lyman break galaxies \citep{angela2005,mountrichas2009,bielby2013}. Finally, results from these analyses have been combined with CMB observations \citep{tegmark2006,alam2017} to measure the full range of cosmological parameters.

As galaxy survey sizes grow and the precision on RSD measurements increases, it is important to keep track of and mitigate potential sources of error. One such source of error, discussed initially by \cite{hirata2009}, is caused by the intrinsic alignment of galaxies due to large-scale tidal fields. Luminous red galaxies (LRGs), which have been targeted extensively by recent surveys, are preferentially aligned along the stretching axis of the local tidal field, which when combined with observational selection effects based on orientation, results in differences in the observed density modes, depending on if the mode is parallel, or perpendicular to, the line of sight. Specifically, a viewing dependent selection effect of observing more galaxies which are pointed toward us can result in an increase in perpendicular k-modes and decrease of parallel $k$-modes. \cite{hirata2009} estimates that this effect could result in the contamination of RSD measurements by 5--10 percent, depending on specifics of the redshift and mass distribution of the galaxy sample in question. Related effects have been studied in the context of other galaxy samples (e.g.\ Lyman-$\alpha$ emitters, \citealt{zheng2011,behrens2014}) and of higher-order statistics \citep{krauss2011}.

Intrinsic alignments of luminous red galaxies are correlations between their respective orientations, shapes, and physical positions. The development of these alignments is a complicated, non-linear process associated with the history of galaxy formation, but the payoff for cosmology if they are understood can be huge \citep{chisari2013}. Proposed alignment mechanisms include linear alignment by large-scale tidal fields \citep{catelan2001,hirata2004}, large clusters of galaxies \citep{thompson1976,ciotti1994}, or tidal torque contributions to galaxy angular momentum \citep{peebles1969}. LRGs specifically have had their intrinsic alignment amplitudes analyzed both in observations \citep{hirata2007,bridle2007,okumura2009,blazek2011,chisari2014} and in simulations \citep{velliscig2015,chisari2015,tenneti2016,hilbert2017}. Analytical models for describing these processes have been developed and expanded upon \citep{blazek2015,blazek2017}. Their effects on weak lensing has been, so far, the greatest motivation for their study \citep{troxel2015}. Intrinsic alignments have been measured using galaxy-galaxy lensing \citep{mandelbaum2006,joachimi2011,huang2016,uitert2017,huang2018,tonegawa2017} for the LOWZ sample \citep{singh2015} which we compare to in this work.

In this paper, we use the fundamental plane of elliptical galaxies to determine, in a statistical way, which galaxies are intrinsically aligned along or across our line of sight. We then divide the BOSS catalogs into subsamples of galaxies in each orientation classification. Our aim is to obtain a measurement of the linear alignment amplitude $B$ using radial alignment measurements, to complement existing shear measurements of the same quantity. We perform this analysis using data from the Baryon Oscillation Spectroscopic Survey (BOSS; \citealt{dawson2013}), which was observed as part of the Sloan Digital Sky Survey - III \citep{eisenstein2011}. Specifically, we use both the ``LOWZ'' and ``CMASS'' catalogs \citep{SDSSDR12} from Data Release 12 (DR12; \citealt{alam2015}). We divide these catalogs into two separate populations depending on each galaxy's estimated orientation relative to us, as assessed using the offset from the Fundamental Plane. Next, we perform clustering measurements to determine RSD parameters for each group, testing the predictions of this effect by \cite{hirata2009}, and comparing our radial measurement of $B$ to a perpendicular measurement using galaxy ellipticities. 

This paper is organized as follows. In Section \ref{sec:theory}, we discuss the theory behind the galaxy intrinsic alignments and how they mimic redshift space distortions. Specifically, in Section \ref{sec:fundyplane}, we discuss the Fundamental Plane, and how we will use it to determine each galaxy's orientation relative to our line of sight. Section \ref{sec:dataset} describes the BOSS DR12 dataset -- both the catalogs themselves, the selection choices we have used for our sample, and our methods of blinding and systematics tests. We discuss in Section \ref{sec:correlation} the methods by which we calculate the clustering statistics on our samples, as well as how we generate our covariance matrices. Finally, in Section \ref{sec:analysis} we discuss how we fit RSD parameters to our samples, and analyze the results. We conclude in Section \ref{sec:conclusion}. 

Unless noted otherwise, we use the following cosmological parameters: $\Omega_{\Lambda}=0.693$, $h=0.68$, $\Omega_{b} = 0.048$, $\Omega_{m} = 0.307$, $n_{s} = 0.96$, $\sigma_{8} = 0.83$, and $T_{\rm cmb} = 2.728\,$K \citep{ade2014}.

%%%%%%%%%%%%%%%%%%%%%%%%%%%%%%%%%%%%%%%%%%%%%%%%%%%%%%%%%%%%%%%%%%%%%%%%%%%%%%%%%%%%%%%%%%%%%%
%%%%%%%%%%%%%%%%%%%%%%%%%%%%%%%%%%%%%%%%%%%%%%%%%%%%%%%%%%%%%%%%%%%%%%%%%%%%%%%%%%%%%%%%%%%%%%
\section{Theory}
\label{sec:theory}

\subsection{Redshift space distortions}
There are two main effects which change the na\"ive linear power spectrum in redshift space. Both of these are due to changes in radial velocities of observed galaxies, although the effects differ in both their sources and their end results. First, peculiar velocities of galaxies cause a random shift in their observed radial velocity, with some galaxies moving toward us and others moving away. This causes a spread in the observed distribution of galaxies on small scales, creating an incoherent ``Finger of God'' effect, where groups of galaxies appear to be pointing toward the observer, when viewed in redshift space. Although this effect has long been understood, a second effect detailing a coherent redshift space distortion was described by \cite{kaiser1987}. On larger scales, galaxies are pulled into local gravitational overdensities, leading to a squashing effect in redshift space, which is a significant correction to the redshift space galaxy correlation function, even on linear scales.

These distortions affect measurements along the line of sight, and so changes to the linear power spectrum are a function of $\mu$, the cosine of the angle between a given direction and the line of sight. We can write the observed overdensities of galaxies in Fourier space as:
\begin{equation}
\delta_{\rm g}(\bmath{k}) = (b+f\mu^{2})\delta_{\rm m}(\bmath{k}),
\end{equation}
where $\delta_{\rm g}$ and $\delta_{\rm m}$ refer to perturbations measured by galaxy tracers and matter, respectively, $b$ is the linear galaxy bias, and $f = {\rm d}\ln G/{\rm d}\ln a$, where $G$ is the growth function. It can be seen that we recover the matter power spectrum multiplied by the tracer bias if we are looking at modes transverse to the line of sight. This equation can also be expressed as the relationships between the matter and galaxy tracer power spectra: 
\begin{equation}
P_{\rm g}(\bmath{k}) = (b+f\mu^{2})^{2}P_{\rm m}(\bmath k)
\end{equation}
It is common to measure the amount of anisotropy by the parameter $\beta = f/b$. 
%%%%%%%%%%%%%%%%

\subsection{Intrinsic alignment effects on RSD measurements}
\label{ss:ia-rsd}

\begin{figure*}
\centering
\includegraphics[scale=0.6]{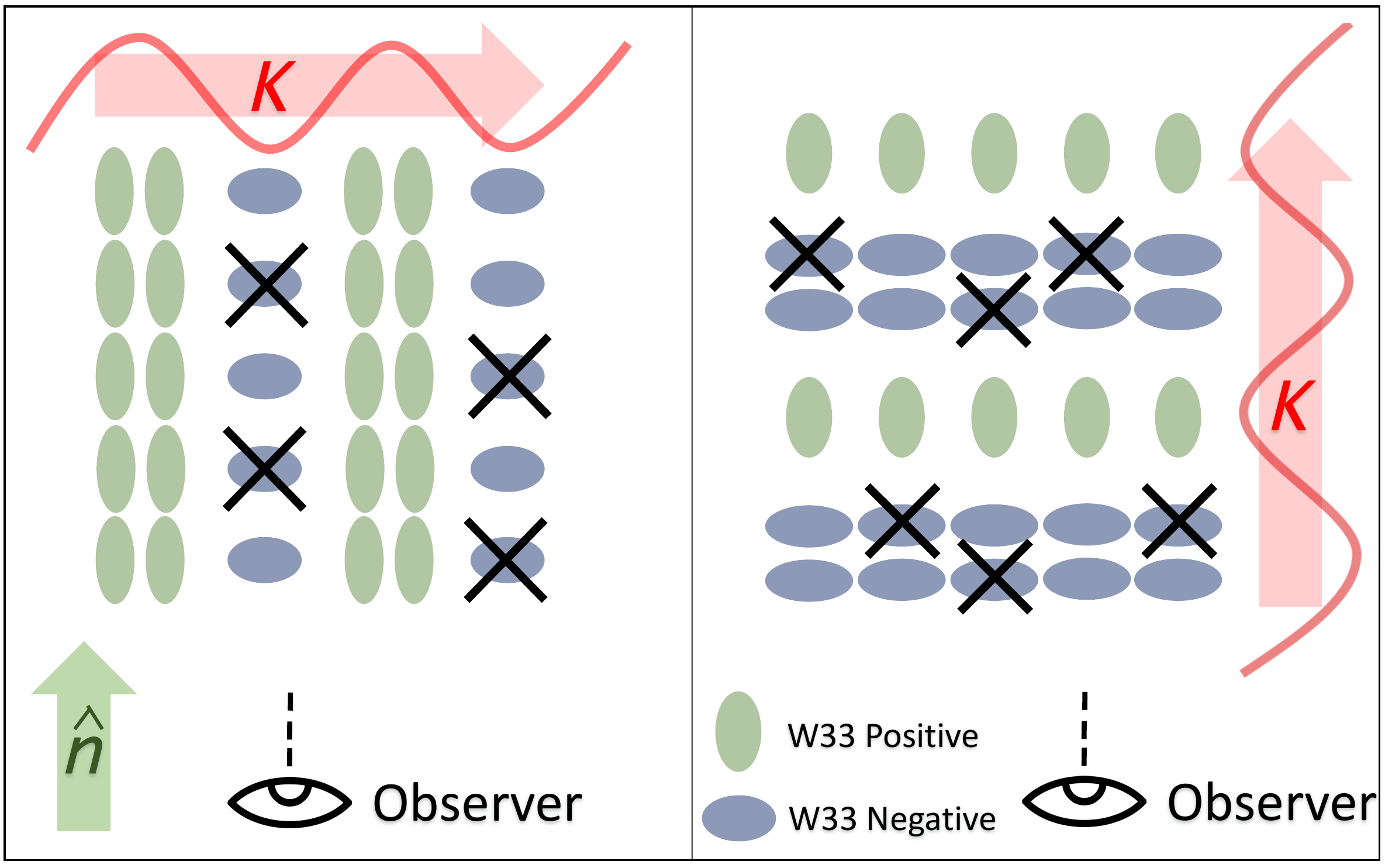}
\caption{Here we show an exaggerated situation demonstrating a selection bias having an anisotropic effect on the resulting $k$-modes. If we assume that our observations are less likely to observe galaxies which are aligned perpendicularly to our line of sight, then some of the galaxies visualized above will not appear to the observer (the crossed-out galaxies). Since galaxies are more likely to be aligned with the higher density modes of the gravitational field, this changes the measured $k$-modes of the observed galaxies. Fourier modes perpendicular to the line of sight will have deeper troughs, and thus a higher inferred amplitude, while $k$-modes parallel to the line of sight will have shallower peaks, and thus a lower inferred amplitude. In this work, we intentionally group galaxies by their orientation, shown here by the color of green or blue. If we were to view all of these galaxies, but were less likely to see galaxies of the `'`blue'' type, then these distortions would dampen the linear Kaiser effect of clustering in our observation. (Note: this figure is similar to Figure 1 of \protect\cite{hirata2009}, but adapted to the situation with two sub-samples.)}
\label{fig:perp_figure}
\end{figure*}

\citet[hereafter H09]{hirata2009} showed that systematic errors to redshift space distortion measurements can occur through the intrinsic alignment of galaxies by large scale tidal fields. It requires two conditions to be met for the galaxy sample in question. First, galaxies must be intrinsically aligned along the stretching tidal field axis (``linear alignment'' in the nomenclature of intrinsic alignment theory). Second, the sample must have a selection effect that depends on the galaxy orientation relative to the line of sight. The systematic error in $f$ depends on the product of these effects.
In this paper, we are deliberately enhancing the selection effects by splitting the BOSS sample into subsamples using a proxy for orientation. This enables us to use the difference in inferred $f$ from the subsamples to probe intrinsic alignments.

It can be seen in a physical sense how alignments can lead to these systematic effects by understanding their resulting changes to observed Fourier density modes, as shown qualitatively in Figure \ref{fig:perp_figure}. The anisotropic selection has a different effect on density modes depending on the angle of the mode with the line of sight directional vector. Selection of galaxies which are aligned perpendicular to the line of sight of the observer leads to a further decrease in the amplitude of troughs in $\bmath k$-modes perpendicular to the line of sight, which serves to amplify those $\bmath k$-modes. Conversely, this selection will lead to a decrease in the amplitude of peaks in $\bmath k$-modes parallel to the line of sight, resulting in a smoothing of these $\bmath k$-modes. This anisotropic effect mimics clustering signals by amplifying modes in different ways along the line of sight than transverse to it. This is the ``physical picture'' which serves to illustrate how redshift-space distortion measurements can be biased high or low, depending on the galaxy selection effect in question.

To describe this effect quantitatively, we give a quick synopsis of the analysis of H09 and highlight the results needed for understanding LRG clustering (we refer the reader to H09 for additional details on the formalism).
%ir resulting equations describing this process. Our notation will closely follow theirs; for a more detailed understanding of the theoretical aspects of this calculation, we encourage a thorough reading of H09. First, we define a function describing the probability of a galaxy being selected, given a viewing direction:
%\begin{equation} \label{eq: epsilon}
%\epsilon (\hat{\textbf{n}} | \textbf{x}) \equiv \int P(\textbf{Q} | \textbf{x}) \Gamma (\textbf{Q} \hat{\textbf{n}},\textbf{x}) d^{3} \textbf{Q}
%\end{equation}
%where we have defined $P(\textbf{Q} | \textbf{x})$ as the conditional probability distribution for a specific orientation of a galaxy, $\bf{Q}$, given that galaxy's position $\textbf{x}$. In the absence of tidal pull on galaxy orientation, this probability distribution would be a constant. $\Gamma (\textbf{Q} \hat{\textbf{n}},\textbf{x})$ is the excess probability of a galaxy being observed, given its orientation and specific geometry. Equation \ref{eq: epsilon} therefore represents the viewing direction dependent selection function. If the galaxy distribution is unaffected by tidal fields, this term integrates to zero; similarly, if there is no excess probability to be observed given a specific orientation, this function vanishes. By incorporating this function into our standard equation for overdensities, and setting the observation direction $\hat{\textbf{n}} = \hat{e_{3}}$, since it is in the radial direction that we see redshift space distortions, we find:
We denote the contribution of anisotropic selection effects to the galaxy overdensity by $\epsilon$ -- that is, the observed number density of galaxies at point $\bmath x$ as seen from direction $\hat{\bmath n}$ is taken to be $1+\epsilon(\hat{\bmath n}|\bmath x)$ times the number density that would be observed if the galaxy orientations were randomized. At linear order,
\begin{equation}
\delta_{\rm g}(\bmath{k}) = (b+f\mu^{2})\delta_{\rm m}(\mathbf{k}) + {\epsilon}(\hat{\bmath e}_{3}|\bmath{k}),
\end{equation}
where we take the convention that the observer's line of sight is the 3-axis.

In the linear tidal alignment model (relevant for LRGs), the intrinsic alignments trace the large-scale tidal field:
\begin{equation}
\epsilon (\hat{\bmath{n}}|\bmath{x}) = A s_{ij}(\bmath{x})\hat{n}_{i}\hat{n}_{j},
\end{equation}
Here $A$ is a biasing parameter, which depends on both the intrinsic alignments and the observational selection effects (we explore this in more detail below), and $s_{ij}$ is the dimensionless tidal field:\footnote{Since $s_{ij}$ is traceless-symmetric, it follows mathematically that $\epsilon (\hat{\bmath{n}}|\bmath{x})$ averages to zero if we average over viewing directions $\hat{\bmath{n}}$.}
\begin{equation}
s_{ij}(\mathbf{x}) = \left[ \nabla_{i}\nabla_{j}\nabla^{-2} - \frac{1}{3}\delta_{ij} \right]\delta_{\rm m}(\mathbf{x}).
\end{equation}
This leads to a modified equation for the power spectrum:
\begin{equation} \label{eq:powerspectrum}
P_{\rm g}(\mathbf{k}) = \left[ b - \frac{A}{3} + (f + A)\mu^{2}\right]^{2}P_{\rm m}(k).
\end{equation}
This equation, although within the confines of the assumptions we have made, is powerful. We have introduced a single parameter, $A$, to describe the effects of intrinsic alignment upon redshift space distortions. Note that the functional behavior of the distortions remains the same, but the intrinsic alignments have shifted the usual parameters in the model. Therefore, any systematics tests aiming to filter out effects of a different functional form will not be able to sift out this specific type of contamination.
%Specifically, if our catalog has an excess of galaxies where the tidal field is pressing galaxies along the line of sight, i.e. $A > 0$, then we see an increase in redshift space distortions, specifically the growth of structure parameter. Conversely, this parameter is suppressed for the galaxy sample selected on galaxies stretched along the line of sight.
In the presence of intrinsic alignment effects, the usual RSD measurement of $f$ can be re-interpreted as a measurement of $f+A$, with the ``standard'' assumption being that $A$ is small.
% For details on estimates of the $A$ parameter for different galaxy samples and measurement techniques, see H09.

In order to extract specifically the tidal alignment effect, we need to factor $A$ into pieces that depend on the observational selection effects (over which we have some control) and pieces that depend on intrinsic alignments (which we seek to measure). This means we need a mathematical description of the intrinsic alignments. Following H09, we treat elliptical galaxies as a triaxial system which is optically thin. For any position ${\bmath s}$ from the center of the galaxy, the volume emissivity is $j(\bmath {s}) = J(\rho)$, where $\rho$ is the ellipsoidal radius, and is defined by:
%\xiao{$ \bmath s \cdot \exp(-{\mathbfss W})\cdot\bmath s$}
%Xiao - I think the current way is correct - see for example Hirata's Tidal alignments paper 2009. Look at the appendix. If we have a dot product, it doesn't make sense to then do a dot product with a scalar, right? 
\begin{equation}
\rho^{2} = \bmath s \cdot \exp(-{\mathbfss W})\bmath s.
\label{eq:r2}
\end{equation}
Here ${\mathbfss W}$ is a $3\times 3$ traceless-symmetric matrix that contains information on the anisotropy of the galaxy, and $J(\rho)$ specifies the radial profile.\footnote{H09 worked only to linear order in ${\mathbfss W}$, and Sec.\ 5.1 of H09 defined $\rho$ with $({\mathbfss I}+{\mathbfss W})^{-1}$ instead of $\exp(-{\mathbfss W})$. For an observational paper working with real galaxies, we need to handle higher orders in ${\mathbfss W}$; the definition in Eq.~(\ref{eq:r2}) is more convenient as the total luminosity of the galaxy extrapolated to $\rho=\infty$ is independent of ${\mathbfss W}$.} For a spherical galaxy, ${\mathbfss W}={\mathbfss 0}$. The radial alignment is encoded in $W_{33}$ (with positive $W_{33}$ for a prolate galaxy pointed at the observer), whereas the sky-projected alignments relevant for weak lensing are encoded in $W_{11}-W_{22}$ and $2W_{12}$.

Note that real elliptical galaxies cannot generally have homologous iso-emissivity contours since this does not explain the well-known isophote twist (see \S\S4.2.3 and 4.3 of \citealt{1998gaas.book.....B} for an overview). This could lead to a difference of the alignment parameters measured by different techniques (e.g.\ Fundamental Plane offset versus ellipticity). However we expect the effect of intrinsic alignments to still be present, and qualitatively similar to the homologous case.

In the linear tidal alignment model, we take the first order approximation in the Taylor expansion of $\langle W_{ij}\rangle$ in the tidal field:
\begin{equation}
\langle W_{ij} \rangle = 2 B s_{ij}.
\end{equation}
This leads to
\begin{equation}
A = 2 (\eta \chi)_{\rm eff} B,
\label{eq:AB}
\end{equation}
where
\begin{equation}
(\eta \chi)_{\rm eff} = \frac{\partial\epsilon(\hat{\bmath e}_3|{\bmath x})}{\partial W_{33}(\bmath x)} = \frac{\partial\ln N_{\rm gal}}{\partial W_{33}}
\end{equation}
is the selection-dependent conversion factor from $W_{33}$ (radial galaxy orientation) to observed number of galaxies.\footnote{In H09, a simple flux threshold was assumed, with various definitions for the flux of an extended object considered. In this case, one could write $A=2\eta\chi B$, where $\eta$ is the slope of the luminosity function and $\chi$ depends on how fluxes area measured for non-spherical objects. In this paper, we implement a more complicated selection algorithm, but we can still think of the conversion factor as an ``effective'' $\eta\chi$.} Here again it is seen that $A$ is only non-zero if we have both non-random orientations $B \neq 0$ {\em and} an orientation-dependent selection effect $(\eta \chi)_{\rm eff} \neq 0$.

The parameter $B$ can be compared to other parameters for the linear intrinsic alignment model used in the literature, such as $b_{\kappa}$ \citep{bernstein2008}, where
\begin{equation} \label{eq:bkappa}
b_{\kappa} = \frac B{1.74}.
\end{equation}

We can also compare to the lensing measurements $A_{I}$ of \cite{singh2015} (derived in Appendix \ref{sec:theory_comp}):
\begin{equation}
B = -0.0233\frac{\Omega_{m}}{G(z)}A_{I}
\end{equation}
where $\Omega_{m}$ is the matter density and $G(z)$ is the redshift-dependent growth function.

%%%%%%%%%%%%%%%%%%%%%%%%%%%%%%%%%%%%%%%%%%%%%%%%%%%%%%%%%%%%%
\subsection{Sub-samples using the Fundamental Plane offset}
\label{sec:fundyplane}

The Fundamental Plane (FP) relation for elliptical galaxies was observed by \cite{djorgovski1987}, who noticed an empirical relationship between a galaxy's luminosity, radius, and projected velocity dispersion. Increasingly detailed measurements of these correlations have been made \citep{jorgensen1995,bernardi2003_2,hyde2009} and there are good theoretical arguments for its existence \citep{bernardi2003}. We use the redshift and luminosity of our observed galaxies in tandem with the FP relation to estimate each galaxy's radius\footnote{We do not use the velocity dispersion information, for several reasons. The velocity dispersion is dependent upon what axis our line of sight takes through the given galaxy, which complicates our assessment of orientation-dependent offsets from the Fundamental Plane. Moreover, the BOSS fibers do not sample the whole galaxy, and we have not explored the subtle correlations that might be imprinted when this effect couples to orientation effects and observational conditions. We avoid these issues entirely, at the cost of some extra statistical error, by fitting for only the redshift and luminosity.}. This radius can then be compared to the radius fit by the SDSS imaging pipeline to estimate the galaxy's orientation, through use of the estimator $W_{33}$ (referenced in Section \ref{sec:theory}).

The key idea is that from our vantage point we do not observe the full 3-dimensional galaxy, but a 2-dimensional projection. In the case of an optically thin galaxy (as for most LRGs), this is simply
\begin{equation}
I(\bmath s_{\perp}) = \int j(\bmath s_{\perp}+s_{3}\hat{\bmath e}_{3})\, {\rm d}s_{3}.
\end{equation}
The projection of a homologous triaxial galaxy gives an image with an effective radius\footnote{For an elliptical image, we define the effective radius as the geometric mean of the semi-major and semi-minor axes of the half-light ellipse. This ellipse has a second moment matrix $r_{\rm es}^2[\exp{\mathbfss W}]_{\rm proj}$, where ``proj'' denotes the $2\times 2$ sub-block of a $3\times 3$ matrix; see the appendix of H09 for a discussion of projections. Equation~(\ref{eq:re}) is the Taylor expansion of the square-root-determinant of this second moment matrix.}:
\begin{equation}
r_{\rm e} =
\exp\left[ - \frac{1}{4}W_{33} - \frac14(W_{13}^2+W_{23}^2) + ...\right]r_{{\rm es}},
\label{eq:re}
\end{equation}
where $r_{\rm es}$ is the effective radius for the spherical galaxy (${\mathbfss W}={\mathbfss 0}$). With this knowledge, we can use Eq.~(\ref{eq:re}) to construct an estimator for $W_{33}$, accurate to linear order:
\begin{equation}
W_{33}^{\rm obs} = -4  \ln \frac{r_{\rm e}}{r_{\rm e0}}.
\label{eq:W33obs}
\end{equation}
where we have distinguished the true value of $W_{33}$ from the linear-order approximation, which we define as $W_{33}^{\rm obs}$. For the rest of the paper, we will refer to the linear-order quantity as $W_{33}$ in order to simplify notation; however, it must be understood that it contains both measurement noise and noise from intrinsic scatter in the FP relationship. Here, $r_{\rm e}$ is the effective radius as measured by BOSS, and $r_{\rm e0}$ is the value of the radius found using the FP. The combination of these measurements gives us an estimate for each galaxy's $W_{33}$ value, and hence each galaxy's orientation.  

\begin{figure}
\centering
\includegraphics[width=\columnwidth,height=!,keepaspectratio]{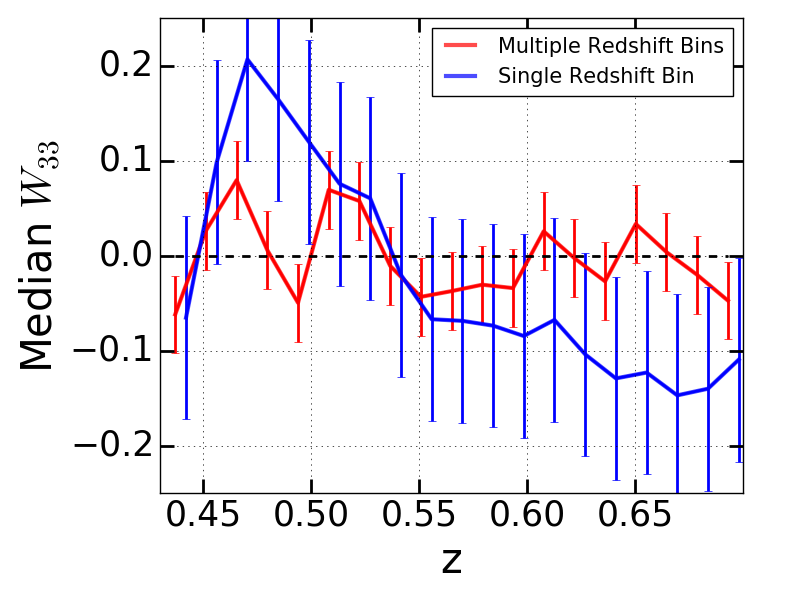}
\caption{For the CMASS samples, we chose to fit Eq. \ref{eq:fitting_eq} separately to galaxies in ranges of $0.43 < z < 0.5$, $0.5 < z < 0.6$, $0.6 < z < 0.65$, and $0.65 < z < 0.7$. This helped mitigate an observed correlation between galaxy redshift and median $W_{33}$ for the CMASS sample. In this plot we compare to the same equation fit over the full redshift range of $0.43 < z < 0.7$ (`single redshift bin'). Here we show the resulting change in median $W_{33}$ value by binned redshift for the CMASS SGC sample. 
}
\label{fig:CMASS_binning}
\end{figure}

Specifically, we find an estimate of the isotropic average size from each galaxy by fitting a simple linear model over all galaxies in the survey for the relation between the logarithmic size and the intrinsic magnitude, as well as the galaxy's redshift:
\begin{equation} \label{eq:fitting_eq}
\log_{10}(r_{{\rm e}0}) = C_{1}M_{rs} + C_{2}z + C_{3},
\end{equation}
where $r_{{\rm e}0}$ is the isotropic average size, $M_{rs}$ is the rescaled galaxy absolute magnitude (to be defined in Eq.~\ref{eq:mag}), and $z$ is the redshift, while the $C$ values are the parameters fit to the model. We fit this model using the Levenberg-Marquardt algorithm \citep{levenberg1944} from the {\sc Scipy} curve-fit algorithm \citep{scipy}. With this model, we estimate each galaxy's angle-averaged size, and hence its orientation, which allows us to separate our catalogs into orientation-divided subsamples. For the CMASS sample, it was found that this fitting procedure resulted in a redshift-biased result for $W_{33}$, as shown in Fig.~\ref{fig:CMASS_binning}. Specifically, higher redshift galaxies were more likely to have fitted radius $r_{e0} > r_{e}$, and therefore a more negative $W_{33}$ value. This implies a change in the slope of the fundamental plane relation with respect to redshift for our CMASS sample. We dealt with this by using four redshift bins in the CMASS sample: $0.43<z<0.5$, $0.5<z<0.6$, $0.6<z<0.65$, and $0.65<z<0.7$. Within each redshift bin, we fit to Eq.~(\ref{eq:fitting_eq}), which helped to mitigate the redshift bias. For the LOWZ sample, we also fit to Eq.~(\ref{eq:fitting_eq}), but for the entire redshift range, $0.15<z<0.43$. Once all $W_{33}$ values were set, we then subtracted the median value for both the LOWZ and CMASS samples, and identified a galaxy's inferred orientation by the sign of its $W_{33}$ value.

%%%%%%%%%%%%%%%%%%%%%%%%%%%%%%%%%%%%%%%%%%%%%%%%%%%%%%%%%
\subsection{Redshift space distortions of sub-samples}
In this paper, we primarily fit for the RSD parameter $f+A$ that is present in our two sub-samples (split based on a proxy for orientation); the {\em difference} $\Delta(f+A)=\Delta A$ is independent of the linear rate of growth of structure $f$, and isolates the anisotropic selection effects. This is the key parameter of our study; however, we also account for two other parameters, $b_{g}$ and $\sigma_{FOG}^{2}$ (a real-space clustering amplitude and Finger of God length parameter), in order to accurately reproduce the correlation function. These are described in more detail in Sections \ref{subsec:FittingMethods} and \ref{subsec:FOG}.

We now outline the central idea of this paper. We split a parent galaxy sample into two sub-samples, using a proxy for orientation (in this case, offset from the fundamental plane); we expect the different subsamples to have different values of $(\eta\chi)_{\rm eff}$. We next compute the difference between the redshift space distortion amplitudes $f_v = f+A$  measured from the two sub-samples. This difference should cancel out the dependence on the true rate of structure growth (via $f$); there should also be some cancellation of the non-linear effects (though this may not be perfect if the sub-sample splitting is sensitive to galaxy properties other than orientation, and these properties are correlated with large-scale environment).

Using Eq.~(\ref{eq:powerspectrum}), we can see the effect that galaxy samples with different values of $A$ will have on the power spectrum. Specifically, separate samples of galaxies, with different values of $W_{33}$, will theoretically have different values of $A$, but the same value of the growth function, and thus the same $f_{v}$. We can therefore find the difference in $A$ between samples as
\begin{equation}
\Delta A \equiv A_{2} - A_{1} = (f + A_{2}) - (f + A_{1}) = f_{v,2} - f_{v,1},
\end{equation}
where $f_{v,X}$ indicates the measurement of $f_{v}$ for sample $X$. Note that the ``measured'' value of the bias (from a fit to the correlation function) is actually
\begin{equation}
b_{\rm meas} = b - \frac A3.
\end{equation}

From our measurement of $\Delta A$ we can then calculate $B$:
\begin{equation} \label{eq:eta_equation}
\Delta A =  2\Delta(\eta \chi)_{\rm eff} B  = 2\left[ \frac{1}{N_{gal -}}\frac{\partial N_{gal -}}{\partial W_{33}} - \frac{1}{N_{gal +}}\frac{\partial N_{gal +}}{\partial W_{33}}   \right] B,
\end{equation}
where a subscript of ``$+$'' indicates the group where $W_{33}$ is greater than the median value (galaxies are brighter/smaller than predicted by the fundamental plane) and a subscript of ``$-$'' indicates the group where $W_{33}$ is less than the median value (galaxies are dimmer/larger than predicted by the Fundamental Plane). Throughout this paper, any sign of the difference in parameters, such as $\Delta f_{v}$, indicates the value in the ``$-$'' group minus the value in the ``$+$'' group.  While $A$ depends on how well our sample splitting procedure traces orientation, this dependence is removed when considering $B$, which depends only on galaxy type and redshift.\footnote{As discussed in Sec.~\ref{ss:ia-rsd}, there are other conventions for the intrinsic alignment amplitude that are related to $B$.} For our sample splits, we have calculated the values of $\Delta(\eta \chi)_{\rm eff}$ to be $-1.111$, $-1.100$, $-1.515$, and $-1.438$ for CMASS NGC, CMASS SGC, LOWZ NGC, and LOWZ SGC respectively.

%%%%%%%%%%
\subsection{Theoretical expectations}
\label{subsec:theoretical_expectations}

One of the foundational principles in cosmology is statistical homogeneity and isotropy. In the context of intrinsic alignments, this means that the intrinsic alignment parameters that relate the radial tidal field $s_{33}$ to the radial alignment $W_{33}$ are the same as those that relate the plane-of-sky tidal fields $(s_{11}-s_{22},2s_{12})$ to the plane-of-sky alignment $(W_{11}-W_{22},2W_{12})$. That is, the measurement of how strongly a galaxy's alignment is affected by its tidal field should not depend on whether we measure it radially, as in this work, or in the plane of the sky, as is done in lensing studies. We explore this concept in our comparison to theory, and previous literature. The caveat, of course, is that given practical observational limitations the radial and plane-of-sky measurements may not be measuring exactly the same thing. Possible issues include non-homologous isophotes (but with a pipeline that assumes a homologous profile); dust extinction; selection effects for the parent samples; and non-linear response (i.e.\ higher-order terms in Eq.~\ref{eq:re}). Plane-of-sky measurements may depend on their own set of assumptions; see \cite{singh2015,troxel2015} for details. 

% \xiao{also including those possible systematics in weak lensing measurements in e.g. Singh 2015?}
% I think that's outside the scope of what we're doing, but I'll include a statement about where to find those. 

We directly compare our CMASS and LOWZ measurements of $B$ to that obtained by \cite{singh2015}, who use correlations of galaxy shapes in the plane of the sky. \cite{singh2015} performs this measurement for the SDSS DR11 LOWZ sample, which we can directly compare to our DR12 LOWZ sample. 

As a further comparison, we can calculate the theoretical intrinsic alignment strength as predicted in \cite{hirata2009}, given a luminosity value for the sample in question. H09 finds that, using catalogs of luminous red galaxies: 
\begin{equation} \label{eq:B_calc}
B = (1.74)(-0.018 \pm 0.006) \left[ \frac{L}{L_{0}}\right]^{1.48 \pm 0.64},
\end{equation}
where $L$ is $K+e$-corrected to $z=0$ and $L_{0}$ is the corrected $r$-band absolute magnitude of $-22$. Our $r$-band absolute magnitudes are listed in Table \ref{table:cmassmed-lowzmed}. Using the tables provided by \cite{wake2006}, who formulate star evolution on models from \cite{bruzual2003}, we find that our $K+e$-corrected $r$-band median absolute magnitudes are $-22.17$ for LOWZ and $-22.295$ for CMASS. We then find that the ratio $L/L_{0}$ referenced in Eq.~(\ref{eq:B_calc}) is 1.170 for LOWZ, and 1.312 for CMASS. This results in a prediction of $B$ equal to $-0.039 \pm 0.010$ for LOWZ, and $-0.046 \pm 0.011$ for CMASS. With these values, we can compare our results to that expected by H09. 

We can also compare the ellipticity-based measurement of \cite{singh2015} to that predicted by H09 for their measurement of $L/L_{0}$. We will use the following equation (derived in Appendix \ref{sec:theory_comp}) to compare values from \cite{singh2015}, who measure the linear alignment parameter $A_{I}$, to our measurement:
\begin{equation} \label{eq:theory_comp_eqn}
B = -0.0233\frac{\Omega_{m}}{G(z)}A_{I},
\end{equation}
where $\Omega_{m}$ and $G(z)$ are the matter density and growth function at a given redshift for their specific parameter choices, which are:
$\Omega_{m,\mathrm{Singh}} = 0.282$ and
$G(0.32) = 0.869$.
Therefore we can calculate that the lensed measurement for the intrinsic alignment magnitude of the LOWZ sample is:
\begin{equation}
B_{\mathrm{pred}} = -0.0348 \pm 0.0038.
\end{equation}
Using the same method, we calculate H09's prediction for the $L/L_{0}$ values found in \cite{singh2015}, which is 0.95. This leads to a theoretical prediction of $-0.029 \pm 0.010$. All of these predictions are compared to our final results in Section \ref{sec:analysis}.

%%%%%%%%%%%%%%%%%%%%%%%%%%%%%%%%%%%%%%%%%%%%%%%%%%%%%%%%%%%%%%%%%%%%%%%%%%%%%%%%%%%%%%%%%%%%%%
%%%%%%%%%%%%%%%%%%%%%%%%%%%%%%%%%%%%%%%%%%%%%%%%%%%%%%%%%%%%%%%%%%%%%%%%%%%%%%%%%%%%%%%%%%%%%%
\section{Survey data}
\label{sec:dataset}

\subsection{Sample definition and characteristics}
Our data is taken from the completed and publicly available SDSS-III \citep{eisenstein2011} BOSS \citep{dawson2013} Data Release 12 (DR12; \citealt{alam2015}) LOWZ and CMASS catalogs. The purpose of BOSS was, through measurements of large scale galaxy clustering, to determine the scale of baryon acoustic oscillations and hence constrain the cosmic distance scale. Using spectroscopic redshift measurements, it recovered the three-dimensional distribution of approximately $1.4 \times 10^{6}$ galaxies, with an effective area of $10,000$ square degrees. Here, we describe the observations that produced the BOSS data and the general properties of BOSS galaxies.

BOSS objects were selected for spectroscopic observation based on SDSS-I/II/III imaging data in five broad bands ($ugriz$) \citep{gunn2006,doi2010}. SDSS I/II \citep{york2000} made use of photometric pass bands \citep{fukugita1996,smith2002,doi2010}, imaging roughly 7606 deg$^2$ in the Northern galactic cap (NGC) and 600 deg$^2$ in the Southern galactic cap (SGC), data of which was released in DR7 \citep{abazajian2009}. For DR8, an increased area of the SGC was surveyed (3172 deg$^2$) \citep{aihara2011}. The full SDSS DR8 dataset includes ``uber-calibration'', which updates the photometric calibrations within SDSS \citep{padmanabhan2008}.
% {\bf fix to be like section 2.1 of http://adsabs.harvard.edu/abs/2018MNRAS.473.4773A, up to DR8 references (BOSS photometry was frozen after DR8); include uber-calibration reference}

Spectroscopy was obtained for BOSS targets using the BOSS spectrograph \citep{smee2013}. Still in operation, it simultaneously obtains 1000 spectra in a given observation. Using the reduction pipeline described in \cite{bolton2012}, redshifts were classified from BOSS spectra with statistical errors from photon noise at a few tens of km s$^{-1}$.

The LOWZ sample selects bright, red objects using the specific color-cuts and $r$-band flux limit defined in \cite{SDSSDR12}. Over the redshift range $0.2 < z < 0.4$, the sample is nearly volume-limited, with a constant space density of $\approx 3 \times 10^{-4} h^{3}\,\mathrm{Mpc}^{-3}$. LOWZ galaxies reside in massive halos, with a mean halo mass of $5.2 \times 10^{13} h^{-1}\,M_\odot$ \citep{parejko2012}. In our sample, we applied a cut to galaxies outside of the range $0.15 < z < 0.43$, since galaxies sufficiently far outside of the target redshift range are more likely to be mis-assigned objects. This cut also follows the procedure adopted first in \cite{anderson2012} and subsequently in other BOSS analyses studying the CMASS and LOWZ sample separately. This makes the two samples more independent by avoiding overlaps in redshift range. Further details on the properties of the LOWZ sample can be found in \cite{parejko2012,tojeiro2014,kitaura2016}.

%The LOWZ sample selects bright, red objects using the specific color-cuts and $r$-band flux limit defined in \cite{SDSSDR12}. Over the redshift range $0.2 < z < 0.4$, the sample is nearly volume-limited, with a constant space density of $\approx 3 \times 10^{-4} h^{3}\,\mathrm{Mpc}^{-3}$. LOWZ galaxies reside in massive halos, with a mean halo mass of $5.2 \times 10^{13} h^{-1}\,M_\odot$. In our sample, we applied a cut to galaxies outside of the range $0.15 < z < 0.43$, since galaxies sufficiently far outside of the target redshift range are more likely to be mis-assigned objects. This cut also follows the procedure of \cite{anderson2012}, and attempts to make the two samples more independent by avoiding overlaps in redshift range. A full analysis of the LOWZ sample can be found in \cite{parejko2012}. 

The CMASS sample applies the specific color cuts and $i$-band flux limit defined in \cite{SDSSDR12} in order to provide a sample with a number density that peaks at $4.3\times 10^{-4}h^{3}\,\mathrm{Mpc}^{-3}$ at $z=0.5$ and is greater than $10^{-4}h^{3}\,\mathrm{Mpc}^{-3}$ within the range $0.43 < z < 0.65$. The selection was designed to obtain a stellar-mass limited sample, regardless of color \citep{SDSSDR12}. This results in a sample of galaxies that are approximately 75\% red in color (e.g., \citealt{ross2013} and references there-in) and elliptical in morphology \citep{masters2011}. The majority ($\sim$ 90\%) of the galaxies are centrals, living in halos of mass $\sim 10^{13} h^{-1}\,M_\odot$ \citep{martin2011}. In our sample, we applied a redshift cut to galaxies outside the range $0.43 < z < 0.7$, again following the procedure of \cite{anderson2012}, and again to make the samples independent.

%The CMASS sample again extends the SDSS I/II LRG cuts to fainter luminosities; however, it also extends them to bluer color space to focus on a higher number density of galaxies in the range $0.4 < z < 0.7$. It aims to be a mass limited sample, regardless of color, although approximately 75\% of galaxies are still red \citep{strateva2001,masters2011}. The majority of the galaxies are centrals, living in halos of mass $\sim 10^{13} h^{-1}\,M_\odot$. In our sample, we applied a redshift cut to galaxies outside the range $0.43 < z < 0.7$, again following the procedure of \cite{anderson2012}, and again attempting to make the samples more independent. For more details on the CMASS sample and selection, see \cite{martin2011} and \cite{parejko2012}. 

The BOSS footprint is divided into two distinct North and South Galactic cap regions, which we refer to as NGC and SGC. These regions have slightly different properties, as described in the appendix of \cite{alam2017} and references there-in. Thus, we analyze the NGC and SGC separately throughout our pipeline, for both LOWZ and CMASS. This yields a total of 4 samples for which we will obtain results.

%%%%%%%%%%%%%%%%%%%%%%%%%%%%%%%%%%%%%%%%%%%%%%%%%%%%%%%%%%%%%%%%%%
\subsection{Data preprocessing}
Separate from our cosmological parameters are the set of galaxy parameters which describe measured quantities of each specific galaxy. The galaxy parameters used in our pipeline fall into two general categories. The ``science'' parameters, specifically right ascension (RA), declination (Dec), redshift ($z$), and de Vaucouleurs fit parameters (magnitudes and radii), are used both in calculating the orientation parameter $W_{33}$ and in finding the multipoles of the correlation function. These parameters are critical to the purpose and results of this project. All other parameters we refer to as ``null'' parameters. These parameters are not used directly for science, but are used to trace possible observational systematics that could impact the clustering results, and are not related to the intrinsic parameters of any specific galaxy. They are the airmass, Galactic reddening $E(B - V)$, sky flux, and point-spread function (PSF) full-width half-maximum (FWHM). We will describe our use of these null parameters in the Section \ref{sec:systematicstesting}. 

This section details our use of the science parameters to calculate the value of $W_{33}$ for each galaxy. It is important to note that for many of our science parameters (and null parameters), we are able to use values specific to each band, whether it be $u$, $g$, $r$, $i$ or $z$. We found that the fundamental plane method had the best-fit results within the $i$-band, and that the $i$ and $r$ bands both had the smallest deviation from the FP fits. Thus, we used the $i$-band for all variables in our analysis. Beyond band choice, there is also the choice of which radius estimator to use. The de Vaucouleurs radius $r$ of a galaxy is found by fitting the surface brightness, $I$, to:
\begin{equation}
I(R) = I_{0} e^{-7.669 \left[ (R/r)^{1/4} - 1 \right]},
\end{equation}
where $I_{0}$ is the surface brightness at the de Vaucouleurs radius \citep{devauc1948}. This law is a specific case of a general S\'ersic profile, with a S\'ersic index of $n=4$. We modify the measured radius by incorporating the asymmetry of elliptical galaxies; we convert from the fitted de Vaucouleurs radius to an effective radius, using:
\begin{equation}
r_{\rm eff} = r_{dev} \sqrt{ab_{\rm{Dev}}},
\end{equation}
where $r_{dev}$ is the de Vaucouleurs radius and $ab_{\rm{Dev}}$ is the ratio of the short elliptical axis to the long elliptical axis, again measured using the de Vaucouleurs fit. A ratio value of $1.0$ returns the de Vaucouleurs radius, while a ratio smaller than one appropriately scales our fit radius to treat highly elliptical galaxies. The values of $ab_{\rm{Dev}}$ for our samples are listed in Table \ref{table:cmassmed-lowzmed}. We next use the spectroscopic redshift from the CMASS and LOWZ tables to calculate the comoving angular diameter distance $D$ to each galaxy. With these distances, we found the size of the galaxy, parameterized by the physical half-light radius:
\begin{equation}
s = D r_{\rm eff}.
\end{equation}
There is a separate size for each galaxy in each band; our analysis uses the $i$-band sizes, and perform our fits using $\log_{10}$ size. The SDSS database does not include extinction correction, so we apply that here, following the maps of \cite{schlegel1998}, as provided by the BOSS catalog; we do {\em not} perform any $K$-corrections \citep{hogg2002}, {\em except} when we compare our results to theoretical expectations in Section \ref{sec:analysis}.

We calculate the rescaled absolute magnitudes using:
\begin{equation} \label{eq:mag}
M_{rs} = m + 5 - 5 \log_{10}D_{A}
= M + 10 \log_{10} (1+z),
\end{equation}
where $M_{rs}$ is the rescaled absolute magnitude, $M$ is the absolute magnitude, $m$ is the observed magnitude, and $D_{A}$ is the angular diameter distance \footnote{We originally intended to use the absolute magnitude $M$ here, but at a late stage discovered that this is what was implemented within the pipeline, which we have called the rescaled absolute magnitude. The rescaled absolute magnitude differs from that using the absolute magnitude $M$ by an added factor of $10\log_{10}(1+z)$. This should not make any substantial difference in our group selection; the difference in the magnitude will be primarily absorbed into the fit coefficients, with a $1.4\%$ ($0.5\%$) $z$-dependent change in the value of $C_{3}$ for CMASS (LOWZ) in Eq.~(\ref{eq:fitting_eq}). We tested this change by comparing groups with a radial magnitude assignment to those constructed with radial distance in Eq.~(\ref{eq:mag}), and found less than $0.3\%$ difference between the definitions of the groups. Note that in the absence of a $K$-correction, the multiples of $\log_{10}(1+z)$ are essentially arbitrary as they correspond to a different slope of the spectral energy distribution that defines the zero-point.}. We perform redshift cuts specific to each survey, restricting LOWZ to $0.15\le z\le 0.43$ and CMASS to $0.43\le z\le 0.70$. There is a clear cutoff in the redshift space distribution of each survey where the color cuts creating the survey were designed to limit the galaxies. However, specifically with the low side of the CMASS survey, there are ``trailing'' galaxies which have escaped the color cuts. We make these redshift cuts for two reasons. First, eliminating these galaxies removes galaxies that were not intended to be included by the color cuts, and prevents outliers in redshift space from dominating our fit. Galaxies on the fringes of the redshift distribution could have properties which tend to group them within the same orientation subdivision, which would then represent a potential systematic error in the sub-sample clustering analyses. Second, the redshift cuts we perform are generally made across the literature for clustering analyses of CMASS and LOWZ to which we compare \citep{ross2016,marin2016,chuang2016}.

At this point, we have values for the size, rescaled absolute magnitude, and redshift of each galaxy. We can then begin the processes described in Section \ref{sec:fundyplane} for finding galaxy orientations for a specific galaxy, with respect to our point of observation. Table \ref{table:surveystats} shows the number of galaxies in each subsample and the number removed due to redshift cuts. The medians and standard deviations of the galaxy parameters for each sample, once cut by redshift, are shown in Table \ref{table:cmassmed-lowzmed}.
\begin{table*}
\caption{Redshift cut statistics for the BOSS CMASS/LOWZ subsamples.
} % title of Table
\centering % used for centering table
\begin{tabular}{c c c c c} % centered columns (4 columns)
\hline\hline %inserts double horizontal lines
Category & CMASS NGC & CMASS SGC & LOWZ NGC & LOWZ SGC \\ [0.5ex] % inserts table
%heading
\hline % inserts single horizontal line
Initial Galaxy Count & 618,806 & 230,831 & 317,780 & 145,264 \\ % inserting body of the table
Redshift Cut (\%) & 8.08 & 9.71 & 21.88 & 21.85 \\
Remaining Galaxies & 568,776 & 208,426 & 248,237 & 113,525 \\
S/N & 30.389 & 26.679 & 83.583 & 74.306 \\[1ex] % [1ex] adds vertical space
\hline %inserts single line
\end{tabular}
\label{table:surveystats} % is used to refer this table in the text
\end{table*}

\begin{table*}
\caption{CMASS and LOWZ medians and standard deviations for both the $W_{33}$--divided and total samples. Note that the median of $W_{33}$ for the full samples is zero by construction.
} % title of Table
\centering % used for centering table
\begin{tabular}{c rrr} % centered columns (4 columns)
\hline\hline %inserts double horizontal lines
\bf Statistic & $W_{33} > 0$ & $W_{33} < 0$ & Total Sample \\ %[0.5ex] % inserts table
%heading 
\hline % inserts single horizontal line
\multicolumn4c{\textbf{CMASS sample}} \\
Redshift & 0.537 $\pm$ 0.065 & 0.546 $\pm$ 0.066 & 0.541 $\pm$ 0.063 \\ % inserting body of the table
Absolute Magnitude (r-mag) & $-$22.184 $\pm$ 0.317 & $-$22.180 $\pm$ 0.307 & $-$22.182 $\pm$ 0.310 \\
Apparent Size (log$_{10}$ kpc) & 1.089 $\pm$ 0.168 & 1.328 $\pm$ 0.169 & 1.213 $\pm$ 0.206 \\
$ab_{\rm{Dev}}$ & 0.669 $\pm$ 0.178 & 0.698 $\pm$ 0.175 & 0.684 $\pm$ 0.175 \\
$W_{33}$ & 0.956 $\pm 0.843$ & $-$1.000 $\pm 0.905$& 0.000 $\pm 1.438$ \\[1ex] % [1ex] adds vertical space
\hline %inserts single line
\multicolumn4c{\textbf{LOWZ sample}} \\
Redshift & 0.315 $\pm$ 0.073 & 0.319 $\pm$ 0.072 & 0.317 $\pm$ 0.075 \\ % inserting body of the table
Absolute Magnitude (r-mag) & $-$22.520 $\pm$ 0.385 & $-$22.516 $\pm$ 0.337 & $-$22.518 $\pm$ 0.399 \\
Apparent Size (log$_{10}$kpc) & 1.117 $\pm$ 0.141 & 1.300 $\pm$ 0.163 & 1.206 $\pm$ 0.172 \\
$ab_{\rm{Dev}}$ & 0.735 $\pm$ 0.169 & 0.734 $\pm$ 0.183 & 0.747 $\pm$ 0.166 \\
$W_{33}$ & 0.700 $\pm 0.865$ & $-$0.793 $\pm 0.637$ & 0.000 $\pm 1.217$ \\[1ex] % [1ex] adds vertical space
\hline %inserts single line
\end{tabular}
\label{table:cmassmed-lowzmed} % is used to refer this table in the text
\end{table*}

\subsection{Sample splitting by orientation}

\begin{figure}
\centering
\includegraphics[scale=0.47]{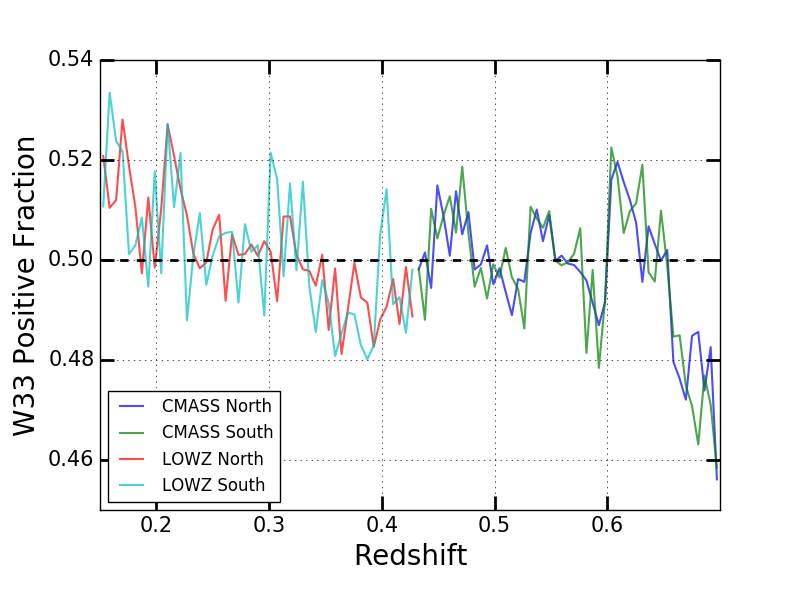}
\caption{For each sample, we show the fraction of positive $W_{33}$ as a function of binned redshift. Note the good agreement of each distribution across the entire redshift range. For the LOWZ samples, this was achieved by using the model within Equation \ref{eq:fitting_eq}. For the CMASS samples, it was necessary to fit this model separately to galaxies in ranges of $0.43 < z < 0.5$, $0.5 < z < 0.6$, $0.6 < z < 0.65$, and $0.65 < z < 0.7$.
}
\label{fig:z_W33comp}
\end{figure}

\begin{figure*}
\centering
\includegraphics[scale=0.47]{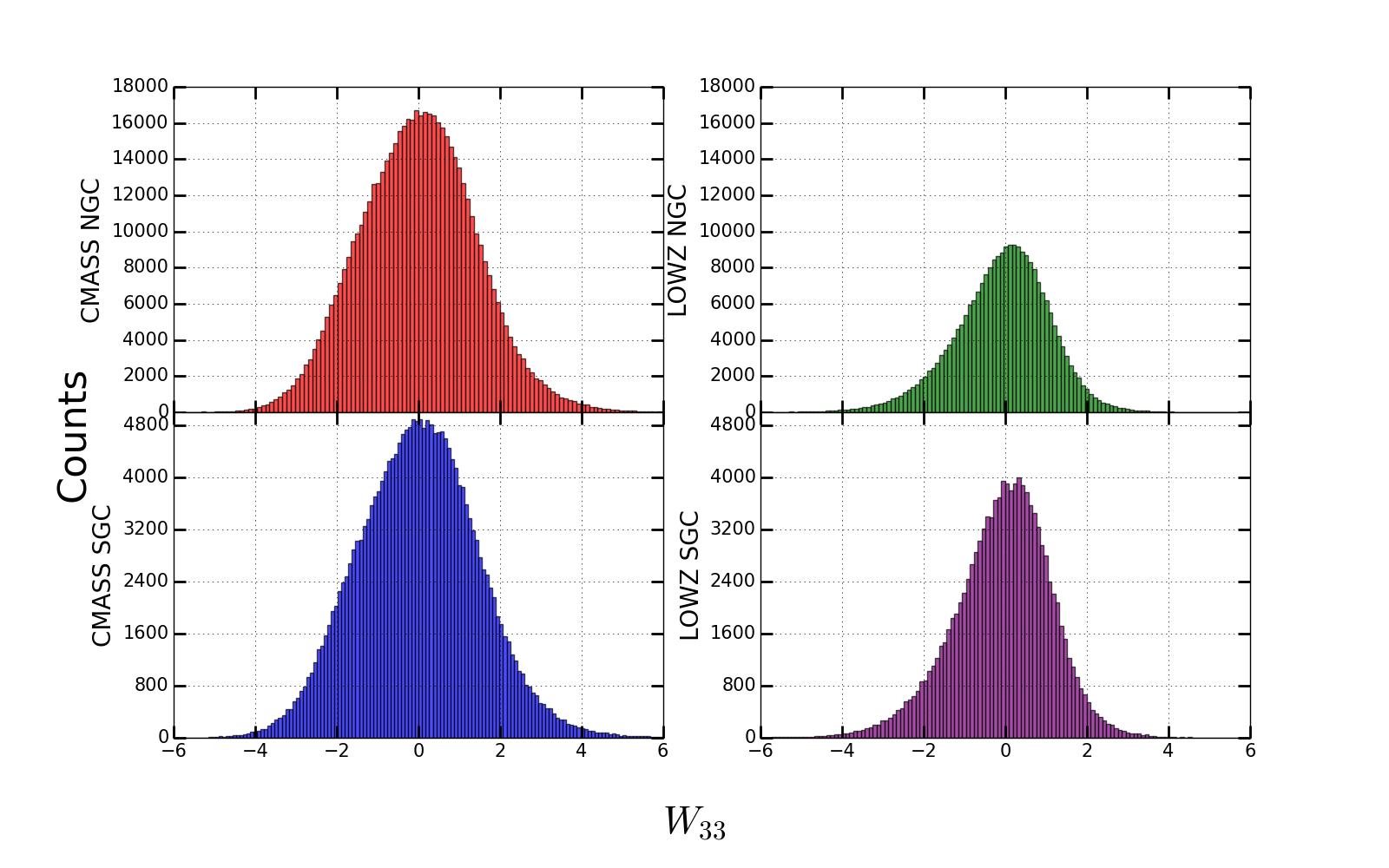}
\caption{Here, we plot the histograms for each survey's distribution of $W_{33}$.
}
\label{fig:W33_histograms}
\end{figure*}

In order to reduce systematic errors in evaluating the $W_{33}$ component of each galaxy, it is necessary to accurately reflect the redshift evolution of the fundamental plane within our sample. We fit all the LOWZ galaxies with a single fit as described in Eq.~(\ref{eq:fitting_eq}). For the CMASS samples, in order to better adjust for changes in the FP due to redshift, we fit this model separately by several bins of redshift. We used four redshift bins of $0.43 < z < 0.5$, $0.5 < z < 0.6$, $0.6 < z < 0.65$, and $0.65 < z < 0.7$. We found that this scheme resulted in galaxy samples with orientations as seen in Fig.~\ref{fig:z_W33comp}. We plot the distributions of $W_{33}$ in Fig.  \ref{fig:W33_histograms}. 

Once values of $W_{33}$ were calculated for each galaxy in our catalog, we then subtracted the median $W_{33}$ from each galaxy, and divided our catalogs into two separate samples, depending on the sign of their $W_{33}$ value. Galaxies with negative $W_{33}$ are larger than we would predict by using the FP method, and we presume that statistically these galaxies are aligned perpendicular to our line of sight. Conversely, galaxies with positive $W_{33}$ are smaller than we would estimate from the FP method, and we presume that statistically these galaxies are aligned parallel with our line of sight.
%As discussed in Section \ref{sec:theory}, the difference between the inferred and ``true'' values of galaxy size can, over the large survey sample, be attributed to differences in galaxy orientation. <-- redundant, so I commented it. C.H.
Although on a galaxy-by-galaxy basis there are other effects which could cause a galaxy to appear bigger or smaller, these effects should cancel out when measuring $\Delta f_v$ for the full sample, assuming that these effects are not correlated with the true value of $W_{33}$ and that all galaxy sub-samples have the same true value of $f$.
%; however, we will gain an aggregate effect from systematic intrinsic orientations due to tidal alignments.

\begin{figure*}
\centering
\includegraphics[scale=0.47]{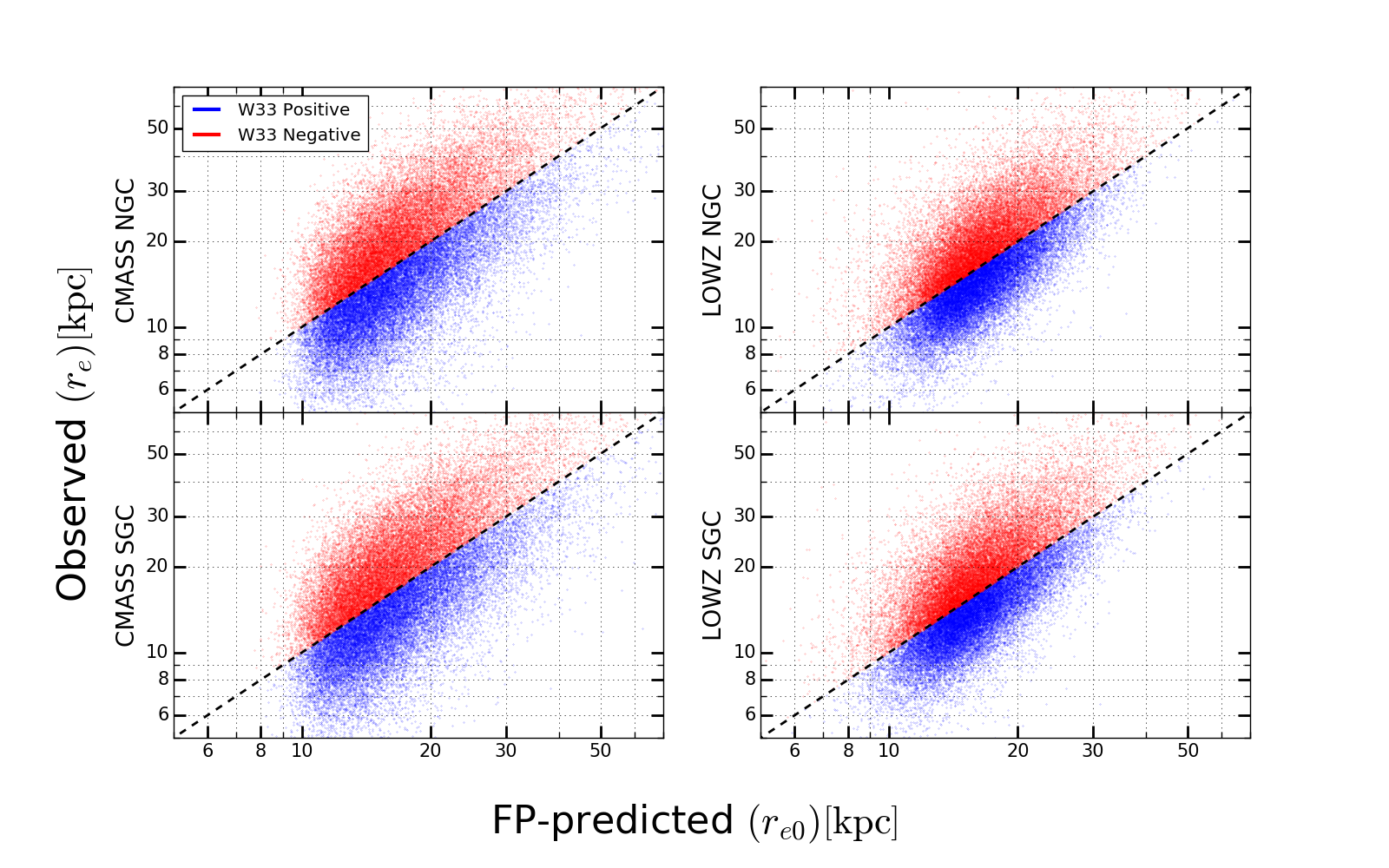}
\caption{For each sample, we use relations of the Fundamental Plane to fit an assumed galaxy radius, based on galaxy luminosity and redshift. Here we plot this assumed radius to the 'true' galaxy radius which has been measured photometrically in the $i$-band. Galaxies are assigned into bins of positive and negative W33, based on whether their fit radius is greater or less than their measured radius.}
\label{fig:re0_W33comp}
\end{figure*}

In Figure~\ref{fig:re0_W33comp}, we compare the photometrically ($i$-band) measured size of each galaxy to predicted size based on each galaxy's redshift and luminosity, using the FP relationship. We plot $25,000$ randomly chosen galaxies from each sample, with a line displaying an exact match between the predicted and observed galaxy size. Our samples of positive and negative $W_{33}$ are defined based on the relationship of these parameters. 

\begin{figure}
\centering
\includegraphics[scale=0.47]{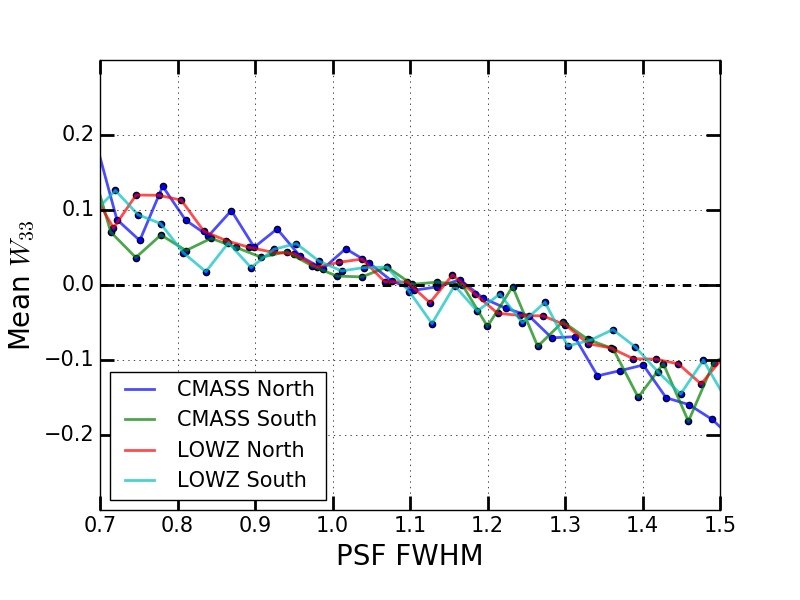}
\caption{Here we show the average $W_{33}$ value, binned by the PSF FWHM. There is a clear correlation between the PSF FWHM value and a galaxy's calculated $W_{33}$ value. This will be analyzed in more detail in Section \ref{sec:systematicstesting}. Further evidence of this can be seen in Table \ref{tab:m_theta_vals}.
}
\label{fig:airmasscorr_W33comp}
\end{figure}

\begin{figure}
\centering
\includegraphics[scale=0.47]{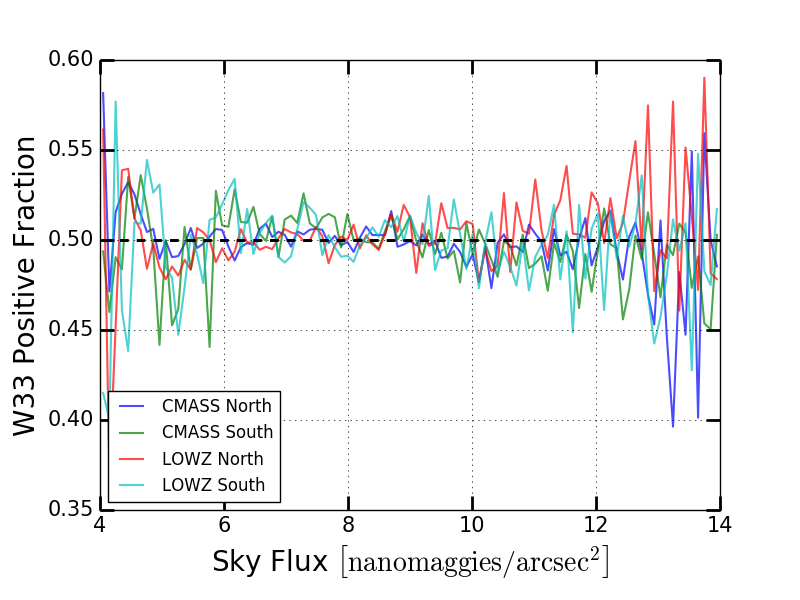}
\caption{For each sample, we show the fraction of galaxies with positive $W_{33}$ as binned by the sky flux at the time of observation. Note that there is no obvious correlation between the sky flux on the night of observation, and the calculated $W_{33}$ value. Toward the edges of the range we can larger deviations from the center, which is due to smaller numbers of galaxies observed with those values of sky flux.}
\label{fig:sky_W33comp}
\end{figure}

Our primary concern with systematic errors is that some parameter related to observation will leak into our estimates of $W_{33}$, and then imprint the spatial structure of the observational effect onto our sub-samples. For example, if galaxies observed at higher airmass have fainter measured magnitudes, then this would offset them from the fundamental plane (they would look ``too large'' for their magnitude and their measured $W_{33}$ would be biased low). In Fig. \ref{fig:airmasscorr_W33comp} we plot the mean $W_{33}$ value in bins of airmass. For an unbiased measurement of $W_{33}$, we would expect this to fluctuate around zero, as it should be uncorrelated with the airmass. The effect of seeing on the observed galaxy number density has been previously documented in \cite{ross2016}, and could lead to a systematic effect in our measurement. Specifically, in the target selection algorithm, the distinction between stars and galaxies can be more blurry in situations of bad seeing. As seeing gets worse, the smallest galaxies (large $W_{33}$ values) would be more likely to be labeled as stars, and less likely to appear in the galactic sample. This cut is related to the comparison of the model and PSF magnitudes of the object in question, and is explained in detail in \cite{SDSSDR12,ross2011}. In Fig.~\ref{fig:sky_W33comp}, we plot the fraction of galaxies with positive $W_{33}$ value in bins of sky flux. In Table \ref{tab:m_theta_vals} we display the slope of a linear fit between the $W_{33}$ value and four potential systematic effects: airmass, extinction, sky flux and the FWHM of the point spread function. A full quantitative treatment of these systematics and how they influence our results is given in Section \ref{sec:systematicstesting}. 

% \begin{table*}
% \caption{Pearson correlation coefficients between a sample's $W_{33}$ values and the listed statistic. Poisson error bars are shown, based on the number of galaxies in each sample.
% \centering % used for centering table
% \begin{tabular}{c c c c c} % centered columns (4 columns)
% \hline\hline %inserts double horizontal lines
% Statistic & \textbf{CMASS North} ($\pm 0.0018$) & \textbf{CMASS South} ($\pm 0.0029$) & \textbf{LOWZ North} ($\pm 0.0025$) & \textbf{LOWZ South} ($\pm 0.0037$) \\ [0.5ex] % inserts table
% %heading
% \hline % inserts single horizontal line
% Airmass: $W_{33} > 0$ & -0.001 & -0.012 & 0.005 & -0.001 \\
% Airmass: $W_{33} < 0$ & -0.001 & -0.001 & -0.004 & -0.004 \\
% Extinction: $W_{33} > 0$ & 0.014 & 0.016 & 0.016 & 0.023 \\
% Extinction: $W_{33} < 0$ & -0.011 & -0.015 & -0.011 & -0.013 \\
% Sky Flux: $W_{33} > 0$ & 0.016 & 0.003 & 0.016 & 0.013 \\
% Sky Flux: $W_{33} < 0$ & -0.034 & -0.034 & -0.023 & -0.024 \\
% PSF FWHM: $W_{33} > 0$ & -0.055 & -0.053 & -0.036 & -0.037 \\
% PSF FWHM: $W_{33} < 0$ & 0.010 & 0.016 & -0.001 & 0.005 \\[1ex]% [1ex] adds vertical space
% \hline %inserts single line
% \end{tabular}
% \label{table:systematics} % is used to refer this table in the text
% \end{table*}
%%%%%%%%%%%%%

\subsection{Systematics tests, random splits, and blinding procedures}
\label{sec:systematicstesting}

The treatment of systematics is an important aspect of this project. We aim to split our sample by the $W_{33}$ values of each galaxy, and to attribute differences among samples as evidence for a radial measurement of intrinsic alignments. However, we want to be sure that the differences in clustering properties of the sub-samples are associated with the targets galaxies themselves, and not due to excess clustering imprinted by observational systematics. We do this by intentionally injecting our orientation measurement with a dependence on different systematic effects, and compare the resulting parameters with the intrinsic uncertainty in our true signal. This section discusses these tests, as well as associated blinding procedures.

First, for what we call Phase I, in Section \ref{subsec:randomsplit} we explain our random division of samples in order to estimate the amount parameters can vary in samples that have no differences in $W_{33}$. There is no specific blinding procedure required at this stage, since we are not computing any clustering properties based on the real sub-sample split (based on $W_{33}$).

Next, in Section \ref{subsec:nullscramble} we describe our injection of systematic effects into the splitting criteria for $W_{33}$. This is still considered a part of Phase I, since there is again no need to blind the output fit parameters, because the values of $W_{33}$ are internally shuffled before being placed in a group. In order to begin Phase II, we require consistency in the $\chi^{2}$ of the fits with our full samples and null samples, as well as final parameter offsets that are consistent with zero, to indicate that systematics are not imitating the effects of $W_{33}$ splitting.

We enter Phase II in Section \ref{subsec:fvinject} by discussing the fitting of our true $W_{33}$-based sample division, while blinding ourselves to the true difference in $f_{v}$. We must be careful here, when checking our samples for errors, to only display combinations of parameters that are independent of the final $\Delta f_{v}$ measurement. The parameter combination $p(\Delta b_{g},\Delta f_{v})$ which achieves this requirement is described in Section \ref{subsec:fvinject}. We also carry out several tests in Phase II for whether differences in the Finger of God properties of the subsamples could bias our measurement of $\Delta f_v$; in these tests, only absolue values of relevant parameters are revealed, so that we remain blind to the direction of any possible correction. In order to finish our calculation and fully un-blind ourselves for Phase III, we require consistency in both samples of this parameter combination, as well as consistency in the $\chi^{2}$/dof of the fits.

\subsubsection{Phase I: Random Splitting}
\label{subsec:randomsplit}

As described in Section \ref{sec:theory}, we want to accurately measure $B$, and therefore $\Delta A$, for our samples which are separated based on galaxy alignment. We first need to estimate the statistical error in $\Delta A$, by measuring the differences in parameters that can occur due to random sample selection alone. In order to achieve this, we create 500 random splits of each survey, and measure the standard deviation of $\Delta A$, that is: $\sigma_{\Delta A}$. We then compare parameter differences for true samples to this value, to decide if true results are significant or if systematic effects need to be mitigated. Blinding is not necessary in Phase I, as we have not divided the sample in any meaningful way, and thus the parameters we are measuring are not affected by intrinsic alignments.

\subsubsection{Null parameter scrambling}
\label{subsec:nullscramble}

We continue Phase I by examining potential systematic effects in more detail. For this step, we do not inject any random parameter values for blinding purposes; however, at one point during this test, we scramble the $W_{33}$ measurements so that they are uncorrelated with their original galaxies, thus destroying any relations between $W_{33}$ and the local tidal field for any particular galaxy; therefore, additional blinding protocols at this stage are not necessary. As discussed above, our list of null parameters includes measurements of airmass, extinction, sky flux and the FWHM of the point spread function. 

For each  survey and null parameter, before we perform any scrambling, we fit a line to a scatter plot between $W_{33}$ and the specific null parameter $\theta$. From this linear fit, we find the slope of the line, $m_{\theta}$. Ideally, if there is no dependence of $W_{33}$ on this null parameter, then $m_{\theta} \approx 0$. The values of $m_{\theta}$ are listed in Table \ref{tab:m_theta_vals}. Next, we scramble the $W_{33}$ values randomly among galaxies, destroying any real signal information for blinding purposes. To each galaxy, we modify the scrambled $W_{33}$ value to become:
\begin{equation}
W_{33} ~~\Rightarrow~~ W_{33} + m_{\theta}\,(\theta_{\rm param} - \bar{\theta}_{\rm param}),
\label{eq:mtheta-inject}
\end{equation}
where $\bar{\theta}_{\rm param}$ is the median value of that parameter within the full sample. This effectively injects the correlation signal with the null parameter into the $W_{33}$ signal. By fitting each sample set for clustering parameters, we find a measurement of $\Delta A$, which represents the potential spurious value if our estimate of $W_{33}$ is contaminated by the null parameter in question, $\theta$. Ideally, we want this difference to be much less than the statistical errors on our test measurements of $\Delta A$.

\begin{table}
\caption{Values of $m_{\theta}$, for each combination of survey and systematic.} % title of Table
\centering % used for centering table
\begin{tabular}{c rrrr} % centered columns (4 columns)
\hline\hline %inserts double horizontal lines
Survey & Airmass & Extinction & Sky Flux & PSF FWHM \\ [0.5ex] % inserts table
%heading
\hline % inserts single horizontal line
CMASS South & $-0.039$ & 0.409 & $-0.007$ & $-0.331$ \\
CMASS South & $-0.103$ & 0.359 & $-0.011$ & $-0.255$ \\
LOWZ North & 0.011 & 0.232 & 0.001 & $-0.261$ \\
LOWZ South & $-0.069$ & 0.318 & $-0.004$ & $-0.219$ \\[1ex]% [1ex] adds vertical space
\hline %inserts single line
\end{tabular}
\label{tab:m_theta_vals} % is used to refer this table in the text
\end{table}

However, even this method is prone to statistical errors. The scrambling step is equivalent to a random assignment of $W_{33}$ values to galaxies, and this random assignment step could result in a parameter shift that has nothing to do with our injected signal, instead dominated by random fluctuations. To beat down these statistical errors, we take two steps. First, we perform the random scrambling step multiple times, so that a true systematic error will become more evident, while statistical fluctuations will tend to cancel out. Second, for each random assignment realization, we perform a ``mirror'' realization where the random groups are the same, but we use the conversion $W_{33} \rightarrow -W_{33}$ before using Eq.~(\ref{eq:mtheta-inject}). This helps to beat down statistical errors even more quickly, as a given realization and its mirror will mostly cancel each other's statistical fluctuations without affecting each other's biased parameter shift. We perform 20 realizations for each null-test, each with their own mirror realization, resulting in 40 total realizations.

\subsubsection{Phase II: $f_{v}$ Blinding}
\label{subsec:fvinject}

\begin{figure}
\centering
\includegraphics[width=\columnwidth,height=!,keepaspectratio]{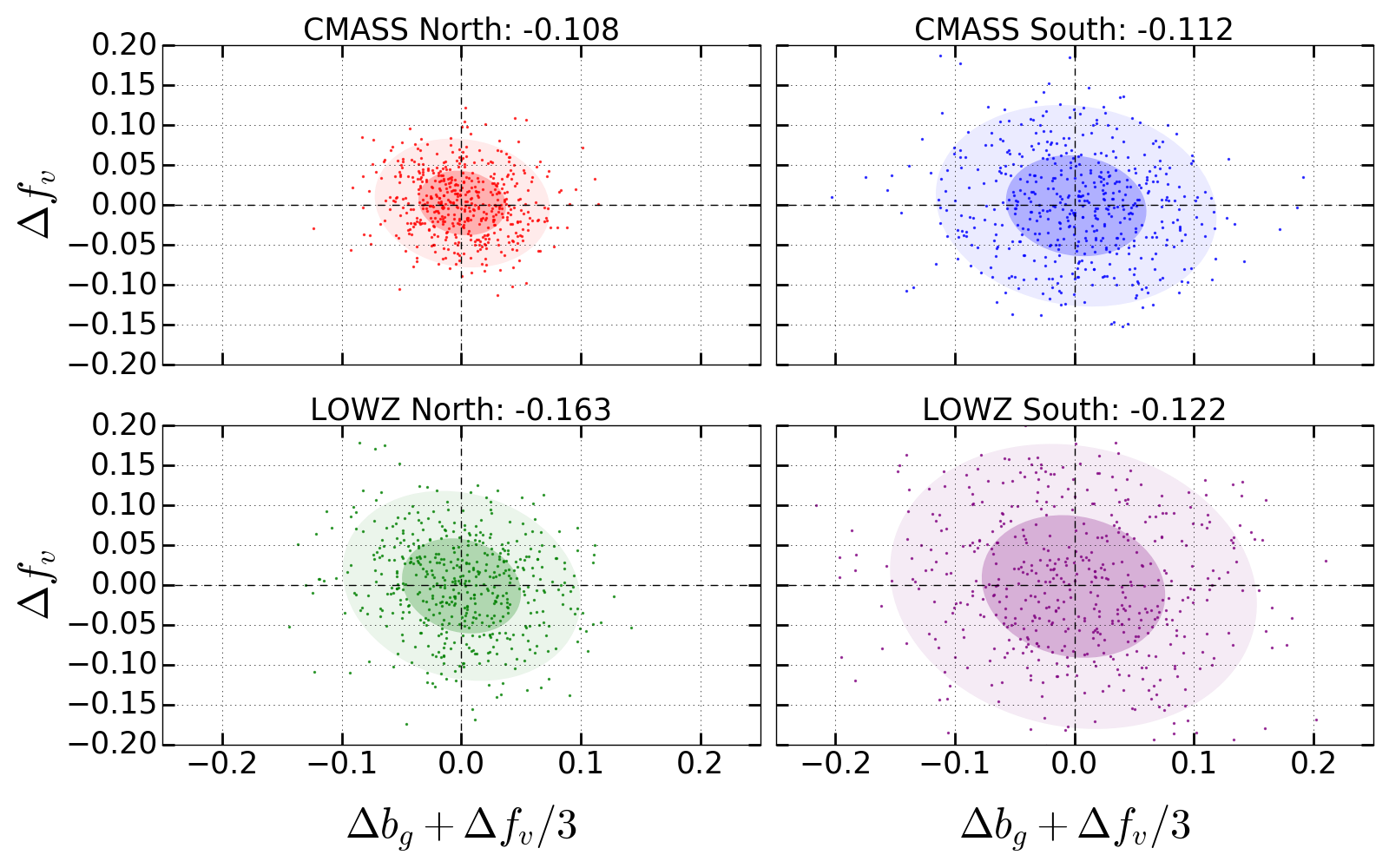}
\caption{Here, we have created 500 random sample separations for each survey. We have compared the parameter $p(\Delta b_{g},\Delta f_{v})$ to $f_{v}$ (Eq. \ref{eq:god_param}). The Pearson correlation coefficient is listed above each plot. We can see that although the coefficient is non-zero, we believe that it is small enough to serve our purpose here as a blinding parameter which is independent of the change due to galaxy orientation.}
\label{fig:god_param}
\end{figure}

As a final test, we disable the scrambling mechanism above, and fit for the true divided samples. However, we blind ourselves to the sample values of $f_{v}$. Instead, we will view the parameter combination which we call $p(\Delta b_{g},\Delta f_{v})$:
\begin{equation} \label{eq:god_param}
p(\Delta b_{g},\Delta f_{v}) \equiv \Delta b_{g} + \frac13\Delta f_{v},
\end{equation}
where $\Delta f_{v}$ and $\Delta b_{g}$ indicate the {\em differences} in the growth parameters and biases for a sample pair. As can be seen from Eq.~(\ref{eq:powerspectrum}), this parameter is independent of our offset due to orientation (it does not depend on $\Delta A$). We also would like this parameter to be uncorrelated with the $f_{v}$ fit for the sample. To test this, we calculated this correlation for all random separations we performed. The results are displayed in Fig. \ref{fig:god_param}. We can see that there is a non-zero correlation value, varying (depending on the sample) between $-0.1$ and $-0.16$. However, as this correlation results in small changes in $f_{v}$ when compared to the random sample separations (see Section \ref{subsec:randomsplitting}), and this step is only necessary to {\em blind} ourselves, not find a final uncorrelated result, we believe that this parameter will be effective as a check at this blinding stage. 

During Phase II, we also run several tests to establish that the Finger of God length is not different enough between the subsamples to affect our measurements of $\Delta f_{v}$. These tests consist fitting our final samples for a smaller range of scales at 20--40 Mpc and reporting $|\Delta(\sigma_{FOG}^2)|$, and also attempting to measure $|\Delta(\sigma_{FOG}^{2})|$ via direct integrals of the correlation function. They are described in detail as part of Section \ref{subsec:FOG}. Note that only absolute values of sub-sample differences are revealed in these Finger of God tests, so that the direction of any correction (were we to attempt one) is not revealed. 

In order to pass this stage, and move on to the unblinded final results of Phase III, we require consistency in the samples for this parameter combination $p(\Delta b_{g},\Delta f_{v})$, and consistency in the $\chi^{2}$ of the fits.  A discrepancy could indicate that our two groups are preferentially selecting galaxy groups of different biases. In Appendix~\ref{sec:biased_subs}, we calculate how this could influence the size of our error bars. Based on our random separation tests, $p(\Delta b_{g},\Delta f_{v})$ has a standard deviation of approximately 0.04 for LOWZ, and 0.08 for CMASS. If any of our subsamples have a value of $p > 0.2$, where $b_{g}$ is the galaxy bias of the full subsample, we will investigate further into the sources of this bias difference, although this still should not affect our measurement of $\Delta f_{v}$. Furthermore, we require our $\chi^{2}$ values, for each subsample, to be within 3$\sigma$ of the mean; specifically, the $\chi^{2}/$dof values must be below 1.50, 1.60, 1.82, and 1.88 for CMASS North, CMASS South, LOWZ North, and LOWZ South respectively.
 
%%%%%%%%%%%%%%%%%%%%%%%%%%%%%
\subsubsection{Phase III: Final Results}
\label{subsec:final_results_unblinding}

Before entering Phase III and unblinding ourselves to the final results, we define what will constitute as a detection, consistency between samples, and consistency between the measurements and theoretical expectations. These requirements are made {\em before} unblinding so that we can be as objective as possible in stating the results of our tests. 

We will check sample consistency by comparing the measured values of $B$ for the north and south subsamples of both CMASS and LOWZ. If these measurements are within 3$\sigma$ of each other, then we will consider them to be consistent. Any discrepancy larger than this will merit a look into what could be causing a difference between the north and south observations. Similarly, if our samples measure $B$ to be greater than 3$\sigma$ from zero, we will consider this a ``detection'' of the intrinsic alignment magnitude. We will consider a measurement of $B$ between 2 and 3$\sigma$ from zero to be ``evidence'' for non-zero intrinsic alignment magnitude. 

Next, we want to analyze how well our results agree with the theoretical expectations. In order to evaluate our results as a whole, we will combine them and their uncertainties with the predictions from \cite{hirata2009}, as listed in Section \ref{sec:systematicstesting}. We will evaluate the ratios of the observed values to the theoretical values:
\begin{equation} \label{eq:theory_frac}
\frac{\mathrm{Obs}}{\mathrm{Theory}} = \frac{\sum_{i}\left[ \frac{B_{i}^{\rm th}}{\sigma_{i}} \right]^{2}\frac{B_{i}}{B_{i}^{\rm th}}}{\sum_{i}\left[ \frac{B_{i}^{\rm th}}{\sigma_{i}} \right]^{2}} \pm \frac{1}{\sqrt{\sum_{i}\left[ \frac{B_{i}^{\rm th}}{\sigma_{i}} \right]^{2}}},
\end{equation}
where $B_{i}$ is our measurement for a specific subsample, $B_{i}^{th}$ is the theoretical prediction for that subsample, and $\sigma_{i}$ is the error on our measurement. A value consistent with $1$ indicates that our results are compatible with the theory. Specifically, we will calculate Eq. \ref{eq:theory_frac} for the LOWZ and CMASS subsamples individually, as well as for all 4 samples in total. We have decided to label a difference from 1 that is less than 2$\sigma$ as `consistent', between 2-3$\sigma$ as in `tension', and as 3+$\sigma$ `incompatible'. We have no specific requirement for the closeness to results from \cite{singh2015}. This is because H09's results are in agreement with that found by \cite{singh2015}, and differences in $L/L_{0}$ will not be a factor in comparing to H09, while they could be a factor in comparison with the specific sample used by \cite{singh2015}.
%%%%%%%%%%%%%%%%%%%%%%%%%%%%%%%%%%%%%%%%%%%%%%%%%%%%%%%%%%%%%%%%%%%%%%%%%%%%%%%%%%%%%%%%%%%%%%
%%%%%%%%%%%%%%%%%%%%%%%%%%%%%%%%%%%%%%%%%%%%%%%%%%%%%%%%%%%%%%%%%%%%%%%%%%%%%%%%%%%%%%%%%%%%%%
\section{Clustering statistics}
\label{sec:correlation}

\subsection{Correlation functions}
For each of our divided catalogs, we use the random galaxy catalogs provided by the BOSS DR12 collaboration in order to normalize geometric sampling effects. Randomly oriented galaxies were given redshifts pulled from the redshift distribution of the matching survey. The random galaxy count is equal to 10 times the real galaxy count. Pair counts were done on the random catalogs in order to construct the correlation function using the Landy-Szalay method \citep{landy1993,peebles1974}. The correlations are calculated as a function of redshift-space separation $s$ and $\mu = \cos\theta$, where $\theta$ is the angle with respect to the line of sight. Specifically, we use: 
\begin{equation}
\xi(s,\mu) = \frac{\mathrm{DD}(s,\mu) - 2\mathrm{DR}(s,\mu) + \mathrm{RR}(s,\mu)}{\mathrm{RR}(s,\mu)},
\end{equation}
where DD refers to the number of pairs of galaxies in the data sample within a specific distance shell, $s \pm\frac12\Delta s$, and within a specific angular range $\mu \pm\frac12\Delta\mu$. RR refers to the same, but for the random sample, while DR refers to the counts of pairs between one data galaxy and one random galaxy. We use 50 logarithmically spaced radial bins from $s = 43$ Mpc to $s = 100$ Mpc. Both DD and RR counts are normalized by the total number of galaxies in that bin, specifically:
\begin{equation}
\mathrm{DD} \rightarrow \frac{\rm DD}{n_{\rm D}(n_{\rm D}-1)}
~~{\rm and}~~\mathrm{RR} \rightarrow \frac{\rm RR}{n_{\rm R}(n_{\rm R}-1)},
\end{equation}
while the DR counts are normalized by:
\begin{equation}
\mathrm{DR \rightarrow \frac{DR}{n_{D} \times n_{R}}}.
\end{equation}

Our correlation function code calculates pairs in 20 $\mu$-bins, from $-1$ to $+1$, with a separation of $\Delta \mu = 0.1$. We do not use a galaxy weighting scheme, for three reasons. Primarily, the importance in our measurement is through the difference in parameters between pairs of samples. As long as the treatment of each sample is the same, the difference in parameters we care about should be unaffected. Second, galaxy weighting has a higher impact on large scales, where our fitting is focused on small to medium scales. Finally, weights are given in BOSS to offset effects such as close pairs of galaxies, where multiple fibers cannot successfully observe both spectra. It is not immediately clear how to treat a weighting scheme where we have divided the samples in a way that does not preserve these close pairs. 

For completeness, we tested a subset of 30 galaxy separations (60 total samples) for both CMASS and LOWZ, finding the correlation functions both with and without weights, and fitting parameters to each, with the weights defined as
\begin{equation}
w_{\rm tot} = w_{\rm systot}(w_{\rm cp}+w_{\rm noz}-1),
\end{equation}
where $w_{\mathrm{systot}}$ accounts for fluctuations in target density with seeing and stellar density, $w_{\mathrm{cp}}$ deals with fiber collisions, and $w_{\mathrm{noz}}$ counteracts redshift dependent bias due to redshift failures. For details on the weighting schemes, see \cite{SDSSDR12}. Within our 30 galaxy testing subsets, we found that for CMASS (LOWZ) the errors in measuring $\Delta f_{v}$ were 0.058 (0.070) for weighted and 0.052 (0.074) for non-weighted, indicating the weighting scheme did not substantially change our final error bars\footnote{The {\em mean value} of $\Delta f_v$ is of course zero for random splits by construction; there is no additional systematic information in the mean value, since the BOSS weights do not know about our sub-sample splits.} (see Sec. \ref{sec:analysis}). The changes in $b_{g}$ and $f_{v}$ due to inclusion of weights are shown in Table \ref{tab:weighted}. It is apparent that the largest change between weighted and non-weighted is in the bias measurements of the CMASS samples. The value of the bias should be independent of the clustering value $f_{v}$, although it may have an effect on error bar size, which we discuss in Appendix~\ref{sec:biased_subs}. 

\begin{table*}\centering
\ra{1.3}
\caption{\label{tab:weighted}The parameter differences between identical samples which we counted using both weighted and non-weighted methods. See Sec. \ref{sec:correlation} for details.}
\begin{tabular}{@{}rrrcrrrr@{}}\toprule
& \multicolumn{2}{c}{$\mathbf{b_{g}}$} & \phantom{abc} & \multicolumn{2}{c}{$\mathbf{f_{v}}$}\\
\cmidrule{2-3} \cmidrule{5-6} 
& \textbf{Weighted} & \textbf{Non-Weighted} && \textbf{Weighted} & \textbf{Non-Weighted} \\ \midrule
CMASS North & 1.882 $\pm$ 0.029 & 1.898 $\pm$ 0.028 && 0.617 $\pm$ 0.025 & 0.619 $\pm$ 0.030 \\
CMASS South & 1.942 $\pm$ 0.036 & 1.986 $\pm$ 0.034 && 0.627 $\pm$ 0.042 & 0.614 $\pm$ 0.034 \\
LOWZ North & 1.773 $\pm$ 0.030 & 1.774 $\pm$ 0.027 && 0.642 $\pm$ 0.038 & 0.641 $\pm$ 0.037 \\
LOWZ South & 1.790 $\pm$ 0.047 & 1.785 $\pm$ 0.045 && 0.597 $\pm$ 0.048 & 0.593 $\pm$ 0.049 \\
\bottomrule
\end{tabular}
\end{table*}

The correlation functions were converted from $\mu$-binned ``wedge'' space to multipole space, for easier plotting and parameter fitting. The formula for conversion is the same as that used in SDSS BOSS analyses \citep[e.g.][]{ross2016}:
\begin{equation}
\xi_{l}(r) = \frac{2l + 1}{2}\sum_{i=1}^{i_{\rm max}}\frac{2}{i_{\rm max}}\xi(r, \mu_{i})L_{l}(\mu_{i}),
\end{equation}
where $\mu_{i} = (i - \frac12)/i_{\rm max}$, $L_{l}$ is the Legendre polynomial of order $l$, $\xi$ is the $\mu$-binned correlation function, and $i_{\rm max}=10$ is the number of positive $\mu$-bins (we simply double the result, since auto-correlations are symmetric under $\mu\leftrightarrow-\mu$). We use the monopole ($l=0$) and quadrupole ($l=2$) of galaxy clustering in our analyses, since they carry most of the information on large-scale redshift space distortions. 

% Our galaxy pipeline, from the galaxy counting to the correlation function calculation, is identical to that used in Martens et al.\ (2018, in preparation). 
% Not included, since this paper is coming out first. 

% To first test our correlation function pipeline, we perform counts in 50 logarithmically spaced bins, from 30 Mpc to 220 Mpc (note the use of Mpc, not $h^{-1}\,$Mpc). Our galaxy pipeline, from the galaxy counting to the correlation function calculation, is identical to that used in Martens et al (2018, in preparation). Here we fit for the full galaxy samples, without doing any divisions based on galaxy orientation. We have combined each survey in Fig.~\ref{fig:demo_Monopole}.

% \begin{figure}
% \centering
% \includegraphics[width=\columnwidth,height=!,keepaspectratio]{DEMO_Monopole.png}
% \includegraphics[width=\columnwidth,height=!,keepaspectratio]{DEMO_Quadrupole.png}
% \caption{{\em Top panel}: Full correlation monopole for each of the four regions: CMASS/LOWZ North/South. The range shown here identifies the BAO peak for clarity, although we fit parameters to a smaller range of 43 to 100 Mpc (approximately 29 to 68 $\mathrm{h^{-1} Mpc}$). For these plots, we counted 2.4 million random galaxies. {\em Bottom panel}: Same as above, but for the quadrupole of the correlation function.}
% \label{fig:demo_Monopole}
% \end{figure}

%%%%%%%%%%%%%%%%%%%%%%%%%%%%%%
\subsection{Covariance matrices}
\label{subsec:covariance}
To provide a best-fit to the observed correlation functions, we assume a Gaussian-distributed likelihood for our vector of measured correlation functions:
\begin{equation}
\mathcal{L}(\bmath{p}) \propto e^{- \chi^{2}(\bmath{p})/2},
\end{equation}
where $\chi^{2}$ is given by: 
\begin{equation} \label{eq:1}
\chi^{2}(\bmath p) = \sum_{\ell,\ell'}\sum_{i,j} (\hat{\xi}^{i}_{\ell}(\bmath{p}) - {\xi}^{i}_{\ell} ) [{\mathbfss C}^{-1}]_{\ell \ell' i j} (\hat{\xi}^{j}_{\ell'}(\bmath{p}) - {\xi}^{j}_{\ell'} ).
\end{equation}
Here ${\bmath p}$ is a vector of parameters; $\ell$ and $\ell'$ are the moments of the correlation function (here equal to 0 or 2); $i$ and $j$ refer to the separation bins; $\hat{\xi}$ is the model correlation function; and ${\mathbfss C}$ is the covariance matrix \citep{sanchez2008,cohn2006}, which we calculate using the method from \cite{grieb2016}, who introduce a theoretical model for the linear covariance of anisotropic galaxy clustering observations, making use of synthetic catalogs. The calculation of the covariances are based on an input linear galaxy power spectrum dependent on both the wavevector and the angle with the line of sight, $P(k,\mu)$. In order to calculate this, we first calculate the linear matter power spectrum using CLASS \citep{blas2011} and then compute the no-wiggle power spectrum from the formulae listed in \cite{eisenstein1998}. This is done at the median redshift of each sample. We next account for redshift space distortion effects to the power spectrum using the procedure outlined in \cite{ross2016}. First we take
\begin{equation}
P(k,\mu) = C^{2}(k,\mu,\Sigma_{s})\left[(P_{\rm lin} - P_{\rm nw})e^{-k^{2}\sigma_{v}^{2}} + P_{\rm nw} \right],
\end{equation}
where we have used
\begin{equation}
\sigma_{v}^{2} = (1 - \mu^{2}) \Sigma_{\perp}^{2}/2 + \mu^{2}\Sigma_{\parallel}^{2}/2
\end{equation}
and
\begin{equation}
C(k,\mu,\Sigma_{s}) = \frac{1+\mu^{2}\beta}{1 + k^{2}\mu^{2}\Sigma_{s}^{2}/2}.
\end{equation}
The parameters used are $\beta = 0.4$, $\Sigma_{s} = 4 h^{-1}$\,Mpc, $\Sigma_{\parallel} = 10 h^{-1}$\,Mpc, and $\Sigma_{\perp} = 6 h^{-1}$\,Mpc, which gives us the the linear matter power spectrum. In order to be used in our covariance matrix calculation, this must be converted to a galaxy power spectrum using a bias appropriate for our specific matter tracers. We follow the results of \cite{marin2016}, who have previously measured full-sample clustering parameters for the LOWZ and CMASS samples, by defining $b_{g} = 1.921$ for our LOWZ sample, and $b_{g} = 1.993$ for our CMASS sample. These power spectra are then used to generate covariance matrices in multipole space, creating covariance matrices for $\xi_{l}(r)$ for $l=0,2$:
\begin{equation}
C^{\xi}_{l_{1},l_{2}}(s_{i},s_{j}) = \frac{{\rm i}^{l_{1}+l_{2}}}{2 \pi^{2}} \int_{0}^{\infty}k^{2} \sigma_{l_{1}l_{2}}^{2}(k) \bar{j}_{l_{1}}(k s_{i})\bar{j}_{l_{2}}(k s_{j})\,{\rm d}k,
\end{equation}
where the multipole-weighted variance integral is
\begin{equation}
\sigma^{2}_{l_{1}l_{2}}(k) = \frac{(2l_{1}+1)(2l_{2}+1)}{V_{s}} \int_{-1}^{1} \left[P(k,\mu) + \frac1{\bar n}\right]^{2} L_{l_{1}}(\mu) L_{l_{2}}(\mu)\,{\rm d} \mu,
\end{equation}
and the bin-averaged spherical Bessel function is
\begin{equation}
\bar{j}_{l}(k s_{i}) = \frac{4 \pi}{V_{si}} \int_{s_{i} - \Delta s /2}^{s_{i} + \Delta s /2} s^{2} j_{l}(ks)\,{\rm d}s.
\end{equation}
Here $V_{si}$ is the volume of the bin for that iteration of $\Delta s$, $V_{s}$ is the volume of the entire sample, $j_{l}$ is the spherical Bessel function of the first kind, $k$ is the wavenumber, $s$ is the distance in redshift space, and $\bar{n}$ is the comoving number density of galaxies for the sample in question.

% In Fig.~\ref{fig:Covariance_Comp}, we plot the absolute value of the ratio between our covariance matrix and that used by \cite{ross2016}, over the relevant scales of interest. They use 1000 mock realizations to generate their covariance matrix. We find relatively good agreement between the matrices, although there is a sign difference in some regions of the off-diagonal sub-matrices. This agreement passes on to the agreement in the error bars surrounding our correlation functions.

% \begin{figure}
% \centering
% \includegraphics[width=\columnwidth,height=!,keepaspectratio]{Covariance_Comp.png}
% \caption{The absolute value of the ratio between our theoretically calculated covariance matrix and the mock-generated covariance matrix of \protect\cite{ross2016}, over the scales of interest. The higher values along the diagonals of the 02 and 20 pieces are due to a zero-crossing in the theoretical algorithm. 
% \label{fig:Covariance_Comp}
% \end{figure}

%%%%%%%%%%%%%%%%%%%%%%%%%%%%%%%%%%%%%%%%%%%%%%%%%%%%%%%%%%%%%%%%%%%%%%%%%%%%%%%%%%%%%%%%%%%%%%
%%%%%%%%%%%%%%%%%%%%%%%%%%%%%%%%%%%%%%%%%%%%%%%%%%%%%%%%%%%%%%%%%%%%%%%%%%%%%%%%%%%%%%%%%%%%%%
\section{Analysis}
\label{sec:analysis}

In this section, we describe both our fitting methods and the results for the multiple Phases. In Sec.~\ref{subsec:FittingMethods}, we explain which parameters we are fitting for, and how the fits are performed. In Sec.~\ref{subsec:fullsample}, we begin the Phase I results. We describe the parameter fits to the full samples of our catalog, without any splitting, and compare them to previous clustering measurements in the literature. In Sec.~\ref{subsec:randomsplitting}, we show the results of our random splitting analysis and explain its significance. We explain the results of our quantitative systematics testing (Phase II) in Sec.~\ref{subsec:systematicsresults}, and in Sec.~\ref{subsec:finalresults} we show our measurements for the true splitting of our samples based on orientation, and describe the result for intrinsic alignments (Phase III).  

\subsection{Parameter fitting methods}
\label{subsec:FittingMethods}

For each subsample, regardless of whether it is a full sample or it is split in some way, we fit with an identical procedure. We calculate the correlation function on scales of 43 to 100 Mpc in 50 radial bins. We fit for the rate of growth $f_{v}$ and the galaxy bias $b_{g}$.

It is important to note that the critical measurements from this paper are {\em not} the values of $f_{v}$ and $b_{\rm g}$ for our samples. Although we have attempted to measure these as accurately as possible, the systematics tests and blinding protocols in this paper are instead aimed at ensuring an accurate measurement of $\Delta A$. Our values of $f_{v}$ and $b_{\rm g}$ should {\em not} be used as references for these fits; instead see papers such as \cite{marin2016} or \cite{chuang2016} for a reference to the sample clustering parameters.

The fit on small scales focuses primarily on parameters due to redshift space distortions; it does not include the BAO peak, which would have little information on the amplitude of intrinsic alignments (but see \citealt{chisari2013}). To calculate the theoretical fit, we use Convolution Lagrangian Perturbation Theory (CLPT; \citealt{carlson2012}), modified on small scales by Gaussian Streaming Redshift Space Distortions (GSRSD). The code for this program was featured in \cite{wang2014}, and is publicly available.\footnote{GitHub link: {\tt https://github.com/wll745881210/CLPT\_GSRSD}. We started from the version that was most recently edited on 29 May 2015.} Note that this code uses $h^{-1}\,$Mpc as its native units whereas this paper uses Mpc; we have checked that our interface contains the correct conversions (but note that some explanatory material in this section uses $h^{-1}\,$Mpc). We modified the GSRSD code to match the specific spatial binning scheme used when we performed the counting algorithms. The code uses CLPT to produce the real-space galaxy correlation function $\xi(r)$, the real-space galaxy pairwise in-fall velocity $v_{12}(r)$, and the real-space galaxy pairwise velocity dispersion $\sigma_{12}^{2}(r,\mu)$. CLPT takes as input a linear matter power spectrum at the redshift of the sample being fitted, which we generate using CLASS \citep{blas2011} set with the cosmological parameters listed in Section \ref{sec:introduction}. These results are then used by GSRSD to calculate the redshift-space correlation function: 
\begin{eqnarray} \label{eq:GSRSD-eqn}
1 + \xi^{s}(s_{\perp},s_{\parallel}) &=& \int \frac{dy}{\sqrt{2 \pi [ \sigma^{2}_{1,2}(r,\mu) + \sigma_{\rm FOG}^{2} ]}}\left[ 1 + \xi(r)\right] \nonumber \\
&& \times \exp \left\{ - \frac{\left[s_{\parallel} - y - \mu v_{12}(r)\right]^{2}}{2\sigma^{2}_{1,2}(r,\mu) + 2\sigma_{\rm FOG}^{2}} \right\}.
\end{eqnarray}
Here $y$ is the real-space pair separation along the line of sight, and $\mu=y/r$. The parameter $\sigma_{\rm FOG}^{2}$ is the increase in variance due to the ``Finger of God'' effect. GSRSD uses a default spacing of $\Delta y = 0.5h^{-1}\,$Mpc in the numerical integral, and a default integration of $\pm 200h^{-1}\,$Mpc around the peak value. Ideally, we require a high enough resolution and range to accurately reproduce the correct correlation functions, but minimize the computation time when running a fit. We found that, in our range of scales and parameter values, a step size $\Delta y = 0.8h^{-1}\,$Mpc and a default integration of $\pm 45h^{-1}\,$Mpc resulted in the same correlation function, but significantly reduced computation time. We made a slight modification to allow the code to accurately treat cases where $\sigma^{2}_{FOG}<0$.
\footnote{Previously, GSRSD set the value of $\sigma_{12}^{2} + \sigma_{FOG}^{2}$ to zero, if it were to extend to negative values. This created a hard cutoff in the $\chi^{2}$ of our fits, since these values created unrealistic correlation functions. We analytically extended Eq.~(\ref{eq:GSRSD-eqn}) to negative values of $\sigma_{12}^{2} + \sigma_{FOG}^{2}$, where $f(x|\sigma^{2}) \rightarrow 2f(x) - \frac{1}{\sqrt{-2\pi\sigma^{2}}}\int dy f(x+y)e^{y^{2}/2\sigma^{2}} $.} This change allowed the parameter space for $\sigma^{2}_{FOG}$ to be extended to include physical situations of a decrease in the spread of red-shift space distortions. In order to produce the redshift-space correlation functions, GSRSD also takes as input the galaxy bias, and the structure growth rate, $f_{v}$. The code outputs the redshift-space correlation function in terms of multipole moments, $\xi_{0,2}(r)$. It is these correlation functions we compare to our measured correlations from the BOSS datasets.

\begin{figure}
\centering
\includegraphics[width=\columnwidth,height=!,keepaspectratio]{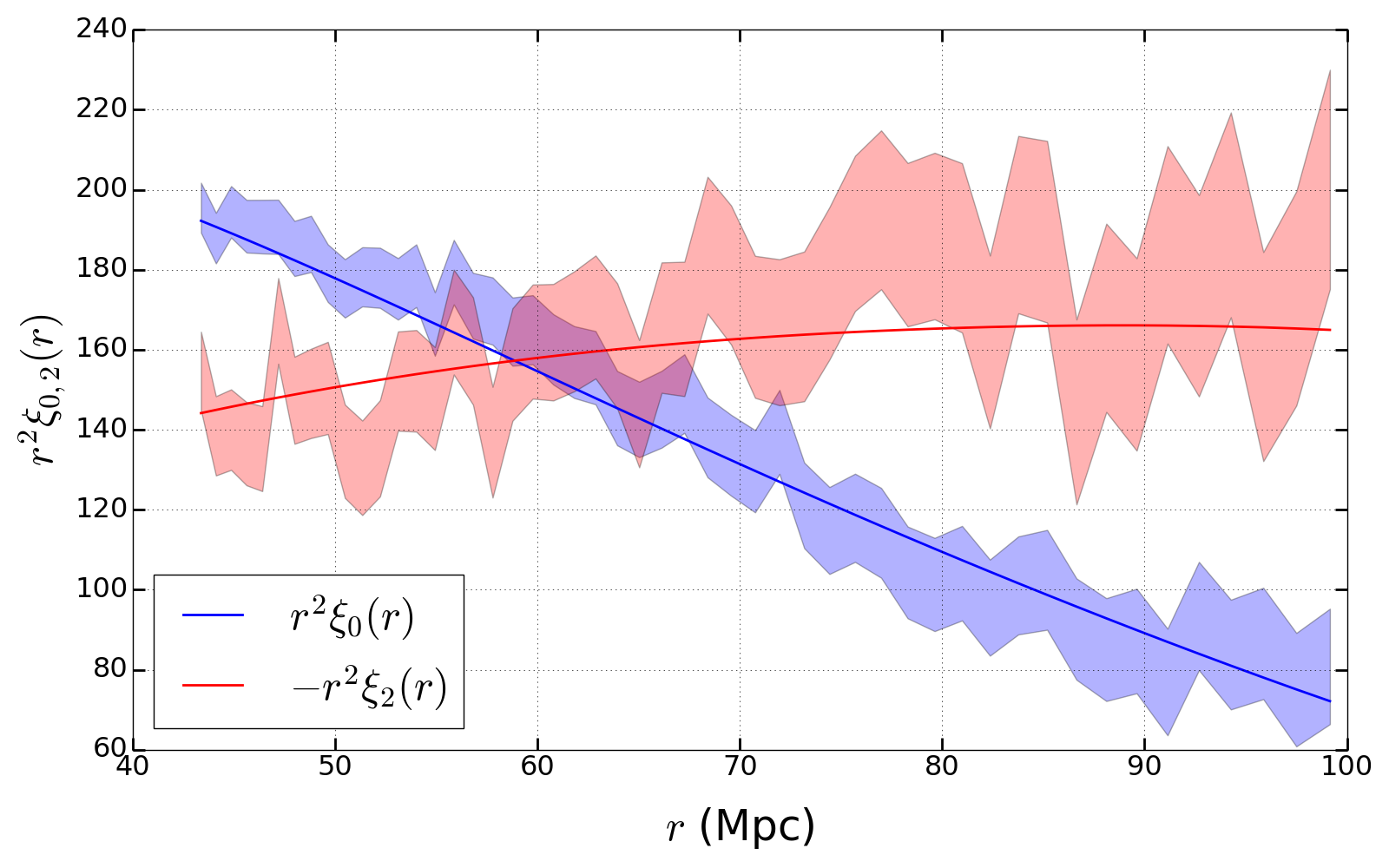}
\caption{Correlation functions from the best-fit parameters, found by the fitting methods in section \ref{sec:analysis}.The shaded areas show the correlation functions from the data with error bars. The dark lines are produced by the best-fit parameters. This is from a random separation of the CMASS NGC catalog. This fit has a $\chi^{2}$ per dof of 1.034. The best-fit parameters are $f_{v} = 0.3712$ and $b_{g} = 1.9486$.}
\label{fig:full_fitplot}
\end{figure}

We use the {\sc Scipy} minimize function \citep{scipy}, specifically the Nelder-Mead method \citep{powell1973}, to find the best-fit parameters by minimizing the $\chi^{2}$ (see Eq.~\ref{eq:1}). Each iteration of the method uses CLPT and GSRSD to create a fitting attempt of the monopole and quadrupole correlation functions, which are then compared to the BOSS values. We find that convergence always occurs within a few hundred iterations, assuming starting values of $f_{v} = 0.7$ and $b_{\rm g} = 1.8$. In Fig.~\ref{fig:full_fitplot} we show an example fit overlaid on the data for a random separation in the CMASS NGC sample.

%%%%%%%%%%%%%%%%%%%%%%%%%%%%%%%%%%%%%%%%%%%%%%%%%%%%%%%%%%%%%%%%%%%%%%%%%%%%%%%%%%%%%
\subsection{Finger of God Effects}
\label{subsec:FOG}

The Finger of God effect, as explained in Sec. \ref{sec:theory}, results in a smearing of galaxy velocities along the line of sight for small scales. It can be seen represented in Eq.~(\ref{eq:GSRSD-eqn}) as $\sigma_{FOG}^{2}$. This parameter can have an important effect on the quadrupole of our correlation function, and ideally should be taken into account as a nuisance parameter. While running our fits over the divided samples, as discussed in Section \ref{sec:systematicstesting}, we found that the value of $\sigma_{FOG}^{2}$ was not well constrained, and likely to vary highly across the parameter space. This indicated a very shallow $\chi^{2}$ surface, relative to this specific parameter choice. We decided then to set $\sigma_{FOG}^{2}$ to a value consistent with our model fits, which also matched the literature. \cite{marin2016} found, depending on the method, a value of $\sigma_{FOG}^2=11.4$ to $22.9 h^{-2}\,$Mpc$^2$ for LOWZ, and a value of $\sigma_{FOG}^2=9.2$ to $13.9 h^{-2}\,$Mpc$^2$ for CMASS. We re-fit 50 subsamples to just two parameters, and found that for LOWZ, the average $\chi^{2}$/dof changed from $1.3745$ to $1.3695$, and for CMASS, from $1.385$ to $1.3695$. This indicated that fitting for $\sigma_{FOG}^{2}$ was not improving our fits, and could be set identically for all subsamples taken from the total data. 

Ideally, we would fit for all three parameters in each subsample; however, for a variety of reasons we felt that it was better to set a value for $\sigma_{FOG}^{2}$ here. First, as stated above, the addition of this parameter did not improve the quality of the fits - in fact, the $\chi^{2}$/dof decreased when $\sigma_{FOG}^{2}$ was fitted for. Second, the important measurement we are conducting is in the difference of $f_{v}$ between two otherwise identical samples. By treating each sample with the same value of $\sigma_{FOG}^{2}$, we reduce the scatter in the resulting values for $f_{v}$, and increase the significance of our measurement of the orientation-based splitting. Finally, we only fit large scales where the Finger of God effect has little impact; hence even large changes in the value of $\sigma_{FOG}^{2}$ result in a small change of $\chi^{2}$.

A possible concern is that when we split into subsamples based on the observed $W_{33}$, the two subsamples may have different satellite fractions and hence different Finger of God lengths. In a fit with fixed $\sigma_{FOG}^{2}$ (hence $\Delta\sigma_{FOG}^{2}$ forced to be zero), this could leak into the measurement of $\Delta f_v$ due to the degeneracy of $\sigma_{FOG}^{2}$ and $f_v$. To estimate the potential impact of this effect, we measured the covariance between the fit values of $\Delta f_{v}$ and $\Delta \sigma_{FOG}^{2}$, and found that
\begin{equation}
\frac{{\rm Cov}(\Delta\sigma_{FOG}^2, \Delta f_v)}{{\rm Var}(\Delta\sigma_{FOG}^{2})} =
\left\{
\begin{array}{lll}
3.7 \times 10^{-3} \, \mathrm{h/Mpc} & & {\rm(CMASS)} \\
3.3 \times 10^{-3} \, \mathrm{h/Mpc} & & {\rm(LOWZ)}
\end{array}
\right..
\end{equation}
This means that an error of $\Delta\sigma_{FOG}^2 = 1 \,({\rm Mpc}/h)^{2}$ propagates into an error of 0.0037 (0.0033) in $\Delta f_v$ for CMASS (LOWZ).
For a ``nightmare'' scenario where our samples were split within groups where one group had a value of $\sigma_{FOG}^2=0$, and the other of $\sigma_{FOG}^2=22 \, ({\rm Mpc}/h)^{2}$, this would constitute a change of $\Delta f_{v} \approx 0.081 (0.072)$ between the two samples for CMASS (LOWZ), which is comparable to the error bars in $\Delta f_{v}$ of approximately 0.05 (0.08). 

We address this concern by calculating a value $\mathcal{I}$ for both the full samples and a selection of 15 randomly split samples, where:
\begin{equation} \label{eq:FOG_Test}
\mathcal{I} = \sum_{i}r^{3}\xi(r,\mu)_{i}.
\end{equation}
This value represents an integral over the correlation function. We sum over a range of $1 \leq r_{\perp} \leq 4$ Mpc and $3 \leq r_{\parallel} \leq 22$ Mpc, where $r_{\perp}^{2} = r^{2}(1 - \mu^{2})$, and $r_{\parallel} = r \mu$. The index `i' refers to all eligible $\mu$ and $r$ bin combinations that meet these criteria within our correlation calculation. The purpose of this is to gain a sense of the FOG size of each subsample, and to understand the variance of this value for a random separation by calculating:
\begin{equation} \label{eq:I_split}
\mathcal{I}_{\rm split} \equiv \frac{\mathcal{I}_{2} - \mathcal{I}_{1}}{\mathcal{I}_{\rm full}}
\end{equation}
where $\mathcal{I}_{i}$ is calculated for the ith subsample split, and $\mathcal{I}_{\rm full}$ is calculated for the full sample. 

We find that for the {\em random} splits the means and standard deviations of Eq.~(\ref{eq:FOG_Test}) are $0.017 \pm 0.026$ for LOWZ South, $0.001 \pm 0.018$ for LOWZ North, $0.003 \pm 0.014$ for CMASS South, and $-0.002 \pm 0.016$ for CMASS North. The means are consistent with zero, as expected for a random split, and the standard deviations decrease for the subsamples with larger numbers of galaxies. With this information, we can test our true subsample splits for a difference in their $\sigma_{FOG}^{2}$ value {\em without} unblinding ourselves to their true fit parameters. From the values of $|\mathcal{I}_{\rm split}|$ for our true subsamples\footnote{As noted in Sec.~\ref{sec:systematicstesting}, only the absolute value is revealed so that if we decided a correction to $\Delta f_v$ were needed, we would not know what direction it would go.} we can calculate the prospective effects on $\Delta f_{v}$, and determine whether or not this issue will affect our measurements. We reveal our results for this test in Section \ref{subsec:unblinding1}.

Finally, we also measured 15 randomly split samples over the ranges of 20--40 Mpc, fitting for $f_{v}$, $b_{g}$, and $\sigma^{2}_{FOG}$. We found that the standard deviations for each sample pair of $\sigma^{2}_{1} - \sigma^{2}_{2}$ were 10.4, 8.4, 11.0 and 7.6 $({\rm Mpc}/h)^{2}$ for LOWZ South, LOWZ North, CMASS South and CMASS North respectively. In Section~\ref{subsec:unblinding1} we reveal the {\em absolute value} of same parameter differences for our true samples, to further analyze if our measurements will be biased by the value of $\sigma^{2}_{FOG}$. (The use of the absolute value ensures that we do not know which direction the potential bias would go.) 

% Therefore, we proceed fixing the value of $\sigma_{FOG}^2$, and fit only for $f_{v}$ and $b_{g}$. 
%%%%%%%%%%%%%%%%
\subsection{Full sample results}
\label{subsec:fullsample}

\begin{figure}
\centering
\includegraphics[width=\columnwidth,height=!,keepaspectratio]{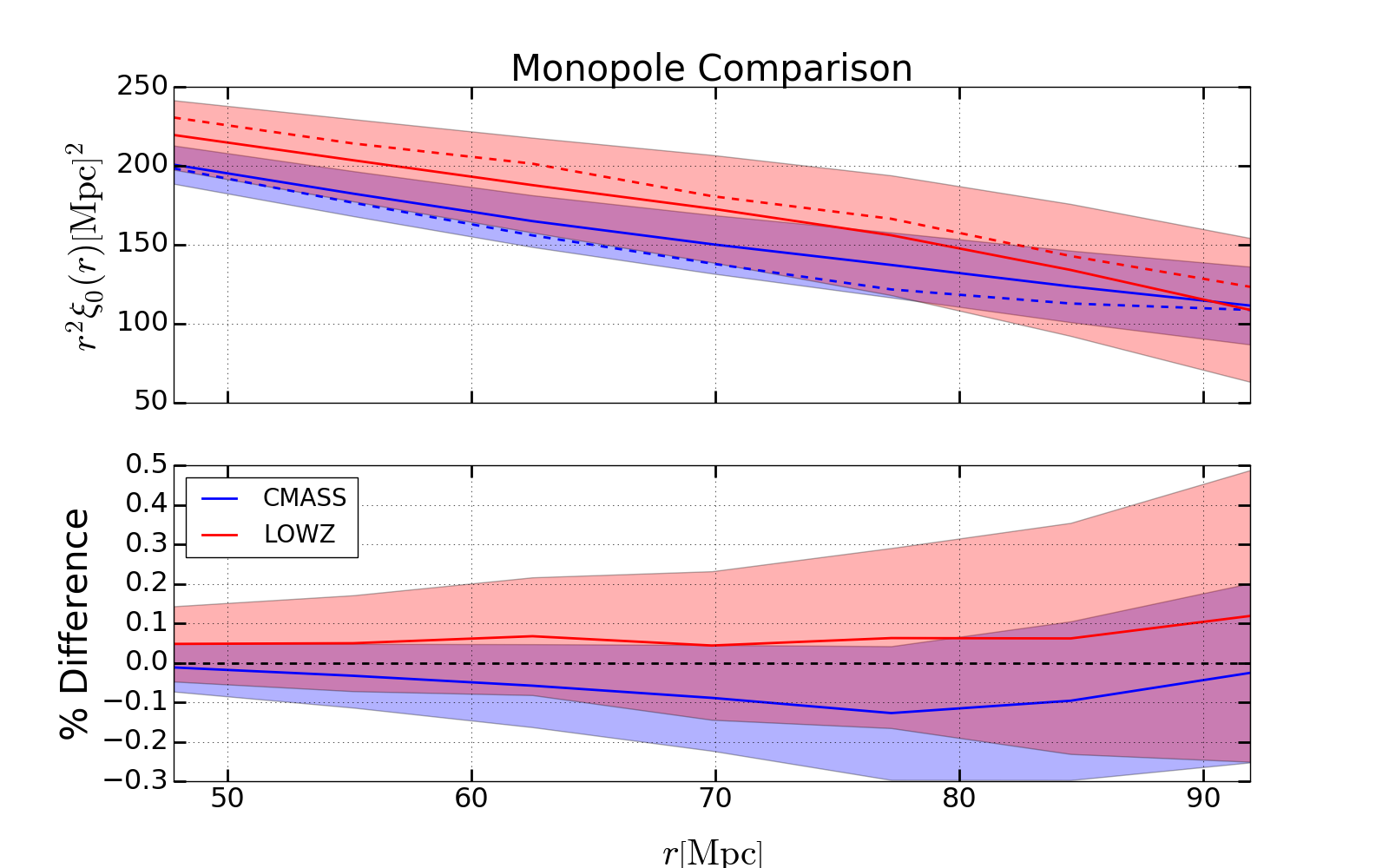}
\caption{{\em Top panel:} We plot the correlation function monopoles for the CMASS and LOWZ. We compare our calculation, with normal lines, to the calculations of \protect\cite{chuang2016}, in dotted lines. {\em Bottom panel:} The percent difference between the sample differences. }
\label{fig:paper_monopole_comp}
\end{figure}

\begin{figure}
\centering
\includegraphics[width=\columnwidth,height=!,keepaspectratio]{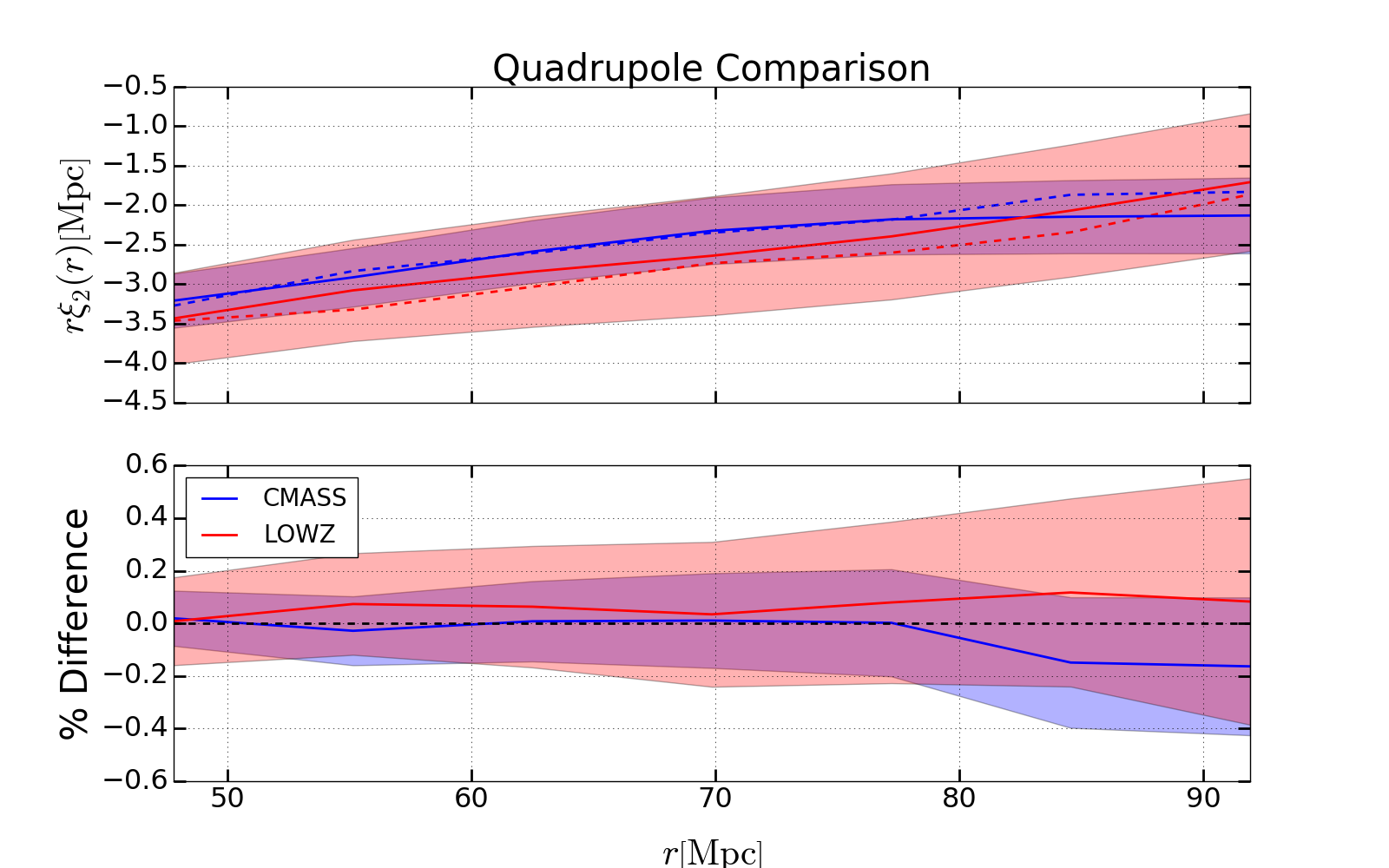}
\caption{{\em Top panel:} We plot the correlation function quadrupoles for the CMASS NGC and LOWZ SGC. We compare our calculation, with normal lines, to the calculations of \protect\cite{chuang2016}, in dotted lines. {\em Bottom panel:} The percent difference between the sample differences.}
\label{fig:paper_quadrupole_comp}
\end{figure}

In order to establish that our fitting process is accurately finding the best-fit parameters, we compare our correlation functions, for the scales of interest, to those calculated by \cite{chuang2016}. We used 2.4 million random galaxies to produce these correlation functions. In Figures \ref{fig:paper_monopole_comp} and \ref{fig:paper_quadrupole_comp}, we plot the monopole and quadrupole of our correlation functions together with those from \cite{chuang2016}, for both the LOWZ and CMASS subsamples, for the range of separations where we fit RSD parameters. We also show the percent difference between the two comparisons. Differences at this stage are primarily due to weighting schemes in the pair counting, which were used in \cite{chuang2016}, while our pair-counts were unweighted (all weights equal to 1). Note that we plot in Mpc, as opposed to the $h^{-1}$ Mpc used in \cite{chuang2016}.

\begin{figure}
\centering
\includegraphics[width=\columnwidth,height=!,keepaspectratio]{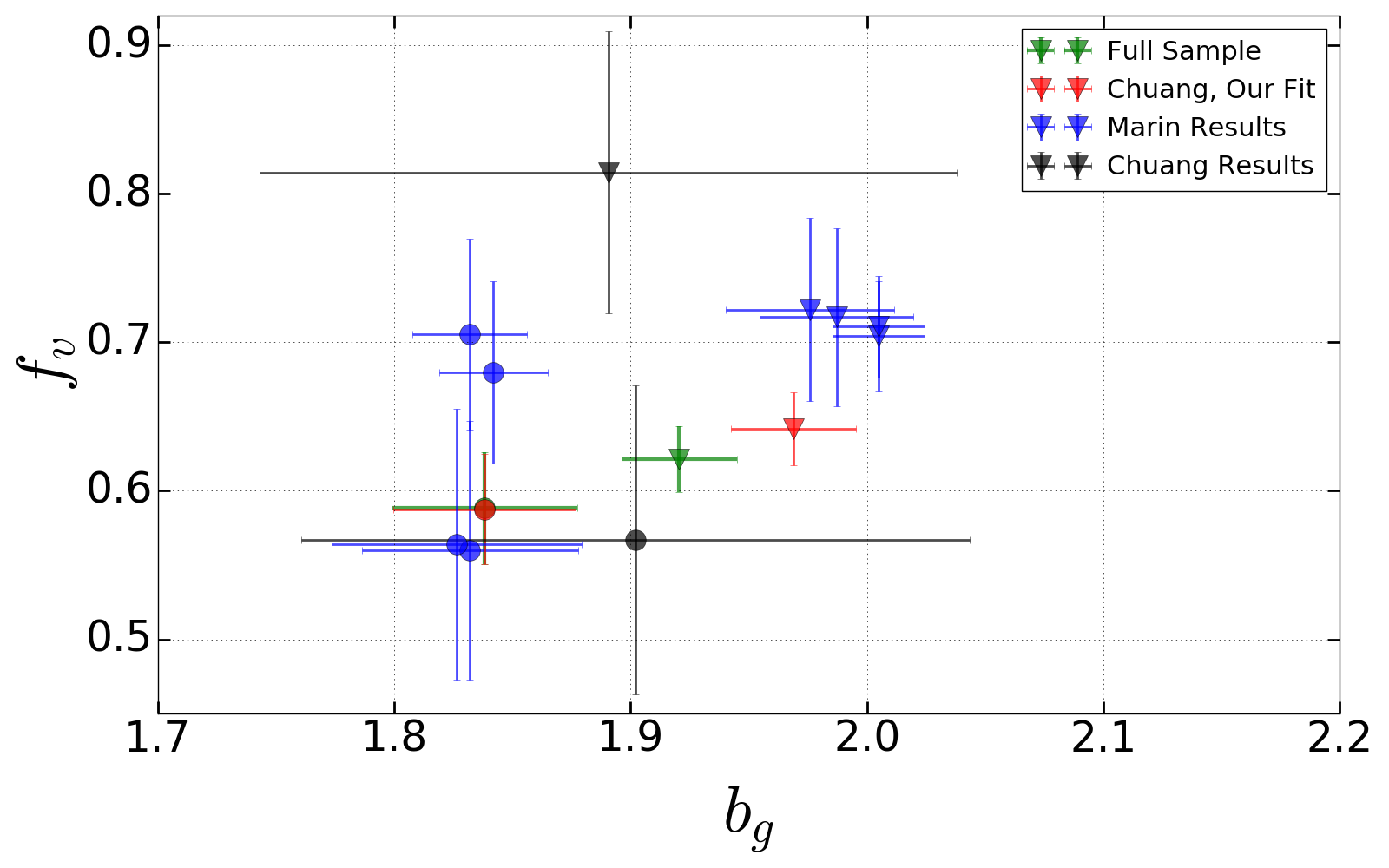}
\caption{Here, we plot the final parameters of $b_{g}$ and $f_{v}$ for a variety of samples. Triangles indicate results from CMASS, and circles indicate results from LOWZ. The green points are the results from our full samples for CMASS and LOWZ. The red points are the results from the correlation functions from \protect\cite{chuang2016} being run through our pipeline. (Note that for LOWZ, the green and red points lie directly on top of each other). The black points are the full results from \protect\cite{chuang2016}. Finally, the blue points are the results from \protect\cite{marin2016}, where we have shown several of their final parameters, which depend on details of the mocks used for fitting.}
\label{fig:ashley_param_comp}
\end{figure}

Next, we used the correlation functions from \cite{chuang2016} in our own pipeline to find best-fit values for $b_{g}$ and $f_{v}$, and compared them to our own full sample fits, as well as the results listed within \cite{chuang2016} and \cite{marin2016}. We plot the results in Fig.~\ref{fig:ashley_param_comp}. Here the we can see how the differences in our results break down between differences in the correlation function, and differences in the fitting process. Green points indicate our results for the full samples, while red points indicate running the correlation function of \cite{chuang2016} through our pipeline. For both samples, these points agree well, which indicate that differences in the calculation of the correlation function are not a major factor when fitting our parameters. The black and blue points refer to the fits presented in \cite{chuang2016} and \cite{marin2016}, respectively. The differences between these points and our own are primarily due to scale differences; we fit on scales of approximately 40 to 100 Mpc, while they fit on a much larger range of scales, between approximately 60 and 260 Mpc. Furthermore, they are fitting and marginalizing over a larger parameter set. 

Our point of comparison here is to highlight the consistency of the pipeline itself; the purpose of our paper is a differential measurement between sub-divisions of galaxies. The actual values of the parameters are not important for our final results, as long as the sub-divisions are taken from the same sample. We do not recommend using our best-fit full sample parameter values for reference outside of this work; instead, use values from \cite{chuang2016} or \cite{marin2016} to exemplify the best clustering parameters to describe the LOWZ and CMASS samples.

\subsection{Phase I: Statistical uncertainties}
\label{subsec:randomsplitting}
Random splittings of the galaxy sample will have a difference in fitted parameters, if only due to statistical fluctuations. In order to evaluate the significance of our orientation-based splitting, we calculate the standard deviation of a random splitting, $\sigma_{\Delta A}$, by dividing each sample 500 times and calculating the distributions of $\Delta A$. This is assuming that the parameters fitted for the difference between two random sub-selections will be Gaussian-distributed. We test for this by calculating the kurtosis of both the $\Delta b_{g}$ and $\Delta f_{v}$ parameters, for each subset. We find that the kurtosis for $\Delta b_{g}$ ($\Delta f_{v}$) are $-0.25$, 0.05, $-0.16$, and $-0.15$ ($-0.17$, $-0.25$, $-0.07$, and $-0.24$) for CMASS North, CMASS South, LOWZ North, and LOWZ South, respectively; for 500 points, the expected error in the kurtosis is $\approx\sqrt{24/500}=0.22$, so these are consistent with zero.

\begin{figure}
\centering
\includegraphics[width=\columnwidth,height=!,keepaspectratio]{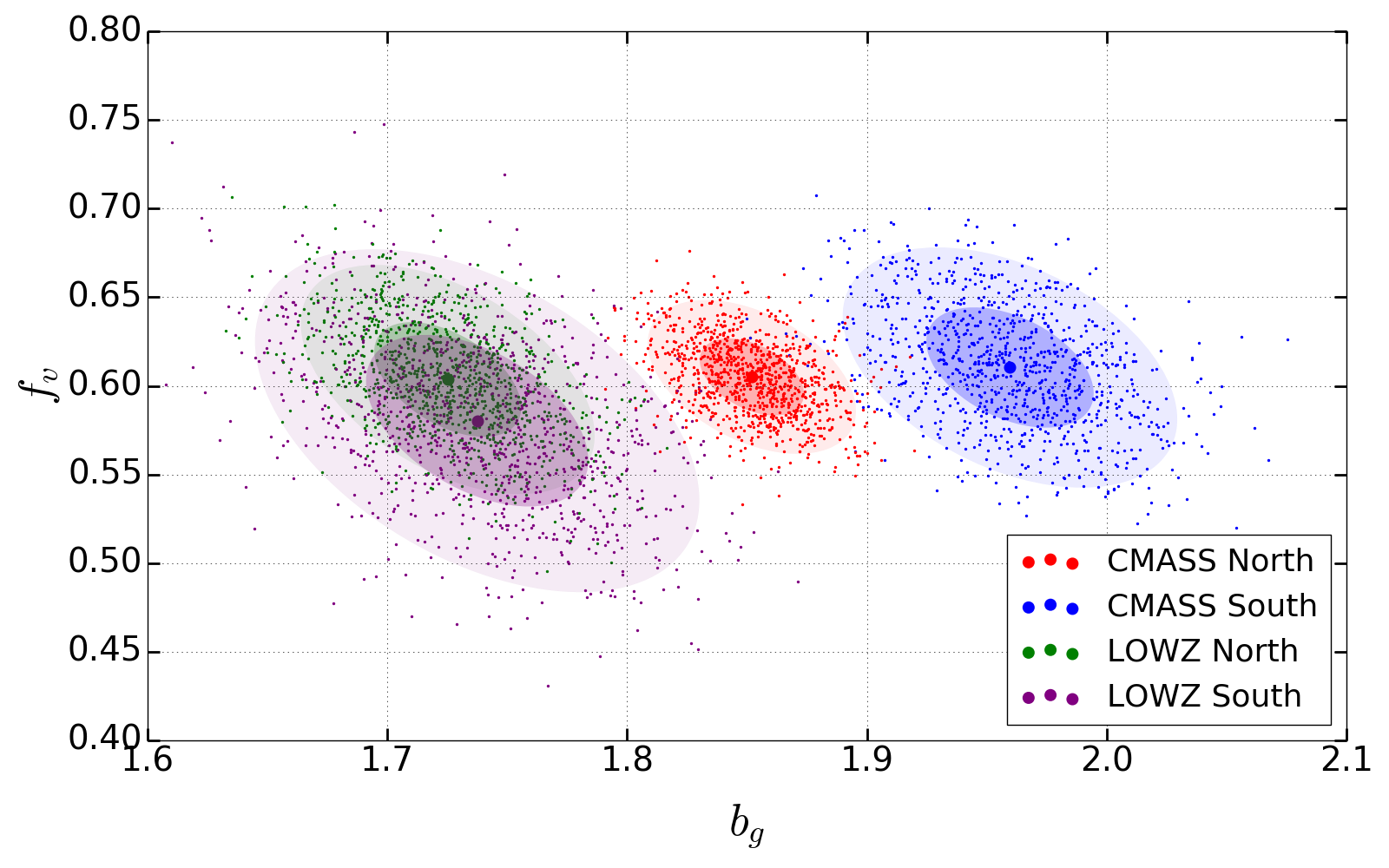}
\caption{We show the results of fitting clustering parameters of $b_{g}$ and $f_{v}$ to each sample of 500 random splits of each survey catalog, resulting in 1000 points.}
\label{fig:sigmaAparams}
\end{figure}

\begin{figure}
\centering
\includegraphics[width=\columnwidth,height=!,keepaspectratio]{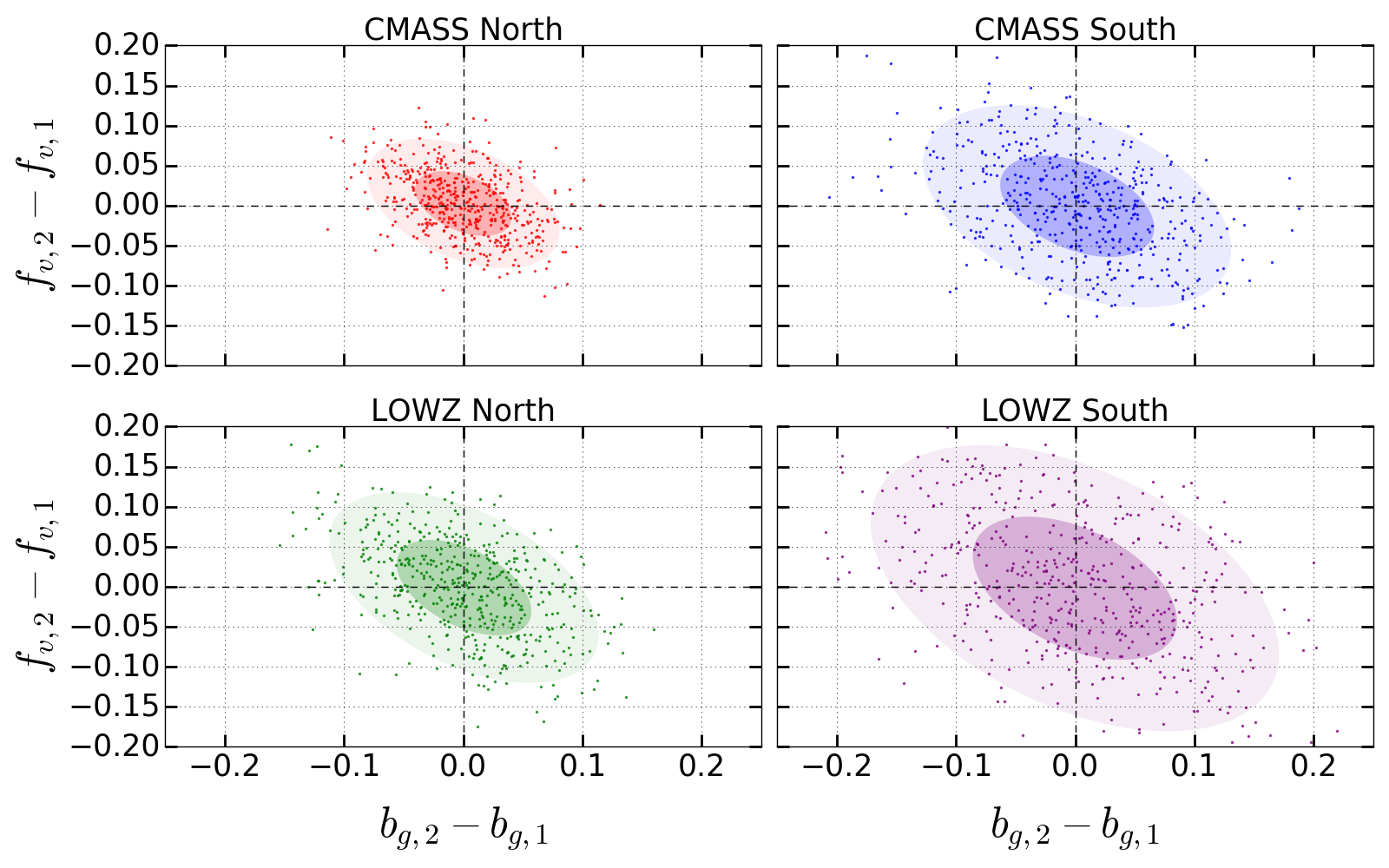}
\caption{The differences in parameter values for 500 random splits of the data.}
\label{fig:delta_sigmaAparams}
\end{figure}

In Fig.~\ref{fig:sigmaAparams}, we show the results of fitting clustering parameters of $b_{g}$ and $f_{v}$ to each half of 500 random splits of each survey catalog, resulting in 1000 plotted points. We show in Fig. \ref{fig:delta_sigmaAparams} the differences in parameters fit for each random split. For each sample, the mean of these differences is centered at zero, as it must be.\footnote{This is more of a code check then a systematics test, since there is no way for a truly random split to give a non-zero average $\Delta b_g$ or $\Delta f_v$.} We are interested in the inferred error bars, specifically the standard deviation of the differences in $f_{v}$, which is equivalent to the standard deviation for our value of $\Delta A$. These values are listed in Table~\ref{tab:sigmaA_aves}. We discuss in Appendix~\ref{sec:biased_subs} how large of an effect $\Delta b_g$ could have on error bars.

\begin{table*}
\caption{Fitting statistics for the 1000 random splits of each survey.
} % title of Table
\centering % used for centering table
\begin{tabular}{c c c c c c} % centered columns (4 columns)
\hline\hline %inserts double horizontal lines
Survey & <$f_{v}$> & <$b_{g}$> & $\sigma_{\delta A}$ & $\sigma_{B}$ & <$\chi^{2}/\rm{dof}$> \\ [0.5ex] % inserts table
%heading
\hline % inserts single horizontal line
CMASS North & 0.605 $\pm$ 0.022 & 1.852 $\pm$ 0.022 & 0.040 & 0.018 & 1.109 \\
CMASS South & 0.610 $\pm$ 0.034 & 1.960 $\pm$ 0.035 & 0.063 & 0.029 & 1.146 \\
LOWZ North & 0.604 $\pm$ 0.032 & 1.725 $\pm$ 0.031 & 0.059 & 0.020 & 1.281 \\
LOWZ South & 0.580 $\pm$ 0.048 & 1.737 $\pm$ 0.046 & 0.089 & 0.031 & 1.323 \\[1ex]% [1ex] adds vertical space
\hline %inserts single line
\end{tabular}
\label{tab:sigmaA_aves} % is used to refer this table in the text
\end{table*}

%%%%%%%%%%%%%%%%%%
\subsection{Systematics-biased subsamples}
\label{subsec:systematicsresults}

Here we show the results for subsamples that have been randomized (to eliminate any real signal) but subsequently biased to reflect selected observational systematics. As mentioned in Section \ref{sec:dataset}, some of our variables describing the data are used solely for systematics checking -- specifically, making sure that observation elements are not leaking into our $W_{33}$ estimator and hence imprinting the systematics maps differently on the two subsamples. We run systematics testing against sky flux, air mass, extinction, and the point-spread function at full-width half-maximum (PSF FWHM). These tests are described in detail in Section \ref{sec:systematicstesting}. Here, we show the resulting fit parameters from these tests, and discuss their relevance with respect to the value of $\sigma_{\Delta A}$ calculated above.

\begin{figure*}
\centering
{\bfseries Airmass
~~~~~~~~~~~~~~~~~~~~~~~~~~~~~~~~~~~~~~~~~~~~~~~~~~~~~~~~~~~~~~~~~~~~~~~~~~~~~~~~~~~~~~~~~~~~~~~~
Extinction
}\\
\includegraphics[width=3in,height=!,keepaspectratio]{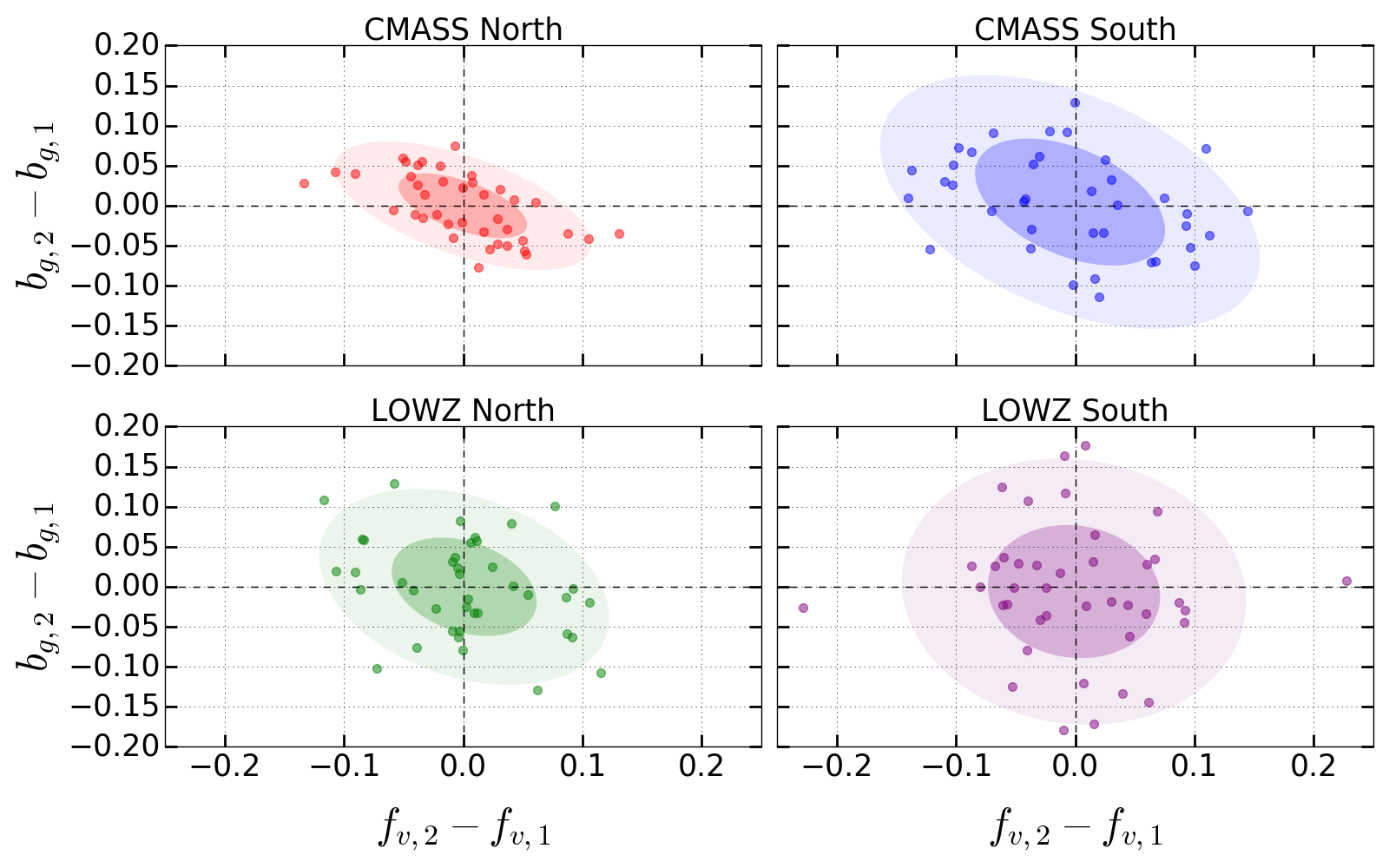}
\includegraphics[width=3in,height=!,keepaspectratio]{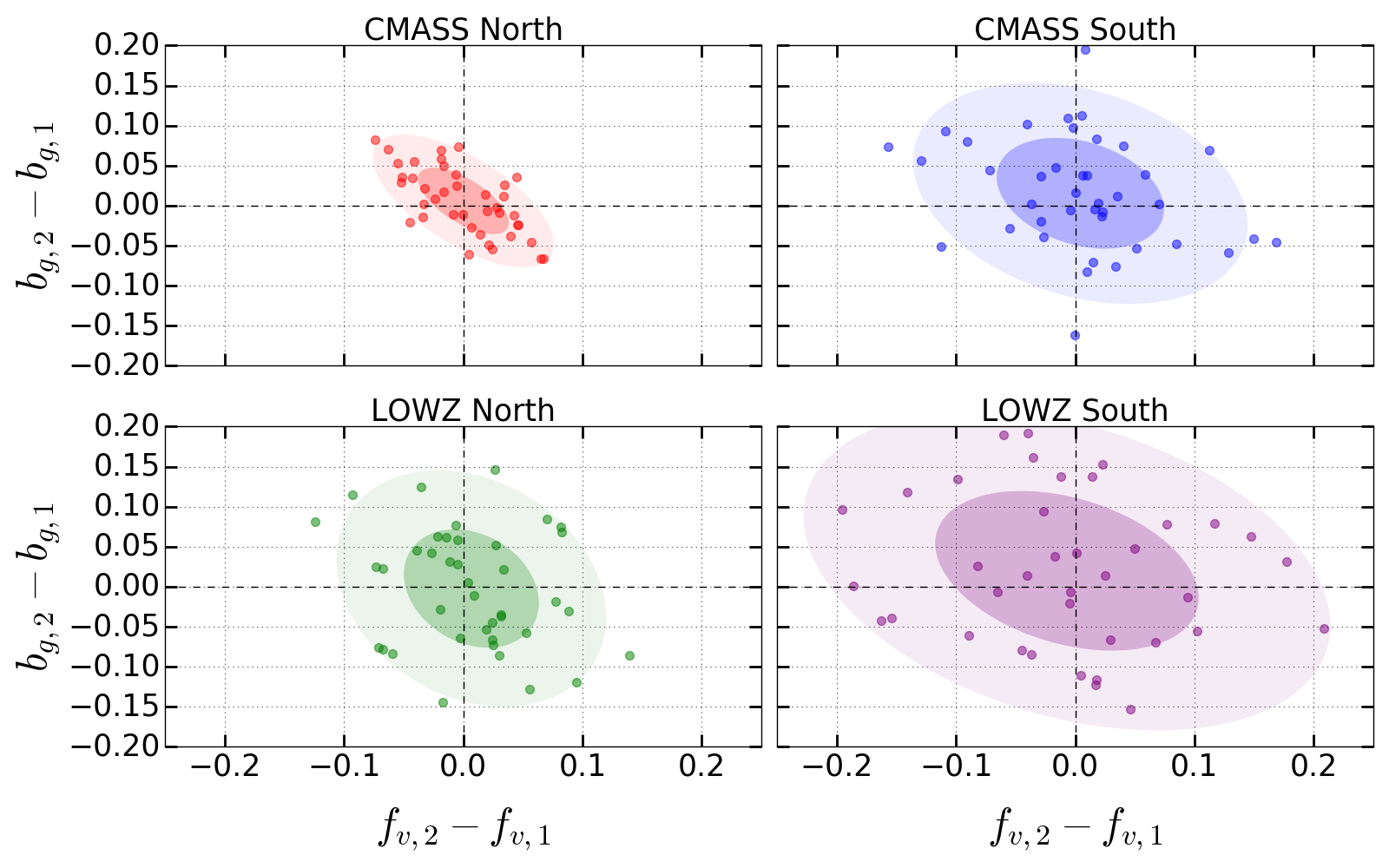}
\\
{\bfseries Sky flux
~~~~~~~~~~~~~~~~~~~~~~~~~~~~~~~~~~~~~~~~~~~~~~~~~~~~~~~~~~~~~~~~~~~~~~~~~~~~~~~~~~~~~~~~~~~~~~~~
PSF FWHM
}\\
\includegraphics[width=3in,height=!,keepaspectratio]{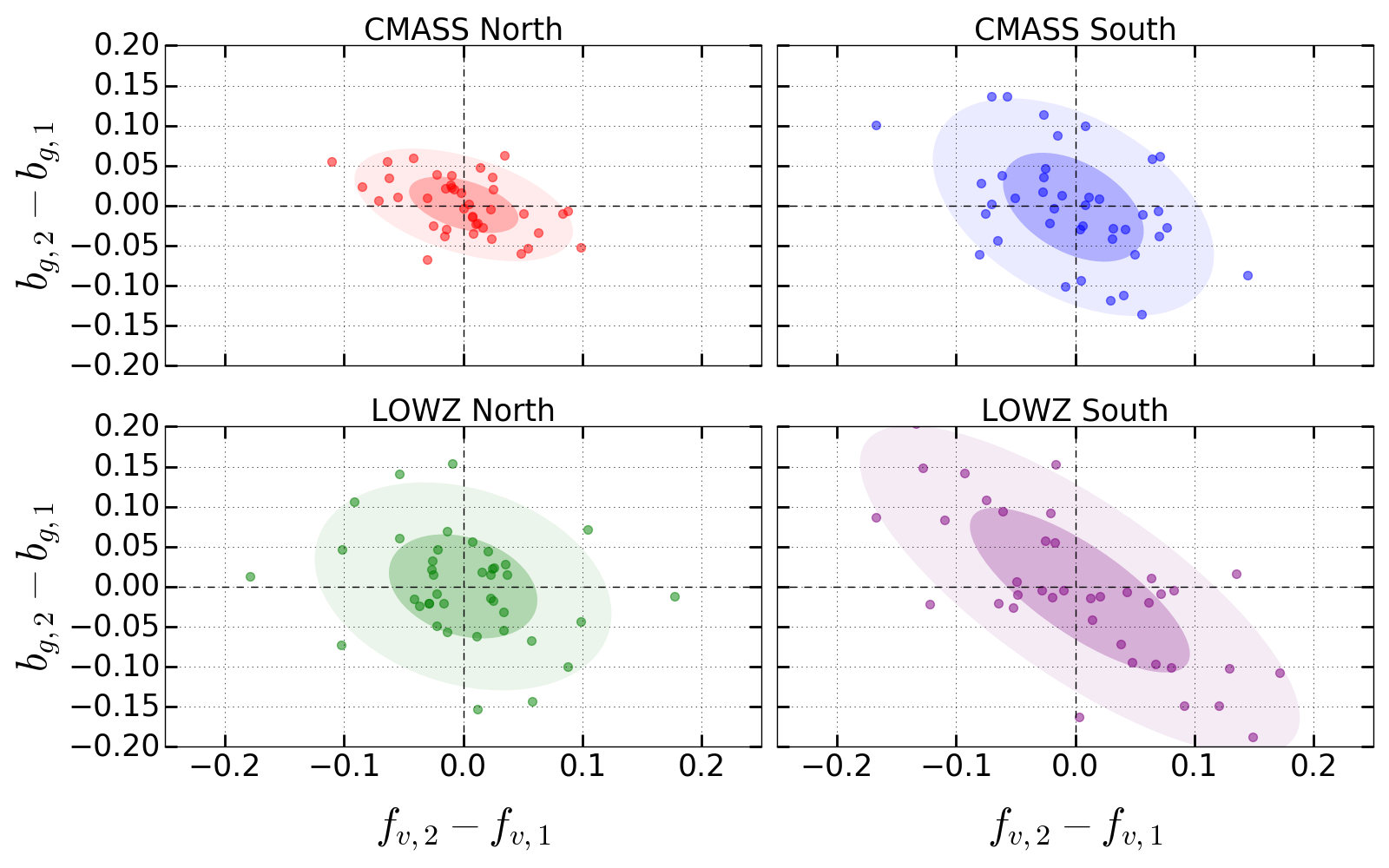}
\includegraphics[width=3in,height=!,keepaspectratio]{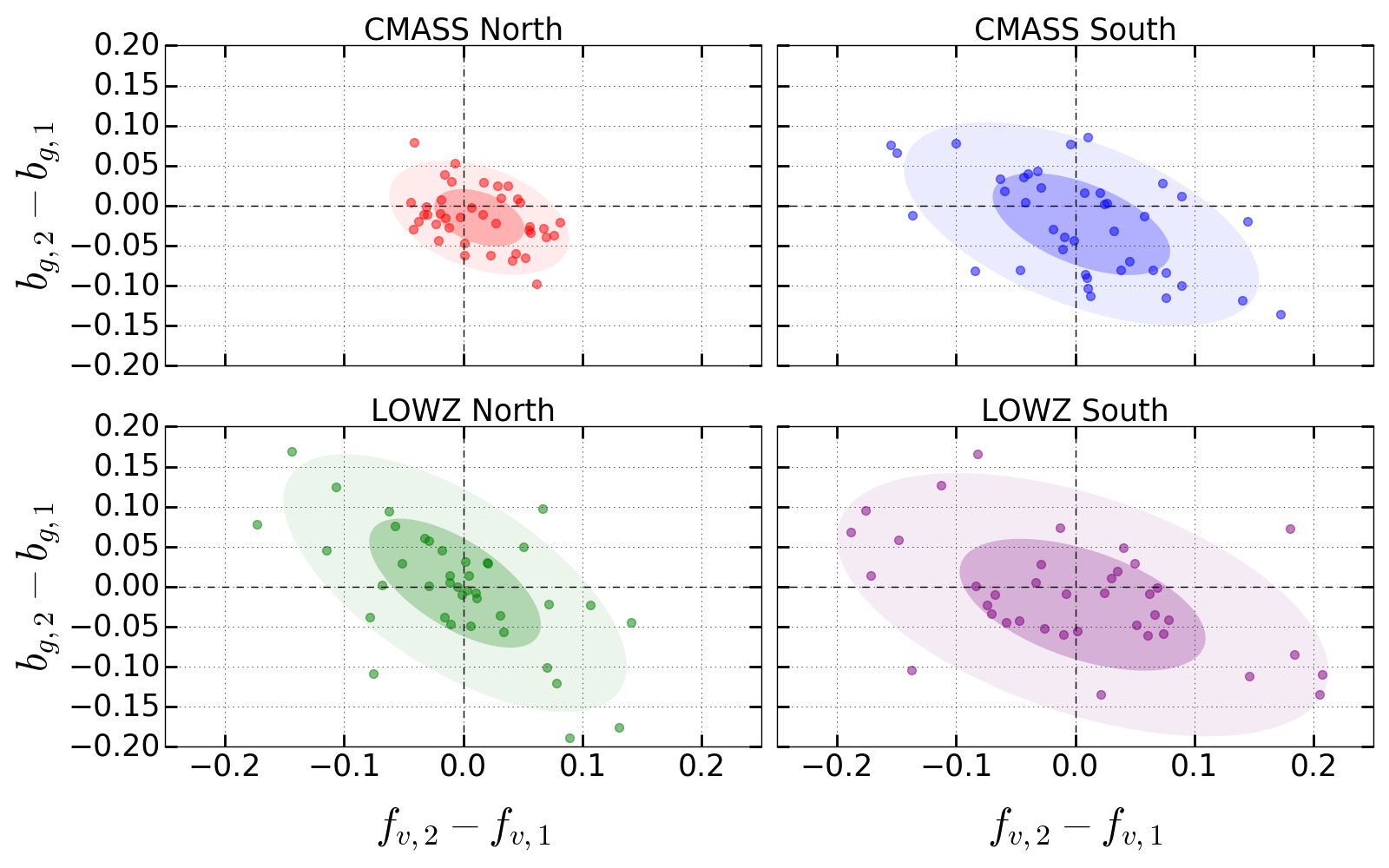}
\caption{For each survey, we show the difference in clustering parameters for each pair of systematics-biased subsamples (i.e.\ with $W_{33}$ scrambled but then each systematic template added in with the best-fit coefficient $m_\theta$). A non-biased result should be consistent with zero, for both parameters. Plotted are the differences in measured bias and measured $f_{v}$ for each survey, where the shaded regions represent 1 and 2 standard deviations. We see here that the systematics due to airmass, extinction, sky flux, and PSF FWHM are consistent with zero.}
\label{fig:_psf_fwhm}
\end{figure*}

Given a specific survey and systematic, we create 40 separations of a positive and negative $W_{33}$ group. For each separation, we calculate the difference in their parameters for bias, $b_{g,-} - b_{g,+}$, and for the growth parameter, $f_{v,-} - f_{v,+}$. In Table \ref{tab:nulltest1} we show the means of these parameter differences for all 40 realizations, as well as the standard deviation of our realizations. In this setting, the mean parameter difference represent the systematic bias potentially added to our signal, while the standard deviations indicate whether or not these biases are consistent with zero. In Fig.~\ref{fig:_psf_fwhm} we visually display all 40 realizations, as well as the mean parameter differences, for the four systematics templates (airmass, extinction, sky flux, and PSF FWHM).

The mean offsets and deviations are also displayed in Table \ref{tab:nulltest1}. It can be seen here that most offsets are consistent with zero. Our largest offsets, within the systematic of the PSF FWHM, reach as large as $-0.022$ for bias and $0.013$ for $f_{v}$ in the LOWZ South sample. When we compare this to the intrinsic uncertainty in the splitting of the LOWZ South sample, at $-0.031$ for bias and $0.089$ for $f_{v}$, we can see that these offsets are well within the standard deviation simply due to random fluctuations. 

\begin{table*}\centering
\ra{1.3}
\caption{The average of parameter differences for our systematic-injected subsamples. Consistency with zero implies that the systematic in question will not disguise itself as galaxy orientation for our sample splitting. Here, the $\sigma_{\Delta f}$ are the uncertainties in the systematic error from each respective systematic, {\em not} the uncertainty in our measurement of $\Delta f_{v}$.}
\label{tab:nulltest1}
\begin{tabular}{@{}rrrcrrrr@{}}\toprule
& \multicolumn{2}{c}{\textbf{Bias}} & \phantom{abc} & \multicolumn{2}{c}{\textbf{Growth}}\\
\cmidrule{2-3} \cmidrule{5-6} 
& $<b_{g,-} - b_{g,+}>$ & $\sigma_{\Delta b}$ && $<f_{v,-} - f_{v,+}>$ & $\sigma_{\Delta f}$ \\ \midrule
\textbf{Extinction}\\
CMASS North & 0.006 & 0.002 && 0.000 & 0.002 \\
CMASS South & 0.016 & 0.007 && 0.004 & 0.006 \\
LOWZ North & -0.002 & 0.004 && 0.007 & 0.003 \\
LOWZ South & 0.02 & 0.008 && -0.007 & 0.007 \\
\textbf{Airmass}\\
CMASS North & 0.000 & 0.002 && 0.000 & 0.002 \\
CMASS South & 0.005 & 0.006 && -0.004 & 0.005 \\
LOWZ North & -0.000 & 0.001 && 0.001 & 0.001 \\
LOWZ South & -0.006 & 0.006 && -0.001 & 0.010 \\
\textbf{Sky Flux}\\
CMASS North & 0.001 & 0.003 && 0.000 & 0.003 \\
CMASS South & -0.002 & 0.008 && -0.002 & 0.006 \\
LOWZ North & 0.000 & 0.002 && -0.000 & 0.002 \\
LOWZ South & -0.004 & 0.006 && 0.004 & 0.005 \\
\textbf{PSF FWHM}\\
CMASS North & -0.015 & 0.007 && 0.013 & 0.006 \\
CMASS South & -0.023 & 0.009 && 0.005 & 0.008 \\
LOWZ North & 0.005 & 0.013 && -0.007 & 0.010 \\
LOWZ South & -0.022 & 0.012 && 0.006 & 0.013 \\
% * <martens.28@osu.edu> 2018-01-27T20:31:00.438Z:
%
% ^.
% * <martens.28@osu.edu> 2018-01-27T20:30:58.683Z:
%
% ^.
\bottomrule
\end{tabular}
\end{table*}

%%%%%%%%%%%%%%%%%%
\subsection{Phase II}
\label{subsec:unblinding1}

\subsubsection{$p(\Delta b_{g},\Delta f_{v})$ Measurement}
At this point, we move into the second stage of the blinding procedure. In this stage, clustering statistics are computed on the {\em true} $W_{33}$-split subsamples, but we blind ourselves to the value of $\Delta f_{v}$, and thus $\Delta A$. The only information from the fits that is revealed in this stage is the parameter $p(\Delta b_{g},\Delta f_{v})$ (which is described in Eq.~\ref{eq:god_param} and contains no dependence on $\Delta A$), and the $\chi^{2}$/dof of the resulting fits.

\begin{figure}
\centering
\includegraphics[width=\columnwidth,height=!,keepaspectratio]{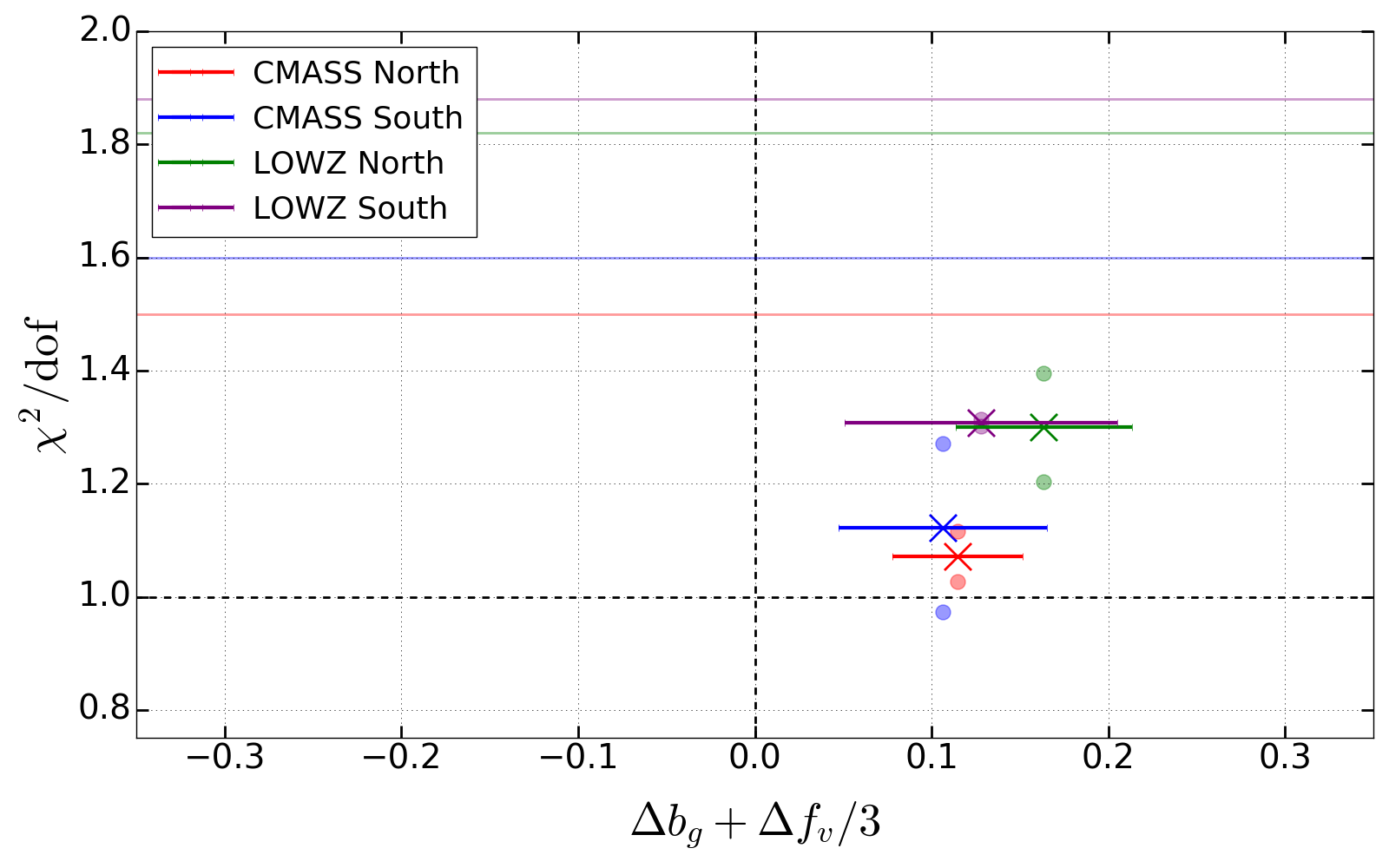}
\caption{Here, we plot $p(\Delta b,\Delta f)$ by the $\chi^{2}$ per degree of freedom of the fits. The solid colored lines indicate the limits of acceptable $\chi^{2}$ for our fits; it can be seen that all of our fits meet this criteria. The 'X' markers indicate the average $\chi^{2}$ for each survey, while the 'O' markers indicate the the $\chi^{2}$ for each subsample. All fits result in $p(\Delta b,\Delta f) < 0.2$, the limit discussed in Sec. \ref{subsec:fvinject}. However, they all show a consistent offset of approximately 0.11 for LOWZ and 0.15 for CMASS.}
\label{fig:Blinded_PLOT}
\end{figure}

In Fig.~\ref{fig:Blinded_PLOT} we display both the $\chi^{2}$/dof of each survey fit, as well as $\Delta b_{g} + \Delta f_{v}/3$. All of the fits have $\chi^{2}$ values that are well within the range of acceptability. However, we do note a consistent positive offset for our blinded parameter, $p(\Delta b,\Delta f)$. Specifically, we find that $p(\Delta b_{g},\Delta f_{v})$ = $0.112 \pm 0.031$ ($0.153 \pm 0.042$) for CMASS (LOWZ). 

It is possible that we are seeing higher clustering in the groups above and below the fundamental plane, which correspond to different measurements of $b_{g}$. As explained in Section \ref{sec:theory}, $W_{33}$ is a proxy for orientation, but contains other information as well, which could lead to a subgroup selection effect that is bias-dependent. For instance, the fundamental plane estimation may have preferentially selected central or satellite galaxies, depending on the subgroup, which could explain a bias difference.
 
The investigation of this bias difference could motivate future work, although it is outside the scope of this paper to explore. A promising route would be Halo Occupation Distribution (HOD) modeling of the subsamples, to identify where the difference arises (including whether it is associated with satellites or centrals). In any case, this result does not change our expected values of $\Delta f_{v}$, and therefore our measurement of intrinsic alignments should remain uncorrupted. As explained in Appendix~\ref{sec:biased_subs}, there could be an effect on the measured size of the error bars, but we estimate this to be below $5\%$ in all realistic scenarios.

\subsubsection{$\sigma_{FOG}^{2}$ Measurement}
As referenced in Section \ref{subsec:FOG}, we are interested in a final test of our subsamples before fully unblinding them. We measured the values of $|\mathcal{I}_{split}|$ (see Eq. \ref{eq:I_split}) for each of our truly divided subsamples. Note that we measure and list here the absolute value of $\mathcal{I}_{split}$ only; we are still blinded to the parameter differences of $f_{v}$ and $b_{g}$, and anything indicating the sign of those values. We find that $|\mathcal{I}_{split}|$ is equal to 0.07001 $\pm$ 0.0257 for LOWZ South, 0.04749 $\pm$ 0.0179 for LOWZ North, 0.01924 $\pm$  0.0143 for CMASS South, and 0.00932 $\pm$ 0.0163 for CMASS North, where the uncertainties are provided by the standard deviations on the random split samples discussed in Section \ref{subsec:FOG}.

It can be seen that the LOWZ values are inconsistent with a measurement of zero, while our CMASS values are consistent with a measurement of zero between 1-2$\sigma$. This is not unexpected; given our results for the values of $p(\Delta b_{g},\Delta f_{v})$ for these subsamples, we expect there to be some level of a potential difference in satellite/central galaxy subscription, which would change the measurement of $\sigma_{FOG}^{2}$ for each sample. However, we want to highlight the magnitude of this discrepancy, which is the important point for our measurement. Our ``nightmare'' scenario discussed a difference of 22 $(Mpc/h)^{2}$ in $\Delta \sigma_{FOG}^{2}$, which corresponds to:
$$
\frac{|\sigma_{FOG,1}^{2}-\sigma_{FOG,2}^{2}|}{\sigma_{FOG,full}^{2}} \approx 2.
$$
where we have assumed that we have one subsample of galaxies with $\sigma_{FOG}^{2} \approx 0$ (no Finger of God) and another with $\sigma_{FOG}^{2} \approx 22 \, (\mathrm{Mpc/h})^{2}$, which when averaged over the full sample appear to be $11 \, (\mathrm{Mpc/h})^{2}$.
From our measurements, we have found:
$$
\frac{|\sigma_{FOG,1}^{2}-\sigma_{FOG,2}^{2}|}{\sigma_{FOG,full}^{2}} \approx 0.07.
$$
This corresponds to a difference in $\sigma_{FOG}^{2}$ of roughly 0.77, which propagates to an uncertainty in $\Delta f_{v}$ of approximately 0.003. This is much less than our statistical uncertainty in $\Delta f_{v}$; in fact, if we are somehow underestimating the measurement of $\sigma_{FOG}^{2}$ with this exercise by a factor of 10, we still have a change in $\Delta f_{v}$ less than our statistical errors in the worst case listed above. 

As discussed in Sec. \ref{subsec:FOG}, we fit our true samples for $f_{v}$, $b_{g}$, and $\sigma_{FOG}^{2}$ on scales of 20--40 Mpc. It can be seen in Table \ref{tab:sigma_FOG} that in all cases, our resultant dispersions are consistent with zero given the errors provided by the random fits. Furthermore, we derive the propagated error on $\Delta f_{v}$, and determine that in all cases this error is less than our statistical error on $f_{v}$.

These two tests (measuring $|\mathcal{I}_{split}|$ and measuring $\sigma_{FOG}^{2}$ on small scales) complement each other in determining the effects of setting the value for $\sigma^{2}_{FOG}$. When testing the value of $|\mathcal{I}_{split}|$ on the subsamples, we have a physical interpretation of what we are measuring, with small enough statistical errors that we have a detection of the differences of this measurement between the subsamples. This difference is at its largest value for LOWZ South, and propagates to an error in $\Delta f_{v}$ of 0.003. However, this measurement specifically probes small scales, and it could be the case that our model is collecting larger-scale effects into its parameter value for $\sigma_{FOG}^{2}$. This issue is resolved by our second test, which is a direct measurement of $\sigma_{FOG}^{2}$ and therefore picks up any and all quasilinear effects that may be contained in the $\sigma^{2}_{FOG}$ parameter. The second test, in the worst case, resulted in $\Delta \sigma^{2}_{FOG}$ of 7.14 $(Mpc/h)^{2}$ for CMASS North, which is consistent with zero, as the standard deviation was 7.56 $(Mpc/h)^{2}$ for our random separations. Given the results from the first test (a propagated error on $\Delta f_{v}$ much smaller than statistical errors) and the second test (a difference in $\sigma_{FOG}^{2}$ that is consistent with zero) we have determined that setting the value of the Finger of God when measuring $f_{v}$ and $b_{g}$ will not significantly bias our results for these parameters. 
\begin{table*}
\caption{Errors on $f_{v}$ due to differences in $\sigma_{FOG}^{2}$ for the true divided subsamples.} % title of Table
\centering % used for centering table
\begin{tabular}{c c c c c} % centered columns (4 columns)
\hline\hline %inserts double horizontal lines
Category & CMASS NGC & CMASS SGC & LOWZ NGC & LOWZ SGC \\ [0.5ex] % inserts table
%heading
\hline % inserts single horizontal line
$|\Delta \sigma_{FOG}^{2}| \left[ Mpc/h\right]^{2}$ & 7.14 & 4.30 & 6.32 & 3.58 \\ % inserting body of the table
Statistical error on $\Delta \sigma_{FOG}^{2}$  & 7.56 & 11.00 & 8.40 & 10.41 \\
Propagated Error to $\Delta f_{v}$ & 0.026 & 0.016 & 0.023 & 0.013 \\
Statistical Error on $\Delta f_{v}$ & 0.040 & 0.063 & 0.059 & 0.089 \\
Propagated Error / Statistical Error & 0.65 & 0.25 & 0.39 & 0.15 \\[1ex] % [1ex] adds vertical space
\hline %inserts single line
\end{tabular}
\label{tab:sigma_FOG} % is used to refer this table in the text
\end{table*}

%%%%%%%%%%%%%%%%%%
%%%%%%%%%%%%%%%%%%
%%%%%%%%%%%%%%%%%%
\subsection{Phase III: Final results}
\label{subsec:finalresults}
\begin{figure}
\centering
\includegraphics[width=\columnwidth,height=!,keepaspectratio]{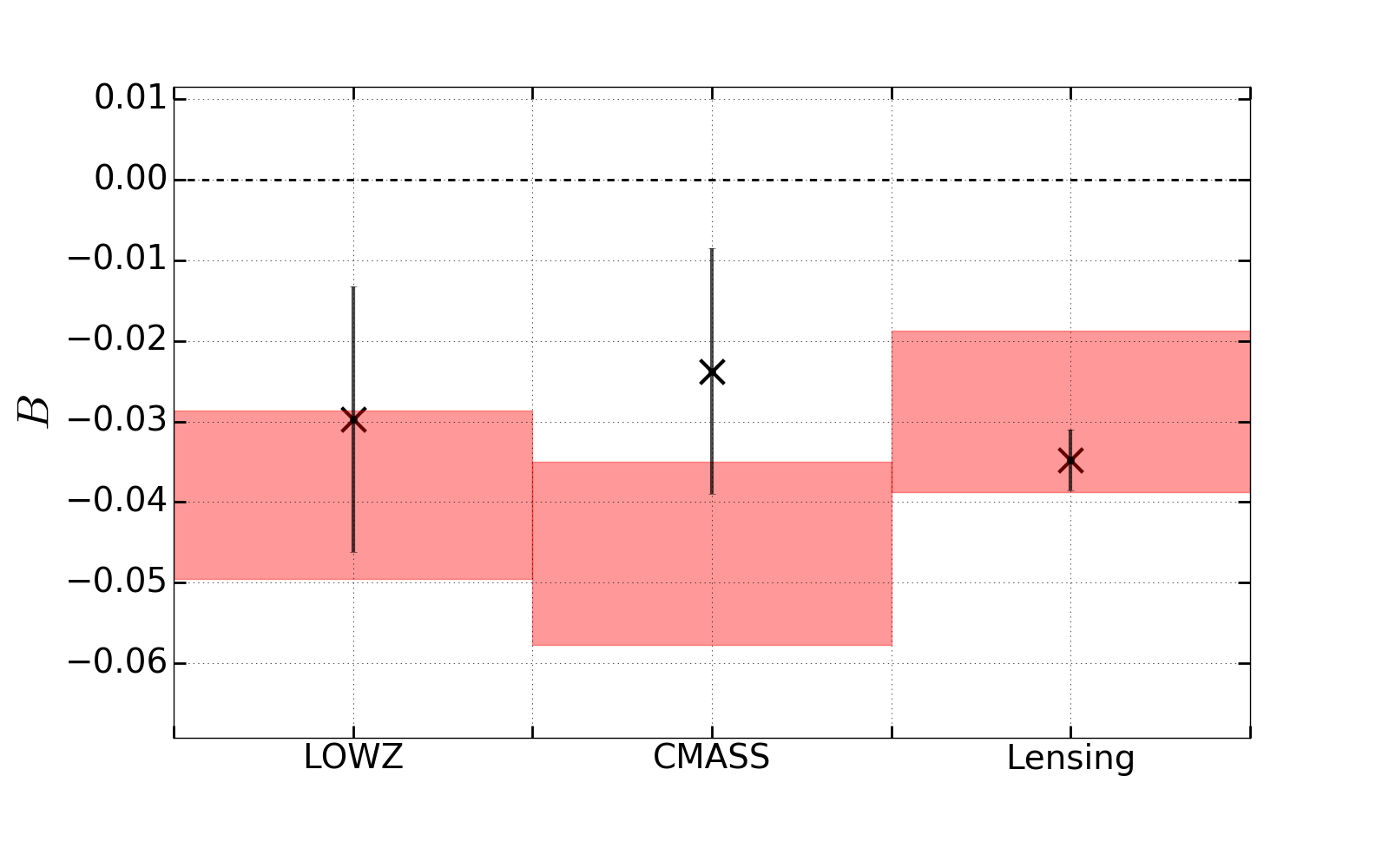}
\caption{Here, we display our final results for the value of B along with theoretical predictions. The data point from \protect\cite{singh2015} was measured from the LOWZ sample using galaxy-galaxy lensing, and is labeled `Lensing.' For each measurement we also show the theoretical prediction of a radial intrinsic alignment measurement at the calculated median luminosities of each respective sample, using the machinery of \protect\cite{hirata2009}. These predictions are shown as red ranges on the plot, overlapping the true measurement with the same sample luminosity. }
\label{fig:FINALPLOT3}
\end{figure}

Throughout this research project, we have been careful to notate and adhere to our specific blinding procedures (Section \ref{sec:systematicstesting}). To be clear, all parts of this paper were written {\em before} unblinding ourselves to the final results of our parameter fits for the true subsamples divided by orientation, {\em except} for the specific parts/changes listed below:
\begin{itemize}
  \item Errors in the plotting scripts for Figures~\ref{fig:ashley_param_comp} and \ref{fig:FINALPLOT3}.
  \item Minor changes to the historical discussion in the introduction.
  \item A typo was fixed in Eq.~(\ref{eq:corrected}).
  \item The paper abstract.
  \item The paper conclusion.
  \item This section of the paper.
  \item The acknowledgements.
  \item Grammatical corrections (e.g.\ punctuation).
\end{itemize}

\begin{figure}
\centering
\includegraphics[width=\columnwidth,height=!,keepaspectratio]{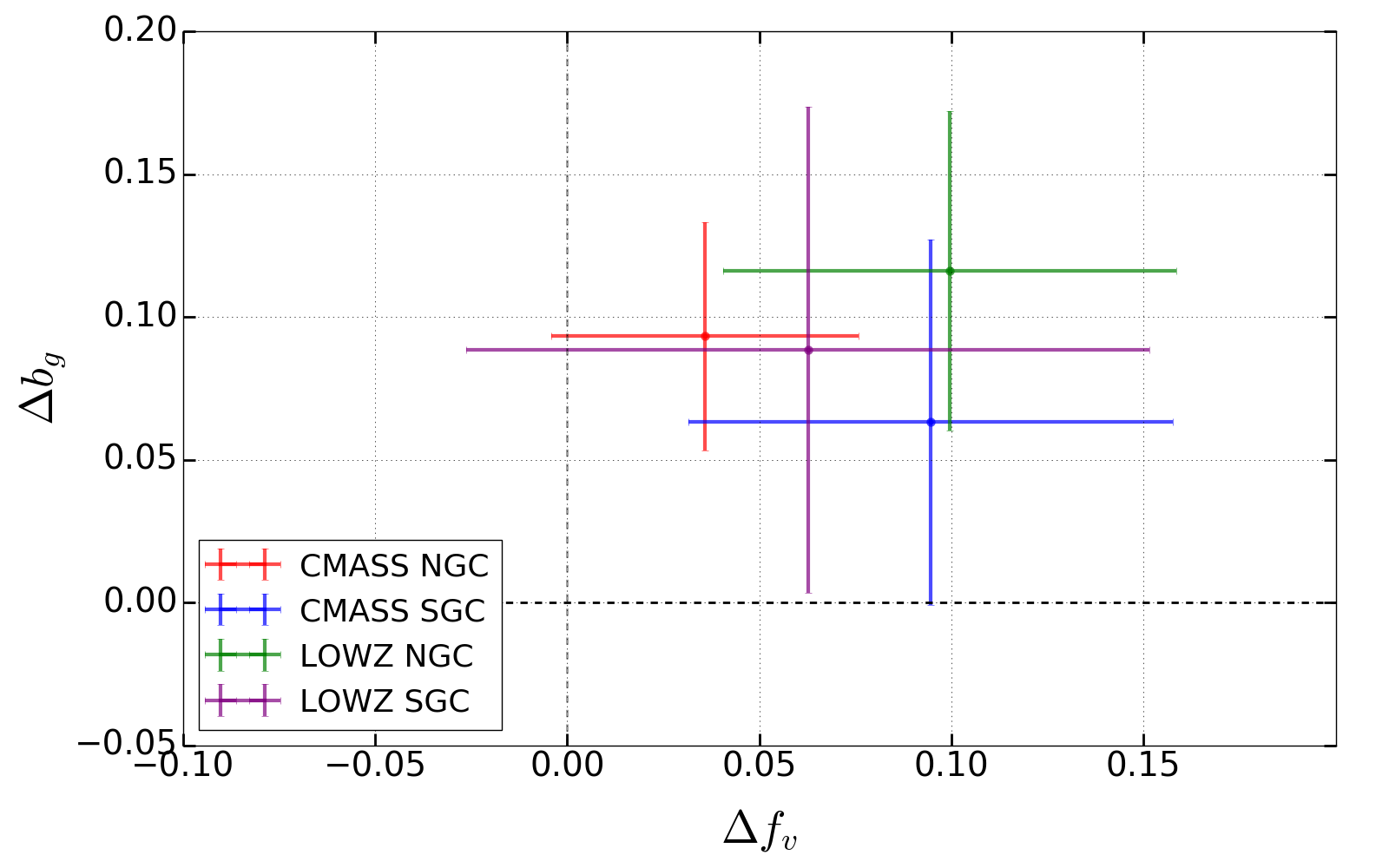}
\caption{For each subsample split, we show the values of $\Delta f_{v}$ and $\Delta b_{g}$.}
\label{fig:FINALPLOT}
\end{figure}

Now we discuss the final unblinded results of our split samples. In Fig. \ref{fig:FINALPLOT}, we show the values of $\Delta f_{v}$ and $\Delta b_{g}$ for the separate subsamples. We see a consistent offset in both the bias parameter (which was expected given our results in Sec. \ref{subsec:unblinding1}) and in the growth parameter, which was the desired result. We display the final result for $B$ in Fig. \ref{fig:FINALPLOT3}, and compare it to the theoretical predictions by \cite{hirata2009}. 

Our final results for the intrinsic alignment values for $B$ are $-0.016 \pm 0.018$ for CMASS North, $-0.043 \pm 0.029$ for CMASS South, $-0.033 \pm 0.019$ for LOWZ North, and $-0.022 \pm 0.031$ for LOWZ South. It can be seen that each individual result for the SGC and NGC groups are consistent with each other, as they are less than 3$\sigma$ apart. We combined our results into a total estimate of $B$ for the LOWZ and CMASS groups individually, and as a total value for both, and compared them to theory, as described in Eq. \ref{eq:theory_frac}. We find an Obs/Theory ratio of $0.51 \pm 0.33$ for CMASS, $0.76 \pm 0.42$ for LOWZ, and $0.61 \pm 0.26$ for the total sample. In all cases, we see that this ratio is in between 1--2$\sigma$ from the value of 1, and we conclude that our results are consistent with the theoretical expectations. Finally, our combined results for $B$ are $-0.024 \pm 0.015$ for CMASS, and $-0.030 \pm 0.016$ for LOWZ, with a total combined signal of $-0.026 \pm 0.011$, which meets our criteria as evidence for the intrinsic alignment magnitude. 

%%
% \begin{table*}
% \caption{Final Results} % title of Table
% \centering % used for centering table
% \begin{tabular}{c c c c c} % centered columns (4 columns)
% \hline\hline %inserts double horizontal lines
% Category & CMASS & LOWZ & Lensing (LOWZ) \\ [0.5ex] % inserts table
% %heading
% \hline % inserts single horizontal line
% $B$ & ??? & ??? & -0.0348  \\ % inserting body of the table
% $\sigma_{B}$  &  & 11.00 & 8.40  \\
% $L/L_{0}$ & 0.026 & 0.016 & 0.023  \\[1ex] % [1ex] adds vertical space
% \hline %inserts single line
% \end{tabular}
% \label{tab:FINAL_RESULTS} % is used to refer this table in the text
% \caption{CAPTION HERE.}
% \end{table*}
%%

%%%%%%%%%%%%%%%%%%%%%%%%%%%%%%%%%%%%%%%%%%%%%%%%%%%%%%%%%%%%%%%%%%%%%%%%%%%%%%%%%%%%%%%%%%%%%%
%%%%%%%%%%%%%%%%%%%%%%%%%%%%%%%%%%%%%%%%%%%%%%%%%%%%%%%%%%%%%%%%%%%%%%%%%%%%%%%%%%%%%%%%%%%%%%
\section{Conclusion}
\label{sec:conclusion}
In this work, we have used the Fundamental Plane to divide galaxies from the CMASS and LOWZ catalogs within SDSS BOSS DR12 into groups depending on their orientation seen by an observer on Earth. By measuring the differences in the clustering parameter $f_{v}$ of these subsamples, we have attempted to quantify the intrinsic alignment magnitude by measuring the dimensionless parameter $B$. We found that for CMASS (LOWZ), the measured $B$ was $-0.024 \pm 0.015$ ($-0.030 \pm 0.016$). We take their ratio relative to our theoretical prediction and combine the results, based on Eq.~(\ref{eq:theory_frac}). We find ${ \rm\ Obs/Theory} = 0.61\pm 0.26$; this constitutes evidence (between 2 and 3$\sigma$) for a measurement of the radial intrinsic alignments, and is consistent with expectations ($<2\sigma$ difference).

Our result gives additional motivation for searches of similar intrinsic alignment effects in samples other than LRGs, such as Emission Line Galaxies (ELGs) or Lyman-Alpha Emitters (LAEs). It is important to note that in those samples, the analysis could become more complicated due to radiative transfer effects influencing the relationship between $\Delta f_{v}$ and the ellipticity of the galaxies.

Given our result, future RSD measurements should aim to minimize orientation-dependent biases in target selection to the extent possible, and understand the possible impact of intrinsic alignments on the full analysis chain.
%Specifically, null tests could be incorporated to check against such ``orientation bias'' by looking at the values of the psf FWHM, as we did for one of our systematic checks.
Even ``null tests'' in which the clustering statistics from various sub-samples are compared can be affected, if the method of splitting into sub-samples is correlated with galaxy orientation. While in this paper such a split was done deliberately to probe intrinsic alignments, in a cosmological parameter analysis using RSDs it is possible that an orientation-dependent split could arise unintentionally, and thus cause a null test to ``fail'' even if selection effects in the full parent sample are well understood.

Several choices were made in our analysis that could be improved for future studies. The scatter in the Fundamental Plane could be reduced by incorporating velocity dispersion information, although the modeling of orientation-dependent effects would be more complex. Multiple bins could be analyzed instead of the two-bin orientation system implemented here, which could improve the aggregate signal-to-noise ratio. Our work could be extended to future surveys: for example the DESI experiment will produce an LRG sample that is 3 times larger than the BOSS sample used in this paper \citep{2016arXiv161100036D}. Increases in both the sample size and the redshift range of target galaxies should allow for more robust measurements of the clustering parameters, and decrease error bar sizes in the resulting measurements of $B$. Using radial methods to measure intrinsic alignments can complement the traditional ellipticity-based approach, in order to gain a better understanding of the behaviour of intrinsic alignments.

% Potential things we could see, and what we would think they mean:
% if non-detection: 
% The assumptions in \ref{subsec:theoretical_expectations} could be affecting our results, including our splitting process and dampening the resulting difference in parameters. Furthermore, the alignments could be more complicated than our current model is prepared for, and so we are not adequately measuring the galaxy orientations. 

% if detection larger than theory:
% This would merit a follow-up investigation of our results, to verify the magnitude. Specifically, using other surveys/subsample divisions. Effects on RSD may need to be revisited.

% tension between north and south:
% Potentially could be related to how reliable the SGC data is, may not be the exact same sample of galaxies. Recommend possibly treating NGC and SGC more differently in that case, with potentially different models/splitting criteria, etc.

% Note that we found a bias difference between the subsamples, and what it is, and that it could be good for future work. 
%%%%%%%%%%%%%%%%%%%%%%%%%%%%%%%%%%%%%%%%%%%%%%%%%%%%%%%%%%%%%%%%%%%%%%%%%%%%%%%%%%%%%%%%%%%%%%
\section*{Acknowledgements}
% Entry for the table of contents, for this guide only
\addcontentsline{toc}{section}{Acknowledgements}
We appreciate the input and helpful conversation from Paulo Montero-Camacho, Niall MacCrann, Ami Choi, Jenna Freudenburg, and Michael Troxel. We especially thank Eric Huff for his role in formulating the idea for this project.

DM, XF, and CMH are supported by the Simons Foundation, the US Department of Energy, the Packard Foundation, the NSF, and NASA.

AJR thanks CCAPP for their support.

Many computations in this paper were run on the CCAPP condo of the Ruby Cluster at the Ohio Supercomputer Center \citep{OhioSupercomputerCenter1987}.

Funding for SDSS-III has been provided by the Alfred P. Sloan Foundation, the Participating Institutions, the National Science Foundation, and the U.S. Department of Energy Office of Science. The SDSS-III web site is http://www.sdss3.org/.

SDSS-III is managed by the Astrophysical Research Consortium for the Participating Institutions of the SDSS-III Collaboration including the University of Arizona, the Brazilian Participation Group, Brookhaven National Laboratory, Carnegie Mellon University, University of Florida, the French Participation Group, the German Participation Group, Harvard University, the Instituto de Astrofisica de Canarias, the Michigan State/Notre Dame/JINA Participation Group, Johns Hopkins University, Lawrence Berkeley National Laboratory, Max Planck Institute for Astrophysics, Max Planck Institute for Extraterrestrial Physics, New Mexico State University, New York University, Ohio State University, Pennsylvania State University, University of Portsmouth, Princeton University, the Spanish Participation Group, University of Tokyo, University of Utah, Vanderbilt University, University of Virginia, University of Washington, and Yale University.


\bibliographystyle{mnras}
\bibliography{sources} % if your bibtex file is called example.bib

%%%%%%%%%%%%%%%%%%%%%%%%%%%%%%%%%%%%%%%%%%%%%%%%%%
%%%%%%%%%%%%%%%%% APPENDICES %%%%%%%%%%%%%%%%%%%%%
\appendix

%%%%%%%%%%%%%%%%%%%%%%%%%%%%%%%%%%%%%%%%%%%%%%%%%%
\section{Effect of a difference in bias on the random-split error bars}
\label{sec:biased_subs}

In Section \ref{sec:analysis} we used random splits of the BOSS samples to determine the error bars on $\Delta b_g$ and $\Delta f_v$. This procedure assumes that the two subsamples have the same clustering amplitude. A possible concern is that if the clustering amplitudes are different, then the error bars on $\Delta f_v$ may have additional cosmic variance contributions and hence may be larger than we calculate based on the random-split method. The purpose of this appendix is to investigate this possibility.

When we separate our sample into two groups based on orientation, we find the correlation functions of, and measure clustering parameters for, each of the subgroups. Since both subgroups are pulled from the same sample, an excess of high-biased galaxies in one subgroup would imply a lack of high-biased in the other, and we would see a difference in the bias parameters measured between the samples, $\Delta b_{g}$. Since the sample measurements are correlated, we must use the covariance matrix of density fluctuations between the two subsamples in order to gain an understanding of the effects on the variance of our measurements.

We consider the issue of whether error bars are increased in Fourier space; the result could be transformed to real space as done in \S\ref{subsec:covariance} if it turned out to be important. For a given density mode ${\bmath k}$, and a difference in bias of the samples $\Delta b$, then the covariance matrix of errors in the bias measurement between those samples can be written as:
\begin{equation}
{\mathbfss C} =
\left(\begin{array}{clcr}
(1 + \frac{\Delta b}{2b})^{2}P(k) + \frac{1}{\bar{n}} & (1 + \frac{\Delta b}{2b})(1 - \frac{\Delta b}{2b})P(k) \\
(1 + \frac{\Delta b}{2b})(1 - \frac{\Delta b}{2b})P(k) & (1 - \frac{\Delta b}{2b})^{2}P(k) + \frac{1}{\bar{n}}   \end{array}\right),
\end{equation}
where $\bar{n}$ is the number density within each sample, assuming they are the same, $P(k)$ is the power spectrum at wavenumber $k$, and $\Delta b$ is the difference in the measured bias of the two samples. The variance in the difference of our estimate of the biases of the two parameters can be written as:
\begin{equation}
\begin{split}
& {\rm Var}(\hat{C}_{11} - \hat{C}_{22}) = {\rm Var}(\hat{C}_{11}) + {\rm Var}(\hat{C}_{22}) - 2 {\rm Cov}(\hat{C}_{11},\hat{C}_{22}) \\
& = 2C_{11}^{2} + 2C_{22}^{2} - 4C_{12}^{2}.
\end{split}
\end{equation}
For two samples with the same bias, we can see that this variance is dependent only upon the shot noise and power in our sample:
\begin{equation}
{\rm Var}(\hat{C}_{11} - \hat{C}_{22})(\Delta b = 0) = \frac{16}{\bar{n}}P(k) + \frac{16}{\bar{n}^{2}}.
\end{equation}
However, for a general difference in bias, we find:
\begin{equation}
{\rm Var}(\hat{C}_{11} - \hat{C}_{22}) = \frac{16}{\bar{n}^{2}} + 8 P^{2}(k)\frac{(\Delta b)^{2}}{b^{2}} + \frac{8}{\bar{n}}P(k)\left[ 2 + \frac{(\Delta b)^{2}}{2b^{2}}\right].
\end{equation}
An effective way to analyze the increase in the size of our error is to look at the ratio between this variance difference with a non-zero bias difference, to that with a bias difference of zero:
\begin{equation} \label{eq:estimation_app}
\frac{{\rm Var}(\hat{C}_{11} - \hat{C}_{22})(\Delta b = 0)}{{\rm Var}(\hat{C}_{11} - \hat{C}_{22})} = 1 + \frac{1}{4}\frac{\bar{n} P(k) (1 + 2 \bar{n}P(k)}{1 + \bar{n}P(k)} \left(\frac{\Delta b}{b}\right)^{2}.
\end{equation}
With this, we can see how our error bars are affected by a given difference in bias of the two samples. For the case relevant to our interests, we state in Sec. \ref{subsec:fvinject} that a change in $\Delta b$ of 0.2 could have an effect on the size of our error bars, and indicate further investigation. Given that for our samples, $b \approx 2$, then this offset would mean $\Delta b/b \approx 0.1$. From Eq. \ref{eq:estimation_app} we can then see that, for values of $\bar{n}P(k) \approx 1$, the modification to the error bar would be $\approx 0.4 \%$. Even given an extreme case where $\bar{n}P(k) \approx 7$, we would see an increase in error of only about $3.3 \%$. 

%%%%%%%%%%%%%%%%%%%%%%%%%%%%%%%%%%%%%%%%%%%%%%%%%%
%%%%%%%%%%%%%%%%%%%%%%%%%%%%%%%%%%%%%%%%%%%%%%%%%%
\section{Comparison of Intrinsic Alignment Values}
\label{sec:theory_comp}
In Section \ref{sec:theory}, we introduced Eq.~(\ref{eq:theory_comp_eqn}) to explain how to compare our intrinsic alignment values to those from \cite{singh2015}. Here, we go through the derivation to that equation. We need to relate Eq.~(53) of \cite{hirata2009} to Eq.~(9) of \cite{singh2015}. The first of those equations of the 2D correlation function is:
\begin{equation}
\label{eq:singh}
\begin{split}
 w_{g+}(r_{p}) =& \frac{A_{I}b_{D}C_{1}\rho_{\rm crit}\Omega_{m}}{\pi^{2}}\int {\rm d}z \frac{W(z)}{D(z)} \int_{0}^{\infty}{\rm d}k_{z} \\
& \times \int_{-\infty}^{\infty}{\rm d} k_{\perp}\frac{k_{\perp}^{3}}{(k_{\perp}^{2}+k_{z}^{2})k_{z}}P(k,z)\mathrm{sin}(k_{z}\Pi_{\mathrm{max}})
\\ & \times
J_{2}(k_{\perp}r_{p})(1+\beta \mu^{2}),
\end{split}
\end{equation}
where $A_{I}$ is the intrinsic alignment measurement, $b_{D}$ is the galaxy bias, $C_{1}\rho_{\mathrm{crit}}$ is fixed at a value of 0.0134, $\Pi_{max}$ is the line of sight separation, $\Omega_{m}$ is the matter density, $D(z)$ is the growth function (G(z) in our notation), and $k_{\perp}$ and $k_{z}$ are the wavenumbers perpendicular and parallel to the line of sight, respectively. $\beta$ is the ratio of $f_{v}/b_{g}$, and $\mu$ is the cosine of the angle between a given direction and the line of sight. Overall, this function $w_{g+}(r_{p})$ is the 2D projected correlation function between the galaxy density field and the galaxy intrinsic shear. $W(z)$ is defined as:
\begin{equation}
W(z) = \frac{p_{A}(z)p_{B}(z)}{\chi^{2}(z)d\chi dz} \left[ \int \frac{p_{A}(z)p_{B}(z)}{\chi^{2}(z)d\chi dz} dz  \right]^{-1},
\end{equation}
where $p_{A}(z)$ and $p_{B}(z)$ are the redshift probability distributions for shape and density sample, respectively, and $\chi(z)$ is the comoving distance to a given redshift $z$. We wish to compare Eq.~(\ref{eq:singh}) to:
\begin{equation} \label{eq:hirata}
w_{\delta +}(r_{p}) = -\frac{b_{\kappa}}{2\pi}\int_{0}^{\infty} P_{m}(k)J_{2}(k r_{p})k \,{\rm d}k,
\end{equation}
where $w_{\delta +}(r_{p})$ is the 2d projected correlation function between matter density and galaxy intrinsic shear, $P_{m}(k)$ is the matter power spectrum at wavenumber $k$, and $b_{\kappa}$ is the measurement of intrinsic alignments. The notational differences of "$g$" and "$\delta$" are due to Eq.~(\ref{eq:hirata}) citing the correlations with the matter power spectrum, which means we must set $b_{D}=1$ when setting the equations equal. Next, we will take the limit of large line of sight distance ($\Pi_{max}$). This results in a quickly oscillating sine function within \ref{eq:singh}, so that we can simplify (using the residue theorem) the integral over $k_z$:
\begin{equation} \label{eq:corrected}
\int_{0}^{\infty}{\rm d}k_{z}\,\frac{k_{\perp}^{3}}{(k_{\perp}^{2}+k_{z}^{2})k_{z}}\mathrm{sin}(k_{z}\Pi_{max}) \rightarrow \frac\pi2k_\perp.
\end{equation}
Next, we know that the window function $W(z)$ integrates to 1. Therefore, Eq.~(\ref{eq:singh}) becomes
\begin{equation}
w_{g+}(r_{p}) = \frac{A_{I}b_{D}C_{1}\rho_{\rm crit}\Omega_{m}}{2 \pi}\frac{1}{D(z)}\int_{0}^{\infty}{\rm d}k\, k P(k,z)J_{2}(k r_{p})
\end{equation}
when considering a narrow redshift range.
By comparing this to Eq.~(\ref{eq:hirata}), we can see that:
\begin{equation}
\frac{A_{I}C_{1}\rho_{\mathrm{crit}}\Omega_{m}}{D(z)} = -b_{\kappa}.
\end{equation}
By inserting the definition of $C_{1}\rho_{\mathrm{crit}}$, changing to our notation of the growth function, and using Eq.~(\ref{eq:bkappa}), we arrive at Eq.~(\ref{eq:theory_comp_eqn}): 
\begin{equation}
B = -0.0233\frac{\Omega_{m}}{G(z)}A_{I}.
\end{equation}
\bsp	% typesetting comment
\label{lastpage}
\end{document}